\documentclass[onefignum,onetabnum]{siamonline220329}

\crefname{section}{Section}{Sections}
\crefname{subsection}{Subsection}{Subsections}

 \usepackage{booktabs}
 \usepackage{multirow}

\usepackage{bm}
\usepackage{dsfont}
\usepackage{subcaption}
\usepackage{xfrac}

\newcommand{\IGNORE}[1]{}   % use to remove comments

\newcommand{\newtext}[1]{{#1}}

\DeclareMathOperator*{\argmin}{argmin}

\usepackage{lipsum}
\usepackage{amsfonts}
\usepackage{graphicx}
\usepackage{epstopdf}
\usepackage{algorithmic}
\ifpdf
\DeclareGraphicsExtensions{.eps,.pdf,.png,.jpg}
\else
\DeclareGraphicsExtensions{.eps}
\fi

\usepackage{enumitem}
\setlist[enumerate]{leftmargin=.5in}
\setlist[itemize]{leftmargin=.5in}

\newsiamremark{remark}{Remark}
\newsiamremark{hypothesis}{Hypothesis}
\crefname{hypothesis}{Hypothesis}{Hypotheses}
\newsiamthm{claim}{Claim}

\newcommand{\centered}[1]{\setlength{\tabcolsep}{0mm}\begin{tabular}{l} #1 
\end{tabular}}

\headers{Efficient Parallel Data Optimization for 4K Image Inpainting}{N. K\"amper, V. 
Chizhov, and J. Weickert}

\title{Efficient Parallel Data Optimization for \\ Homogeneous Diffusion Inpainting of 4K Images
	\thanks{
		\funding{This work has received funding from the European 
			Research Council (ERC) under the European Union's Horizon 2020 
			research and innovation programme (grant agreement no. 741215, ERC 
			Advanced Grant IN\-CO\-VID)}}}

\author{Niklas K\"amper \thanks{Mathematical Image Analysis Group, Faculty of 
Mathematics and Computer Science, Saarland University, 66041 Saarbr\"ucken, 
Germany (\email{kaemper@mia.uni-saarland.de}, 
\email{chizhov@mia.uni-saarland.de}, \email{weickert@mia.uni-saarland.de})}
	\and Vassillen Chizhov\footnotemark[2]
	\and Joachim Weickert\footnotemark[2]}

\usepackage{amsopn}

\begin{document}

\maketitle              % typeset the header of the contribution

% REQUIRED
\begin{abstract}
Homogeneous diffusion inpainting can reconstruct missing image areas with 
high quality from a sparse subset of known pixels, provided that their 
location as well as their gray or color values are well optimized. 
This property is exploited in inpainting-based image compression, which is 
a promising alternative to classical transform-based codecs such as JPEG and 
JPEG2000.
However, optimizing the inpainting data is a challenging task. 
Current approaches are either fairly slow or do not produce high quality  
results. 
As a remedy we propose fast spatial and tonal optimization algorithms 
for homogeneous diffusion inpainting that efficiently utilize GPU parallelism,
with a careful adaptation of some of the most successful numerical concepts.
We propose a densification strategy using ideas from error-map dithering 
combined with a Delaunay triangulation for the spatial optimization.
For the tonal optimization we design a domain decomposition solver that 
solves the corresponding normal equations in a matrix-free fashion and 
supplement it with a Voronoi-based initialization strategy.
With our proposed methods we are able to generate high quality inpainting 
masks for homogeneous diffusion and optimized tonal values in a runtime that 
outperforms prior state-of-the-art by a wide margin.
\end{abstract}

% REQUIRED

% REQUIRED

%%%%%%%%%%%%%%%%%%%%%%%%%%%%%%%%%%%%%%%%%%%%%%%%%%%%%%%%%%%%%%%%%%%%%%%%%%%%%%
\section{Introduction}
\label{sec:intro}

Inpainting-based image compression methods 
\cite{AG94,BHK10,Ca88,DMMH96,GWWB08,GLG12,JS23,LSWL07,PHNH16,RPSH20,
SPME14,SCSA04,WZSG09,XSW10,ZD11, GL14} can be promising alternatives to 
classical transform-based codecs. 
During encoding, they store a small subset of pixels. In the 
decoding phase, the missing parts of the image are reconstructed 
from the stored pixels by an inpainting operator. 
Earlier approaches~\cite{AG94,BHK10,Ca88,DMMH96,GLG12,LSWL07,ZD11} mostly 
rely on image data at semantic image features such as edge locations. 
Many later approaches~\cite{GWWB08,HMWP13,PHNH16,SPME14} 
instead carefully optimize the locations of the known data, without any 
restrictions to semantic image features. 
In terms of quality, inpainting-based compression methods have been shown 
to be able to outperform the widely used JPEG~\cite{PM92} and 
JPEG2000~\cite{TM02} for images with small to medium amounts of 
texture~\cite{PHNH16,SPME14}.

Surprisingly, already with simple linear homogeneous diffusion~\cite{Ii62} 
one can obtain very good inpainting 
results~\cite{BBBW08,BLPP17,BHR21,CRP14,HSW13,MHWT12} under the condition 
that the inpainting data are highly optimized.
For piecewise constant images, such as cartoon images, depth maps or flow 
fields, one can even achieve state-of-the-art 
results~\cite{GLG12,HMWP13,LSJO12,MBWF11} and outperform HEVC~\cite{SOHW12}. 
In contrast to nonlinear inpainting methods, homogeneous diffusion inpainting 
is parameter-free and its discretization yields a sparse system that benefits 
from its linearity.
In~\cite{KCW23} it has been shown that we can solve 
this linear system in real-time for 4K color images.

Achieving good reconstruction quality with homogeneous diffusion inpainting 
requires, however, a careful optimization of the inpainting data that we 
store in the encoding phase. 
This consists of selecting the subset of pixels to be stored 
(i.e.\ \textit{the inpainting mask}) and optimizing the gray or color values 
at the stored pixels (i.e.\ \textit{the tonal data}). Usually, these 
optimizations are computationally much more expensive than the inpainting 
itself, and even though the encoding phase often does not have to be 
performed in real-time, it is still desirable to have highly efficient 
methods. 

Most approaches for spatial optimization either focus on high quality 
masks~\cite{CRP14,HSW13,MHWT12} at a high computational cost, or lower 
quality masks~\cite{BBBW08} at a much lower cost. In practice one typically 
does not aim at either extreme, and we are instead interested in methods that 
can achieve good quality in reasonable time~\cite{CW21,DAW21,KBPW18,SPKW23}. 
Our GPU-based implementation in the present work is orders of magnitude 
faster than the spatial optimization in~\cite{CW21,DAW21,KBPW18}, which makes 
it a practical approach for mask optimization of 4K images. 
A GPU-based algorithm also targeted at 4K 
spatial optimization, and comparable in terms of efficiency, is the recent
neural approach by Schrader et al.~\cite{SPKW23}. We outperform the latter 
in terms of quality and speed, while having much lower memory requirements. 

For the tonal optimization problem all methods can theoretically achieve the 
same reconstruction quality, as it is a linear least squares problem with a 
unique solution for non-empty masks~\cite{MBWF11}, and thus 
there are no risks of getting stuck in local minima. Approaches based on 
inpainting echoes~\cite{MHWT12} or Green's functions~\cite{H17} are 
unsuitable for large images due to the memory constraints imposed by 
the arising dense systems, or the large computational cost of their 
matrix-free variants. For this reason we do not discuss them in this 
paper beyond a short overview of the literature. Other viable alternatives 
include~\cite{CRP14,HW15_2,H17,MHWT12,PSAW22}, however, all of those 
are orders of magnitude slower than our proposed GPU-based algorithm.
Even the recent approach by Chizhov and Weickert~\cite{CW21}, which is one 
of the fastest on masks of 5\% density, takes approximately 10 minutes for 
the tonal optimization of a 4K color image. 
While these methods can be improved with parallel GPU implementations, and 
we do implement several of those on the GPU for comparison, they are still 
not able to reach the efficiency of our method, which is specifically designed 
for highly parallel architectures such as GPUs.

%-----------------------------------------------------------------------------
	
\subsection{Our Contribution}
\label{sec:contribution}

% In the first part of our companion paper~\cite{KCW23} we have proposed a full 
% multigrid \textit{optimized restricted additive Schwarz (ORAS)} inpainting 
% solver for homogeneous diffusion inpainting. It achieves real-time 
% decoding with more than $60$ frames per second for 4K color images, by 
% exploiting the potential of highly parallel architectures such as GPUs. Our 
% goal in the second part is to complement our inpainting solver with a fast 
% data optimization framework, in order to also make the encoding for image and 
% video compression with homogeneous diffusion viable for 4K image resolutions.
% To this end, we propose fast and highly parallel methods to optimize the 
% stored data.

\newtext{
The goal of our paper is to develop a state-of-the-art data optimization 
framework for homogeneous diffusion inpainting that
complements the fast inpainting solver from~\cite{KCW23}. To this end, 
we devise fast and highly parallel methods to optimize the data to be stored. 
This allows us to construct the first viable data optimization approach 
for sparse 4K image inpainting in high quality that also exploits 
tonal optimization.}

For the spatial optimization we extend the densification approaches by 
Daropoulos et al.~\cite{DAW21}, and Chizhov and Weickert~\cite{CW21}. We 
employ a Delaunay-based densification, which we supplement with a strategy 
for choosing the initial inpainting mask, based on the analytic mask 
optimization framework of Belhachmi et al.~\cite{BBBW08}. Our methods 
offer state-of-the-art performance.
We achieve fast runtimes by efficiently computing the Delaunay triangulation 
in each densification iteration from an approximation of the Voronoi 
tessellation, obtained with the  
\textit{jump flood algorithm (JFA)}~\cite{RT06}.

For our tonal optimization approach, inspired by the nested \textit{conjugate 
gradients (CG)} approach by Chizhov and Weickert~\cite{CW21}, we solve the 
corresponding normal equations in a matrix-free fashion. 
\newtext{Instead of relying on CG, we construct a novel highly parallel 
solver for tonal optimization adapted to our sparse inpainting problem. 
It carefully adapts and integrates ideas from multigrid and domain 
decomposition, see~\Cref{fig:tonal-ras-schematic}.}
As an outer solver, we adapt a \textit{restricted additive Schwarz} domain 
decomposition method~\cite{CS99}, which is able to exploit the fact that the 
influence area of each known pixel is localized in practice.  
In addition, we devise a computationally inexpensive initialization strategy 
allowing for a much earlier termination of the tonal optimization. 
We derive the latter from a theoretically motivated surrogate problem, which 
also subsumes the initialization from the work of Gali\'c et 
al.~\cite{GWWB08}. 

% Since our spatial and tonal optimization methods both rely on the computation 
% of multiple inpaintings, we integrate our fast inpainting solver proposed in 
% the first part of our companion paper~\cite{KCW23}. 
Since our spatial and tonal optimization methods both require the computation 
of multiple inpainting solutions, we use the multigrid ORAS inpainting 
solver proposed in~\cite{KCW23}. This is to the best of our knowledge the 
fastest homogeneous diffusion inpainting solver for parallel architectures 
such as GPUs.
Thanks to all of our contributions, we are able to achieve runtimes of less 
than a second for the full data optimization of 4K color images
with an optimal linear scaling behavior with respect to the image resolution.
% To the best of our knowledge, this is currently the fastest high-quality data 
% optimization for homogeneous diffusion inpainting. 
\newtext{Our tonal optimization is currently the fastest method for 
homogeneous diffusion inpainting, taking less than half a second 
for a 4K RGB image. Our spatial 
optimization is the second fastest available mask optimization approach, 
also taking less than half a second for the construction of outstanding  
quality masks for a 4K RGB image.
The only faster spatial optimization strategy is the method by 
Belhachmi et al.~\cite{BBBW08}, however, it results in inpaintings 
that can have a PSNR of up to $6$ dB lower compared to our approach, 
see \Cref{fig:spatial-comp-density}.}
%-----------------------------------------------------------------------------

\subsection{Related Work}
\label{sec:related}

In the following we discuss prior works on spatial and tonal optimization and 
how they are related to our framework. 

%-----------------------------------------------------------------------------

\subsubsection{Spatial Optimization}
\label{sec:spatial_optimization}
Finding a subset of pixels to be stored that result in a high quality 
reconstruction is a challenging combinatorial optimization problem. Here we 
only focus on approaches for homogeneous diffusion-based inpainting in 2D. There are also a number of related works, for example the free knot problem for spline interpolation~\cite{SS03}. 
%\vassillen{If you need extra citations 
%we can add the paper "A COMPARATIVE STUDY OF SOME GREEDY PURSUIT ALGORITHMS FOR
%SPARSE APPROXIMATION", it has a nice and short overview of various 
%matching pursuit algorithms.} 

%-----------------------------------------------------------------------------

\paragraph{Analytic Approach} 
Belhachmi et al.~\cite{BBBW08} developed a framework for mask optimization 
based on a continuous shape analytic perspective of homogeneous diffusion 
inpainting, which was later extended by Belhachmi and 
Jacumin~\cite{BJ22,BJ23} for noisy images.
The practically relevant result is that the mask density should 
be chosen as an increasing function in the absolute value of the Laplacian 
of the target image. 
In the discrete setting they suggest applying a dithering method to the 
absolute value of the smoothed Laplacian image, e.g.\ Floyd-Steinberg 
dithering~\cite{FS76}, in order to generate a mask that matches the desired 
density. While this approach is very fast, as it does not require computing 
any inpaintings, in practice the dithering approximation leads to a limited 
reconstruction quality~\cite{MHWT12}. 
In the current work we aim at a considerably higher quality of the 
inpaintings, and therefore integrate the Belhachmi approach only as a partial 
initialization in our algorithm.

%-----------------------------------------------------------------------------

\paragraph{Gradient-based Methods}
With non-smooth optimization strategies such as primal-dual solvers and 
optimal control, one can obtain high-quality non-binary inpainting 
masks~\cite{BLPP17,CRP14,HSW13,OCBP14}. For compression purposes, however, 
the masks have to be binarized, which reduces the quality~\cite{HW15_2}. 
Nevertheless, these methods produce state-of-the-art masks but have a 
prohibitively large runtime even for small images. Since we aim at fast 
algorithms that scale to 4K images, we instead consider greedier strategies 
that produce good results in short runtimes.

%-----------------------------------------------------------------------------

\paragraph{Sparsification}
Mainberger et al.~\cite{MHWT12} have proposed a \textit{probabilistic 
sparsification (PS)} algorithm that starts with a full mask and iteratively 
removes pixels from the mask until the desired density is reached. In each 
iteration, a set of candidate pixels from the mask is chosen and temporarily 
removed. An error map is computed from the inpainting with the reduced mask. 
Then a fixed number of the candidates are reintroduced back into the image: 
the ones with the highest pointwise errors. While this approach can be easily
adapted to different inpainting operators, such as PDE-based 
ones~\cite{MHWT12} or linear splines interpolation over 
triangulations~\cite{DDI04}, it is computationally fairly expensive. 
Additionally, since PS starts from a full mask, reaching low densities takes 
many more iterations than starting from an empty mask and densifying. 
Last but not least, PS is prone to getting stuck in suboptimal local minima 
both because it uses only pointwise errors as an oracle, and because the 
error itself in the initial iterations is a result only of a very localized 
neighborhood that is not predictive of the influence of mask points in later 
iterations. These considerations suggest studying densification approaches 
with an oracle that adapts to the density of the mask in each iteration.

%-----------------------------------------------------------------------------

\paragraph{Densification}
Densification approaches start from an empty mask and iteratively insert new 
pixels. In practice, this also means that fewer iterations need to be 
performed compared to sparsification-based methods, since the usual mask 
densities for inpainting-based compression are below 10\%.
\newtext{In compression settings, densification has been successfully applied 
to constrained data structures such as subdivision 
trees~\cite{DNV97,GWWB08,PHNH16,SPME14}.}
In the current work we are interested in the more general case of 
unconstrained masks.
Unconstrained densification strategies have been used successfully for 
diffusion-based~\cite{BJ22a,CW21,DAW21,JCW23,TPM20} and  
exemplar-based~\cite{KBPW18} inpainting, as well as as linear interpolation 
on Delaunay triangulations~\cite{Ad13,EA17}. 
The works of Daropoulos et al.~\cite{DAW21}, Chizhov and 
Weickert~\cite{CW21}, 
and Jost et al.~\cite{JCW23} consider not just pixel-wise errors, but 
aggregate the error over mask-adaptive neighborhoods based on the Voronoi 
diagram~\cite{DAW21,JCW23} or the Delaunay triangulation~\cite{CW21}, which 
improves the reconstruction quality compared to a pixel-wise densification.
The work by Jost et al.~\cite{JCW23} even generalizes the optimization 
to masks prescribing derivatives and local integrals. Currently, Voronoi- 
and Delaunay-based densification approaches provide the best 
quality-to-runtime ratio for mask construction, so we use those as a basis 
for the mask generation algorithms we develop in the present work.

%-----------------------------------------------------------------------------

\paragraph{Non-local Pixel Exchange}
Even though densification-based methods can perform better than 
sparsification-based ones, they are still greedy and can thus get stuck in 
local minima. To alleviate this, a global relocation strategy called 
\textit{nonlocal pixel exchange (NLPE)} has been devised by 
Mainberger et al.~\cite{MHWT12}. 
It stochastically chooses a mask pixel and a subset of non-mask pixels with 
high pointwise errors, and iteratively computes the errors for the 
reconstructions resulting from relocating the mask pixels to the candidate 
non-mask pixel locations. The best swap (potentially none) is kept. 
While this approach can escape from local minima and achieve significant 
improvements, it requires a huge number of iterations (i.e.\ computational 
resources) to converge. We also note that for linear spline interpolation on 
triangulations a similar approach has been used to optimize 
masks~\cite{MMCB18}. 

%-----------------------------------------------------------------------------

\paragraph{Neural-based Approaches}
Besides model-based spatial optimization algorithms, a number of neural mask 
generation frameworks have been proposed in recent years. 
Dai et al.~\cite{DCPCWK19} developed a deep learning method for adaptive 
sampling that uses an inpainting network and a mask optimization network, 
which are trained separately. In order to improve the quality, 
Peter~\cite{P23} jointly trains inpainting and spatial optimization with 
Wasserstein GANs. 
Alt et al.~\cite{APW22} train a mask generator specifically for homogeneous 
diffusion inpainting. In order to be able to perform a fast backpropagation 
for the mask network they propose to replace the inpainting with a learned 
surrogate solver that approximates homogeneous diffusion inpainting, which
is learned together with the mask network in a joint fashion. 
The efficiency has been improved further by Peter et al.~\cite{PSAW22} with a 
modified network architecture.
The aforementioned networks were, however, only trained on small images which 
leads to a poor performance for 4K images. Due to memory and computational 
constraints, directly training on 4K images is infeasible on standard 
hardware. 

To scale neural methods to 4K images, Schrader et al.~\cite{SPKW23} proposed 
a coarse-to-fine approach that divides the input image into small patches
and creates an inpainting mask for each patch with a mask generation 
network. 
Each patch uses a mask density estimated by the average Laplacian magnitude 
according to the analytic approach by Belhachmi et al.~\cite{BBBW08}. This 
allows efficient scaling of the mask construction to higher resolutions, and 
is able to generate masks for 4K color images in less than $0.5$ seconds on a 
high-end GPU.

While neural approaches have the advantage of not needing to perform several 
inpainting steps, model-based optimization is transparent, and the ideas 
within generalize to a larger class of problems. Therefore, in the current 
work we rely on non-neural optimization techniques. Thanks to using the 
fast inpainting solver, proposed in~\cite{KCW23}, and the 
efficient optimization algorithms developed in the current work, we are 
even able to achieve similar runtimes to the method by Schrader et 
al.~\cite{SPKW23} with much lower memory usage, while producing 
higher quality masks.

%-----------------------------------------------------------------------------

\paragraph{Video Coding}
Most approaches for inpainting-based video compression, such 
as~\cite{APKMWH21,PSMM15}, optimize the inpainting masks for each frame 
independently.
Breuß et al.~\cite{BHR21} instead optimize inpainting masks only for key frames, with the optimal control approach of Hoeltgen et al.~\cite{HSW13}. For the intermediate frames, they use optical flow to interpolate the masks.

%-----------------------------------------------------------------------------

\subsubsection{Tonal Optimization}
After generating a mask, e.g.\ using one of the above algorithms, one can 
improve the reconstruction quality further by modifying the stored color 
values at mask pixels. This is done by minimizing some error metric with 
respect to the reference image. The mean squared error (MSE) is an especially 
appealing choice, since the global optimum of the optimization problem can be 
found by solving a linear system of equations (the normal equations). 
Consequently, we focus on methods that minimize the MSE.

%-----------------------------------------------------------------------------

\paragraph{Matrix Structure}
The system matrix for the normal equations is of the form 
$\bm{B}^{\top}\bm{B}\in\mathbb{R}^{m\times m}$, where $m$ is the number of 
mask pixels~\cite{H17, MHWT12}. The latter is a symmetric positive 
definite matrix. Such matrices are advantageous from a numerical 
perspective, since algorithms such as the Cholesky decomposition~\cite{Hi02} 
or the conjugate gradient (CG) method~\cite{HS51, Sa03} can be used to 
efficiently compute a solution. 
We note, however, that neither $\bm{B}^{\top}\bm{B}$ nor $\bm{B}$ are sparse. 
Despite this, the matrix-vector product $\bm{B}\bm{x}$ can be carried out 
through a sparse matrix-vector product followed by solving a sparse linear 
system, both of which can be implemented in a matrix-free fashion 
(i.e.\ without explicitly forming or storing the matrices).

%-----------------------------------------------------------------------------

\paragraph{Direct vs Iterative Methods}
Linear system solvers can generally be split in two 
classes: direct ones and iterative ones. Classical direct methods, such as 
Cholesky, LU, or QR factorizations~\cite{Hi02}, are suited for dense and not 
too large matrices, and their runtime is typically independent of the 
condition number of the matrix (the latter can affect the numerical 
stability, however). 
Iterative solvers, such as CG~\cite{HS51, Sa03}, on the other hand, scale 
favorably to large sparse problems and allow one to stop early, but their 
convergence speed is typically dependent on the condition number. In practice 
a matrix-free matrix-vector product can be integrated easily in iterative 
methods, and they are more readily parallelizable.

In our target case of tonal optimization for 4K images, a standard mask 
density of $5\%$ results in a system matrix of size $m\times m$ where 
$m \approx 400000$. 
As this matrix is dense, we would need approximately $640$ GB of memory if we 
wished to form it and store it. On the other hand, as discussed, the product 
$\bm{B}\bm{x}$ can be implemented in a matrix-free fashion. In practice, this 
constrains us to iterative approaches, but we nevertheless discuss some tonal 
optimization frameworks based on direct methods, since e.g.\ they have been
shown to be able to outperform iterative approaches on masks with few 
pixels~\cite{H17}.

%-----------------------------------------------------------------------------

\paragraph{Inpainting Echoes}
For smaller images one can compute the matrix $\bm{B}$ explicitly and store 
it. The columns in this matrix are termed 
\textit{inpainting echoes}~\cite{MHWT12}. Mainberger et al.~\cite{MHWT12} 
precompute the echoes and store them, and thus essentially form and 
store the matrix $\bm{B}$. They note that one could solve the normal 
equations involving $\bm{B}^{\top}\bm{B}$ with an LU solver~\cite{Hi02}, 
but that this is too slow even for $256\times 256$ images. This supports 
our statement that for large images an iterative solver has to be used.
They employ a modified  successive over-relaxation (SOR) solver~\cite{Sa03} 
with an under-relaxation weight and random mask points traversal. 
In practice, storing $\bm{B}$ is infeasible for larger images due to memory 
constraints, and recomputing an inpainting echo (i.e.\ a column of $\bm{B}$) 
each time a mask value needs to be modified in an iterative approach is also 
prohibitively expensive. In our method, we use a much more efficient way to 
evaluate products $\bm{B}\bm{x}$ without ever explicitly forming or storing 
any matrix.

%-----------------------------------------------------------------------------

\paragraph{Green's Functions}
Homogeneous diffusion inpainting can equivalently be formulated in terms of a 
linear combination of Green's functions associated with mask 
points~\cite{H17, HPW15, KN22}. The Green's functions are a 
kind of spectral counterpart to the inpainting echoes. We will denote the 
matrix of the Green's functions, for a specific mask, with $\bm{G}$. Then 
the corresponding normal equations take the form $\bm{G}^{\top}\bm{G}$ 
once again~\cite{H17}. The main difference to the inpainting echoes is that 
products $\bm{G}\bm{x}$ can be evaluated efficiently and in a matrix-free 
fashion using the fast orthogonal DCT-II and DCT-III transforms~\cite{RY90}. 
Hoffman~\cite{H17} forms $\bm{G}^{\top}\bm{G}$ explicitly and uses a Cholesky 
decomposition on it. He demonstrates that for a low number of mask points the 
Choleksy approach applied to the Green's functions formulation can be more
efficient than conjugate gradients applied to the normal equations of the 
classical formulation. Since we are dealing with many more mask points due 
to considering $4K$ images (in Hoffman's case this was just $4\%$ of a 
$256\times 256$ image), it is clear that the above approach is impractical 
in our setting.

For 1D signals, Plonka et al.~\cite{PHW16} have proposed a method, based on the 
Green's functions approach, that jointly optimizes the locations of the mask 
pixels as well as their tonal values. As this approach only considers 1D 
signals it is not applicable in our setting, and since it uses Green's 
functions it suffers from similar restrictions as the approach by Hoffman.

%-----------------------------------------------------------------------------

\paragraph{Nested Conjugate Gradients}
A potential remedy of the memory constraint problem was proposed by Chizhov 
and Weickert~\cite{CW21} for their finite elements framework. They use a 
CG solver~\cite{HS51, Sa03} on the normal equations, but instead of 
explicitly computing the system matrix they evaluate the matrix-vector 
products $\bm{B}\bm{x}$ and $\bm{B}^{\top}\bm{x}$ in an efficient manner. 
They rewrite $\bm{B} = \tilde{\bm{A}}^{-1}\tilde{\bm{C}}$, where 
$\tilde{\bm{C}}$ is sparse, and where $\tilde{\bm{A}}$ is sparse and 
symmetric positive definite. 
Therefore, they use a conjugate gradients solver whenever they have to compute 
$\tilde{\bm{A}}^{-1}\bm{y}$ allowing them to exploit the sparsity of the 
problem.
With this so-called nested CG approach, they achieve a memory and runtime 
efficient tonal optimization. In our approach, we use a similar nested 
strategy in our framework, where the nested iterations are replaced by an
adapted version of the fast inpainting algorithm from~\cite{KCW23}.

We also note that the mesh construction and system matrix assembly for the 
finite element framework are not easily parallelizable, and the iterations 
cannot be implemented in a matrix-free fashion due to the irregular finite 
elements mesh. Thus, for fairness in the experiments, we consider a finite 
differences formulation that is easily parallelizable and allows implementing 
all matrix-vector products in a matrix-free fashion.

%-----------------------------------------------------------------------------

\paragraph{Non-binary Mask Optimization}
Hoeltgen and Weickert have shown~\cite{HW15_2} that a thresholded non-binary 
mask spatial optimization~\cite{BLPP17,CRP14,HSW13,OCBP14} can also be 
interpreted as a joint tonal and binary mask spatial optimization. 
Non-binary masks can be optimized with sophisticated non-smooth optimization 
strategies such as primal-dual algorithms, which then indirectly perform a 
tonal optimization. 
An extension to color images is, however, not straightforward. Additionally, 
while the above methods achieve very good results, they are already quite 
computationally intensive even for small images. 

%-----------------------------------------------------------------------------

\paragraph{Neural-based Approaches}
Besides model-based methods, neural networks can also be efficiently used to 
implement tonal optimization. Recently, Peter et al.~\cite{PSAW22} proposed 
the first tonal optimization neural network for homogeneous diffusion 
inpainting, which results in a memory efficient and fast method. While it is 
trained to minimize the MSE, it does not solve the least squares problem,
which means that it does not necessarily obtain the best MSE. Furthermore, 
the network was only trained on small image resolutions and does not perform 
well for 4K resolution images. A na\"ive extension to 4K images would be 
impractical due to memory and computational limitations of the training 
process. 

%-----------------------------------------------------------------------------

\paragraph{Localization-based Acceleration}
Since the influence zone of a single mask pixel is mostly limited for many 
inpainting operators, localization strategies are an option to accelerate 
tonal optimization. 
This works especially well for perfectly localized inpainting 
operators such as Shepard interpolation with truncated 
Gaussians~\cite{AAS17,P19}, or smoothed particle 
hydrodynamics~\cite{DAW21}. \newtext{For less local operators, there are also 
methods that artificially limit the influence with a
segmentation~\cite{HMWP13,JPW21}.} Similarly, one can also restrict the 
support of inpainting echoes in homogeneous diffusion~\cite{H17}. This 
essentially constructs an approximation $\tilde{\bm{B}}$ of $\bm{B}$ that is 
not fully dense anymore. Nevertheless, the number of restricted echoes which 
have to be computed and stored is still very large for 4K images. 

In the current work we consider domain decomposition methods instead, which 
are a much more sophisticated way to localize computations (and thus reduces 
accesses to global GPU memory) without artificially imposing locality 
constraints on non-local operators such as homogeneous diffusion.

%-----------------------------------------------------------------------------

\paragraph{Quantization}
In the context of image compression, quantization is often a factor which 
limits the number of available gray or color values. Including this 
quantization directly into the tonal optimization yields better quality than 
simply applying the quantization afterwards. For this discrete optimization 
problem several approaches have been proposed, such as projection 
methods~\cite{PHNH16} or stochastic strategies~\cite{MMCB18, SPME14}.
Since we only deal with the continuous optimization problem in our paper,
we do not include any of these strategies. In a full compression setting, 
however, one could easily integrate one of these methods. 

%-----------------------------------------------------------------------------

\paragraph{Singularity Suppression}
Homogeneous diffusion inpainting exhibits logarithmic singularities inherited 
from the continuous boundary value problem formulation in 2D~\cite{MM06}. 
Those are especially visible at lower resolutions, so while this is not 
really a concern for us because we consider 4K images, we nevertheless note 
that this issue has been tackled previously.
Interpolation swapping~\cite{SPME14} removes discs around the known data 
after the initial inpainting and uses a second inpainting to fill in the 
discs. This reduces the effect of the singularities at the mask pixels, but 
it modifies the inpainting operator itself. While we could also apply this in 
our approach, it often reduces the overall reconstruction quality due to the 
smoothing around the mask pixels. 

%-----------------------------------------------------------------------------

\paragraph{Error Balancing}
An early predecessor of tonal optimization has been proposed by 
Gali\'c et al.~\cite{GWWB08} in 2008. It modifies the tonal values of each 
mask pixel based on the inpainting error of the neighboring pixels. This 
reduces the effect of singularities at mask pixels, while also improving the 
inpainting quality. We extend this idea both practically and theoretically, 
and use it as a fast way to get a good initialization for our tonal 
optimization algorithm.

%-----------------------------------------------------------------------------

\subsection{Paper Structure}
\label{sec:organisation}

We give a short review of homogeneous diffusion inpainting in 
\cref{sec:inpainting} and in \cref{sec:inp-solver} we give an overview of  
the Multigrid ORAS inpainting solver that we use in our spatial and tonal 
optimization methods. 
In \cref{sec:spatial}, we present our Delaunay-based 
densification method for the spatial optimization. \Cref{sec:tonal}  
provides details on our tonal optimization methods as well as our strategy 
for constructing a good initial guess. We evaluate our spatial and tonal 
optimization methods experimentally in \cref{sec:experiments}. Finally, 
we conclude our paper in \cref{sec:conclusions} and discuss our ongoing work. 

%%%%%%%%%%%%%%%%%%%%%%%%%%%%%%%%%%%%%%%%%%%%%%%%%%%%%%%%%%%%%%%%%%%%%%%%%%%%%%

\section{Homogeneous Diffusion Inpainting}
\label{sec:inpainting}

In this section we briefly review homogeneous diffusion 
inpainting~\cite{Ca88}, which is our inpainting method of choice and is used 
for all results in this paper. Homogeneous diffusion inpainting is a simple 
and very popular method for inpainting-based compression. It is 
parameter-free, allows to establish a comprehensive data selection theory, 
and can lead to particularly fast algorithms. 
It has been shown that, if the inpainting data is carefully chosen, it can 
achieve very good results~\cite{HMWP13,JPW21,MBWF11,MHWT12}.

%-----------------------------------------------------------------------------

\subsection{Mathematical Formulation}
Let us consider a continuous grayscale image $f: \Omega \to \mathbb{R}$, 
defined on a rectangular image domain $ \Omega \subset \mathbb{R}^2$.
We store only a fraction of the full image data on a small subset 
$K \subset \Omega$.
Homogeneous diffusion inpainting aims at restoring $f$ in the inpainting domain 
$\Omega \setminus K$, by solving the Laplace equation on $\Omega\setminus K$ 
with reflecting boundary conditions on $\partial\Omega$, and with the known 
data $f|_{K}$ used as an interpolation constraint $u|_K = f|_K$:
\begin{alignat}{2}
	\label{eq:inp-continuous-1}
	-\Delta u (\bm{x}) &= 0, &\quad &\bm{x} \in \Omega
	\setminus K, \\
	\label{eq:inp-continuous-2}
	u (\bm{x}) &= f (\bm{x}), &\quad &\bm{x} \in K, \\
	\label{eq:inp-continuous-3}
	\partial_{\bm{n}} u (\bm{x}) &= 0 &\quad &\bm{x} 
	\in 
	\partial \Omega,
\end{alignat}
where $\partial_{\bm{n}} u$ denotes the derivative of $u$ in outer normal 
direction $\bm{n}$. 

%-----------------------------------------------------------------------------

\paragraph{Discrete Setting}
For a discrete image $\bm{f} \in \mathbb{R}^N$, with $N$ pixels, we 
discretize \cref{eq:inp-continuous-1,eq:inp-continuous-2,eq:inp-continuous-3} 
with finite differences. 
Using an indicator function $\bm{c} = \mathds{1}_K \in\{0,1\}^N$, also called 
the \textit{inpainting mask}, the reconstruction $\bm{u}$ with homogeneous 
diffusion inpainting is given by the following linear system of equations:
\begin{equation}
	\label{eq:inp-discrete-lsi}
	\underbrace{(\bm{C}+(\bm{I}-\bm{C}) \bm{L})}_{=:\bm{A}} \bm{u} 
	= \underbrace{\bm{C} \bm{f}}_{=:\bm{b}},
\end{equation}	
where $\bm{I} \in \mathbb{R}^{N\times N}$ is the identity matrix, 
$\bm{C} := \text{diag}(\bm{c}) \in \mathbb{R}^{N\times N}$ is a 
diagonal matrix with the components of the inpainting mask on the diagonal, 
and $\bm{L}$ is the $5$-point stencil discretization of the negated Laplacian 
with reflecting boundary conditions.
\newtext{
If the inpainting mask is non-empty, i.e. $\bm{c}\ne \bm{0}$, the inpainting problem has a unique solution since $\bm{A}$ is non-singular in that 
case (for more details on the matrix structure see~\cite{MBWF11}).}
For an RGB color image the inpainting is performed on each of the three 
channels separately. Thus, we get three non-coupled linear systems of this 
type (the system matrix is the same for all of them). 

% %-----------------------------------------------------------------------------
% \paragraph{Uniqueness of the Solution}
% \newtext{If $\bm{c}\ne \bm{0}$ then the problem 
% (\ref{eq:inp-discrete-lsi}) has a unique solution since 
% the matrix $\bm{A} = (\bm{C}+(\bm{I}-\bm{C})\bm{L})$ is non-singular in that 
% case (for more details on the matrix structure see~\cite{MBWF11}).
% If the mask is empty $\bm{c}=\bm{0}$ (or $K=\emptyset$ in the continuous 
% setting), the system $\bm{A}\bm{u} = \bm{L}\bm{u} = \bm{0}$ has infinitely 
% many solutions consisting of flat images with different mean gray 
% values. 
% In practice $\bm{c}=\bm{0}$ would mean that we have no mask pixels, i.e., 
% we have no budget to store even a single pixel from the image, so 
% this case is not of practical relevance.
% On the other hand, if we have one mask pixel, then as desired 
% we get the flat solution with mean gray value the single mask 
% pixel's color.} 

%-----------------------------------------------------------------------------
\paragraph{Homogeneous Diffusion Inpainting and Image Edges}

\newtext{
Edges are important semantic and mathematical features in images. 
Thus, a simple and appealing idea is to reconstruct an image from its 
edge information. 
When the Laplacian $\bm{L}\bm{f}$ is considered as an edge detector, 
this idea is realized as homogeneous diffusion inpainting~\cite{Ca88}. 
If $\bm{L}\bm{f}$ is available exactly, the image can be recovered up 
to mean gray value as $\bm{L}^+(\bm{L}\bm{f})$, where $\bm{L}^+$ is the 
Moore-Penrose inverse. Since $\bm{L}$ has a high-pass profile, $\bm{L}^+$ 
has a low-pass character and (\ref{eq:inp-discrete-lsi}) fills 
in information smoothly between mask points.}

\newtext{
In practice, we do not want to store $\bm{L}\bm{f}$ 
exactly, leading to the sparse approximation problem with the inpainting 
from (\ref{eq:inp-discrete-lsi}).
Earlier approaches~\cite{AG94,BHK10,Ca88,DMMH96,GLG12,LSWL07,ZD11} 
for homogeneous diffusion inpainting used the pixels left and right of 
the edges as known data, in order to preserve them in the reconstruction. 
While modern approaches usually do not directly use the edges to determine 
the inpainting data, the optimized inpainting masks still cluster
around prominent image edges.
For an illustration of this effect, see \Cref{fig:spatial_visual_comp}.
Further theoretical justifications for choosing reconstruction 
data based on the Laplacian can be found in the work of 
Belhachmi et al.~\cite{BBBW08}.}

\newtext{We also note that the choice of reflecting boundary conditions 
(\ref{eq:inp-continuous-3}) 
is standard in image processing, as it produces the best results for 
the compression of natural images. 
In fact the matrix $\bm{L}$ has eigenvectors the rows of the 
DCT-II transform~\cite{St99}, which is known to have the best 
energy compaction properties for natural images, out of all the 
DFT/DST/DCT variants.}

%-----------------------------------------------------------------------------
\paragraph{Alternative Inpainting Operators}
\newtext{
An alternative class of sparse inpainting operators use locally 
supported positive definite kernels~\cite{DAW21,P19,We10}.
While the latter has considerable advantages since it does not have 
to deal with global interactions, there are  non-trivial challenges 
associated with this approach. The kernels come with various parameters 
that have to be optimized per mask point. This necessitates multiple 
iterations to produce a single inpainting~\cite{DAW21}, that is neither 
over-blurred -- due to the kernel supports being too large, nor has 
holes -- due to the kernel supports being too small.}

\newtext{
Other options include nonlinear inpainting methods such as anisotropic 
diffusion inpainting~\cite{We97}. Such methods can achieve 
state-of-the-art quality~\cite{GWWB08}, however, that comes at a 
considerable computational cost. We have left the study of these 
alternatives as future work, and have focused instead on constructing 
the first complete framework for data optimization applicable to sparse 
4K image inpainting.}

%%%%%%%%%%%%%%%%%%%%%%%%%%%%%%%%%%%%%%%%%%%%%%%%%%%%%%%%%%%%%%%%%%%%%%%%%%%%%%
\section{Multigrid ORAS Inpainting Solver}
\label{sec:inp-solver}

\newtext{
To keep our manuscript self-contained, we describe the Multigrid ORAS 
inpainting solver that we use for the inpainting problems that occur in our 
proposed data optimization methods.
It consists of a domain decomposition method embedded into a multigrid 
framework. For more details on the Multigrid ORAS inpainting solver, we refer 
the reader to~\cite{KCW23}.
}
%-----------------------------------------------------------------------------

\subsection{Optimized Restricted Additive Schwarz Method}

\begin{figure}[tb]
	\centering
	\includegraphics[width=4cm]{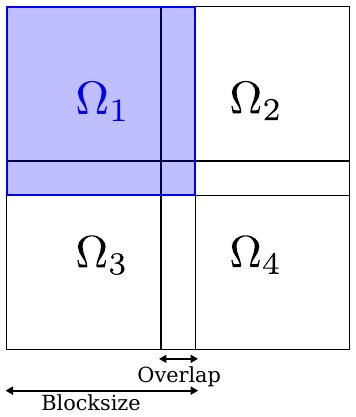}
	\caption{\textbf{Domain Decomposition Example.} The domain is 
            divided into four overlapping subdomains. The subdomain 
            $\Omega_1$ is highlighted in blue.}
	\label{fig:subdivision}
\end{figure}\textbf{}

\newtext{
The \textit{optimized restricted additive Schwarz (ORAS)} method~\cite{SGT07}, 
in its discrete formulation, is an iterative technique for solving linear 
systems that arise from the discretization of boundary value problems, such as 
\cref{eq:inp-discrete-lsi}.
It is one of the simplest domain decomposition methods~\cite{DJN15, TW05} and 
is well suited for parallelization. As a first step, the image domain 
$\Omega$ with $N$ pixels is divided into $k\in \mathbb{N}$ overlapping 
subdomains $\Omega_1,...,\Omega_k \subset \Omega$, such that 
$\cup_{i=1}^k\Omega_i=\Omega$. An example of such an overlapping division 
into four subdomains is shown in \cref{fig:subdivision}.}

\newtext{Starting with an initial approximation $\bm{u}^0 \in \mathbb{R}^N$ that
fulfills the interpolation condition $\bm{C} \bm{u}^0 = \bm{C} \bm{f}$ at 
the known locations, in each iteration $n$, we try to correct the current 
approximation $\bm{u}^n$ by locally computing correction terms for every subdomain.
We compute the corrections by solving local variants of the global 
linear system}

\begin{equation}
\bm{A}_i 
\bm{v}_i^n = \bm{R}_i \bm{r}^n,
\end{equation}
\newtext{where the right-hand side consists of the global residual 
$\bm{r}^n = \bm{b} - \bm{A} \bm{u}^n$ restricted to the subdomains. 
We use matrices $\bm{R}_i\in \mathbb{R}^{|\Omega_i| \times N}$  
to restrict global vectors to the local domains by extracting the values 
from the subdomain $\Omega_i$ and ignoring all values outside of it.
The local system matrices $\bm{A}_i $ are given by restricting the 
global system matrix $\bm{A}$ to the subdomains $\Omega_i$ with additional 
boundary conditions at the subdomain boundaries which are 
not part of the global image boundary. At these so-called inner subdomain 
boundaries, we impose mixed Robin boundary conditions, which improve the 
convergence speed compared to simpler Dirichlet boundary 
conditions~\cite{SGT07}.}

\newtext{The next iterate $\bm{u}_{n+1}$ is given by extending the local 
corrections $\bm{v}_i$ from the subdomain $\Omega_i$ to the global 
domain $\Omega$ and adding them to the old iterate $\bm{u}_{n}$:}

\begin{equation}
    \bm{u}^{n+1} = \bm{u}^n + \sum_{i=1}^{k} 
		\bm{R}_i^\top \bm{D_i} \bm{v}_i^n
\end{equation}
\newtext{For the extension, we use the transposes of the restriction matrices 
$\bm{R}^\top_i$. 
However, in order to achieve convergence, we also have to weight the local 
corrections in the overlapping regions before adding them to the global 
solution~\cite{EG03,Ga08}, to compensate for the fact that we add multiple 
corrections in the overlapping regions.
To this end we define diagonal nonnegative weight matrices $\bm{D}_i \in 
\mathbb{R}^{\lvert\Omega_i\rvert \times \lvert\Omega_i\rvert}$, such that}
\begin{equation}
\label{eq:partition-of-unity-discrete}
\bm{I} = \sum_{i=1}^{k}  \bm{R}_i^\top \bm{D}_i \,
\bm{R}_i,
\end{equation}
\newtext{where $\bm{I} \in\mathbb{R}^{N\times N}$ is the identity matrix of size 
$N \times N$}. 

%-----------------------------------------------------------------------------

\subsection{Embedding ORAS in Multigrid}

\newtext{Since the ORAS method only allows communication between neighboring blocks,
many iterations are necessary to spread information over large distances.
This means that high frequency components of the error are reduced much faster 
than low frequency components, which leads to an overall slow convergence rate.
As a remedy, domain decomposition methods are usually used together with a 
two-level scheme that computes a correction on a very coarse grid to couple 
all subdomains together~\cite{DJN15, TW05}.
In our case, where the influence area of each mask pixel is already 
approximately localized, such a global coupling would be to coarse and result 
in a poor performance.
Instead, the more fine-grained coupling of multigrid 
methods~\cite{BD96, Br77,BHM00, MF18, Hac85, TOS01, Wes92} can be more 
beneficial.
% Since the influence area of each mask pixel is approximately localized,
% a global coupling is not really necessary 
% and instead the more fine-grained 
% coupling of multigrid methods~\cite{BD96, Br77,BHM00, Hac85, TOS01, Wes92} are more 
% beneficial.
Multigrid methods transfer the problem to coarser grids where the low 
frequencies appear as higher frequencies, which can be reduced more 
efficiently with iterative solvers such as Jacobi~\cite{Sa03}, 
Gauss-Seidel~\cite{Sa03}, or the ORAS method~\cite{SGT07}.}

%-----------------------------------------------------------------------------
\paragraph{Two-Grid Cycle}

\newtext{
We first give an overview over the simple two-grid cycle, which we then 
extend to use multiple resolution levels. Our two grids are the original
fine grid with grid spacing $h$, and a coarser grid with spacing $H=2h$.}

\newtext{
We start the cycle $k+1$ with an approximation $\bm{u}_h^{k}$ of the solution 
$\bm{u}_h$ of the problem $\bm{A}_h\bm{u}_h = \bm{b}_h$, where $\bm{A}_h$ is 
the original system matrix and $\bm{b}_h$ the right-hand side on the 
original grid.
The goal of the two-grid cycle is to find a good approximation of the true  
error $\bm{u}_h-\bm{u}_h^k$, by decimating different frequencies of the error 
over different grids.
We reduce the high frequencies by performing a few iterations 
$\vartheta_\text{pre}$ with a \textit{smoother}: an iterative solver that 
dampens the high frequencies efficiently - e.g.\ damped Jacobi, Gauss-Seidel, 
or in our case ORAS. This results in the approximation of the solution} 
\begin{equation}
\bm{u}_h^{k+1/3} = 
\text{pre-smooth}
(\bm{A}_h,\bm{b}_h,\bm{u}_h^{k},\vartheta_\text{pre}).
\end{equation}
\newtext{With enough iterations $\vartheta_\text{pre}$, the error will have only 
negligible high-frequency components and we can transfer it almost perfectly 
to the coarser grid with a restriction matrix $\bm{R}^H_h$.
While we do not know the true error in practice, we know the residual and 
its relation to the error:}

% Provided we have used enough iterations $\vartheta_\text{pre}$, the error 
% $\bm{e}_h^k = \bm{u}_h-\bm{u}^{k+1/3}_h$ will have only negligible 
% high-frequency components. We can then represent it almost perfectly on the 
% coarser grid. The transfer between the grids can be formalized using a 
% restriction matrix $\bm{R}^H_h$ and a prolongation matrix $\bm{P}^h_H$. 
% The two are designed so that sufficiently low frequency components 
% are reproduced exactly, while higher frequency components are smoothed 
% out to avoid aliasing.

% A problem of the above formulation is that we do not know the error 
% in practice, so we cannot transfer it to the coarser grid. 
% However, we know the residual and its relation to the error:
%
\begin{equation}
    \bm{r}^k_h = \bm{b}_h-\bm{A}_h\bm{u}_h^{k+1/3} = 
    \bm{A}_h\bm{u}_h - \bm{A}_h\bm{u}_h^{k+1/3} = 
    \bm{A}_h\left(\bm{u}_h-\bm{u}_h^{k+1/3}\right) = 
    \bm{A}_h\bm{e}_h^k.
\end{equation}
\newtext{
Instead of solving this equation on the fine grid, we transfer it to the coarser grid and solve it for the coarse error $\bm{e}_H^k$ as}
\begin{equation}
    \label{eq:coarse_grid_solve}
    \bm{A}_H\bm{e}_H^k = \bm{r}^k_H.
\end{equation}
\newtext{
Here $\bm{A}_H$ is an analogue of $\bm{A}_h$, obtained by discretizing the 
inpainting problem on the coarser grid, and 
$\bm{r}^k_H = \bm{R}^H_h \bm{r}^k_h$ is the coarse residual, obtained by 
restricting the original residual to the coarser grid.
Since the matrix $\bm{A}_H$ is twice smaller in each dimension compared to 
$\bm{A}_h$, the computational cost is also reduced.
We can then transfer $\bm{e}_H^k$ to the fine grid by using a 
prolongation matrix $\tilde{\bm{e}}^k_h = \bm{P}^h_H\bm{e}^k_H$ 
and use it to correct our approximation of the solution:}
\begin{equation}
    \bm{u}_h^{k+2/3} = \bm{u}^{k+1/3} + \tilde{\bm{e}}^k_h = 
    \bm{u}^{k+1/3} + \bm{P}^h_H\bm{e}^k_H.
\end{equation}
\newtext{
A subsequent post-smoothing with $\vartheta_\text{post}$ iterations is 
applied to smooth high frequency error components potentially introduced by 
the correction step:}
\begin{equation}
    \bm{u}^{k+1}_h = 
    \text{post-smooth}
    (\bm{A}_h, \bm{b}_h, \bm{u}_h^{k+2/3}, \vartheta_\text{post}).
\end{equation}

%-----------------------------------------------------------------------------

\paragraph{Extension to More Resolution Levels}

\newtext{
We extend the two-grid cycle to multiple levels, which allows to 
efficiently dampen different frequencies of the error at different 
resolutions. 
The solve on the coarse level is replaced by another correction step on an 
even coarser level. We iterate this process recursively until the desired 
number of resolution levels are reached. This type of hierarchical 
application of the two-grid cycle is termed a \textit{V-cycle}.}    

\begin{figure}[H]
	\centering
	\includegraphics[scale=0.70] 
	{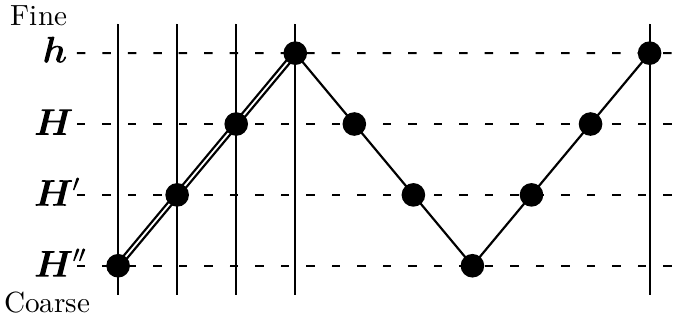}
	\caption{\textbf{Reduced Full Multigrid Scheme.} 
        The initial guess is constructed in a coarse-to-fine 
        manner, also known as one-way or cascadic 
        multigrid~\cite{BD96}. 
        Then we continue with additional V-cycle correction steps 
        (a single V-cycle is visualized above).
        }
	\label{fig:multigrid_reduced}
\end{figure}

\newtext{
Additionally we speed up the computation even further by providing a
good initialization. The latter is obtained by computing an inpainting 
solution on a coarser grid and prolongating it to the fine resolution.
We extend this to to a complete coarse-to-fine initialization strategy. 
While this is usually used with a complete V-cycle on each resolution level, 
we instead skip the V-cycles on the lower levels and just use a single 
pre-smoothing iteration in each lower level. The resulting reduced 
\textit{full multigrid (FMG)} scheme, is depicted in 
\cref{fig:multigrid_reduced}. 
It decreases the amount of costly restriction and prolongation operations, 
while still offering sufficient convergence for homogeneous diffusion 
inpainting.}

%-----------------------------------------------------------------------------

\subsection{Implementation Details}

\newtext{In our ORAS method we decompose the image domain 
into multiple overlapping blocks of size $32\times 32$ pixels with an 
overlap of $6$ pixels. These values are optimized for our 
\textit{Nvidia GeForce GTX 1080 Ti} GPU, which results in $36852$ local 
problems in each iteration of the ORAS method for a 4K color image on the 
finest resolution layer. 
The local problems are solved using a simple conjugate gradient 
algorithm~\cite{Sa03}. We stop the local CG iterations on a block when 
the squared 2-norm of the residual has been reduced to a fixed fraction 
of the squared global residual norm. Compared to a stopping criterion 
based on the local relative residual norm, this leads to a more uniform 
convergence, since all blocks have the same squared residual norm after 
each ORAS iteration. This also avoids using unnecessarily many CG 
iterations in blocks that are already sufficiently converged.} 

\newtext{
For our multigrid embedding, we use a single V-cycle on the 
finest level with a single ORAS iteration as pre- and post-smoothing.
The number of resolution levels is chosen, such that the coarsest level 
consists of a single ORAS block.
For the restriction of the inpainting mask we consider a coarse pixel to be a
known pixel, if at least one of its four corresponding fine resolution pixels
is a known pixel. For each coarse mask pixel its value is obtained as an 
average over the corresponding fine resolution pixel values. This method 
offers the advantage of increasing the mask density with each level, which 
in turn leads to faster convergence at the coarser levels.}

%%%%%%%%%%%%%%%%%%%%%%%%%%%%%%%%%%%%%%%%%%%%%%%%%%%%%%%%%%%%%%%%%%%%%%%%%%%%%%

\section{Spatial Optimization}
\label{sec:spatial}

The problem of spatial optimization or mask optimization for homogeneous 
diffusion inpainting consists of finding a binary mask $\bm{c} \in\{0,1\}^N$, 
that results in a reconstruction minimizing some error metric such as the 
$L_2$ error $\|\bm{u}(\bm{c},\bm{f})-\bm{f}\|_2$, under the constraint 
$\|\bm{c}\|_1 = \lfloor d\cdot N \rfloor$ on the number of mask pixels. 
Here $\|\cdot\|_1$ denotes the 1-norm of $\bm{c}$ (i.e.\ the number of mask 
pixels), and $d$ is a user-specified target density, which in a compression 
setting will be a function of the compression ratio.

%-----------------------------------------------------------------------------

\subsection{Densification}
Our spatial optimization method is based on the Voronoi-based densification 
strategies by Daropoulos et al.~\cite{DAW21} and Jost et al.~\cite{JCW23}, 
and the Delaunay-based strategy by Chizhov and Weickert~\cite{CW21}, which 
combine densification with ideas from error map dithering~\cite{KBPW18}.

Densification methods for spatial optimization iteratively add mask pixels in 
areas with high inpainting errors, starting with some initial mask. In each 
iteration, an inpainting $\bm{u}$ is computed from the current mask, and an 
error map is constructed as $\bm{e} = |\bm{u}-\bm{f}|^2$~\cite{KBPW18}. 
As our goal is to minimize the $L_2$ error, we use the squared error. 
Then one tries to balance the error at each iteration by introducing new pixels 
wherever the error is highest in the current error map. An issue with a 
na{\"i}ve implementation of such an approach is that if we introduce more 
than one pixel per iteration, the new pixels may clump around a high error 
area. The problem lies in the fact that if new pixels are introduced at the 
locations of highest pointwise error, then every new pixel fails to account 
for the effect of all other new pixels. 

In practice, it is paramount that we introduce multiple pixels per iteration 
in order to keep the number of inpaintings and thus computational cost low. 
A remedy to this is to introduce a mask-adaptive partition, that approximates 
the influence region of potential new pixels, and to insert only a single 
pixel per cell. We can then think of the densification as an error balancing 
process, that tries to constructs a partition such that the error is close to 
equal in each cell.

%-----------------------------------------------------------------------------

\begin{figure}[tb]
	\begin{subfigure}[b]{0.325\linewidth}
		\centering
		\centerline{\includegraphics[width=1.0\linewidth]
			{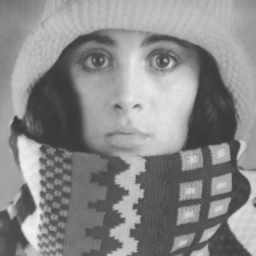}}
		\subcaption{original image \textit{trui} }
	\end{subfigure}
	\hfill
	\begin{subfigure}[b]{0.325\linewidth}
		\centering
		\centerline{\includegraphics[width=1.0\linewidth]
			{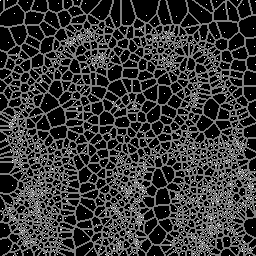}}
		\subcaption{\centering Voronoi tessellation}
	\end{subfigure}
	\hfill
	\begin{subfigure}[b]{0.325\linewidth}
		\centering
		\centerline{\includegraphics[width=1.0\linewidth]
			{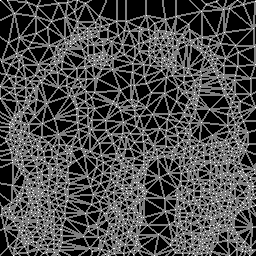}}
		\subcaption{\centering Delaunay triangulation}
	\end{subfigure}
	\smallskip
	\caption{\textbf{Voronoi and Delaunay Example on 
        \textit{trui} with a 2\% Mask Density}.}
	\label{fig:trui_voronoi_delaunay}
\end{figure}

\paragraph{Mask-Adaptive Partition}
The Voronoi tessellation~\cite{AKL13} offers a reasonable approximation of 
the influence zones of the already added mask pixels, as discussed 
in~\cite{DAW21}, but in densification we are interested in the influence 
regions of potential new pixels. This means that ideally the partition of the 
space would have the current mask pixels as vertices, and the cells would be 
bounded by those. The dual of a Voronoi diagram - the Delaunay 
triangulation~\cite{AKL13} - can be used to partition the domain in such a 
way. Given the mask pixels as vertices, it produces the triangulation that 
maximizes the minimum of all angles in the triangulation. This has the 
desirable effect that triangles are not elongated, and thus provide better 
approximations to the influence areas of isotropic operators such as 
homogeneous diffusion.
\Cref{fig:trui_voronoi_delaunay} shows an example of a Voronoi diagram and its 
corresponding Delaunay triangulation for an optimized inpainting mask with a 
density of 2\%.

%-----------------------------------------------------------------------------

%-----------------------------------------------------------------------------

\begin{figure}[tb]
\setlength{\tabcolsep}{1mm}
\begin{tabular}{cccc}
    \centered{\rotatebox{90}{\small \textit{Delaunay triangulation}}} &
    \centered{\includegraphics[width=0.3\textwidth]
        {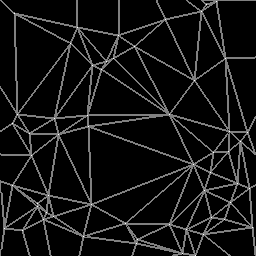}} & 
    \centered{\includegraphics[width=0.3\textwidth]
        {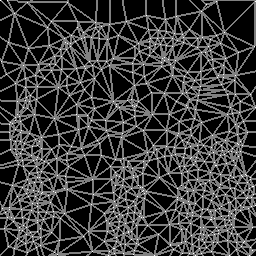}} &
    \centered{\includegraphics[width=0.3\textwidth]
    {images/spatial/delaunay_example/delaunay_vis_v2_i19.png}}
    \vspace{0.2mm} \\
    
    \centered{\rotatebox{90}{\small \textit{inpainted}}} &
    \centered{\includegraphics[width=0.3\textwidth]
        {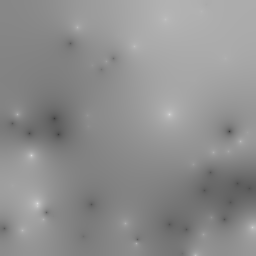}} & 
    \centered{\includegraphics[width=0.3\textwidth]
        {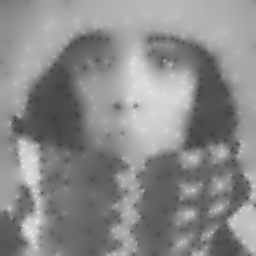}} &
    \centered{\includegraphics[width=0.3\textwidth]
        {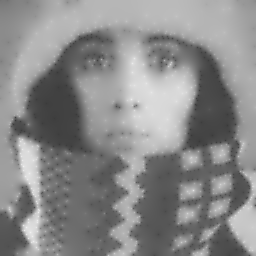}}
    \vspace{0.2mm} \\

    & iteration 1 & iteration 10 & iteration 20 \\
    
\end{tabular}

\caption{\textbf{Delaunay Densification Example on \textit{trui} 
         with a 2\% Mask Density.}}
\label{fig:delaunay_dens_example}
\end{figure}

\paragraph{Algorithm}
To start the densification, we need a non-empty initial mask $\bm{c}$ with a small 
amount of mask pixels in order to compute the first inpainting. We may obtain 
such an initial guess by using uniform random sampling or with another 
spatial optimization method, for example the Belhachmi approach~\cite{BBBW08}.
In each iteration, we compute an inpainting 
$\bm{u} = \bm{A}^{-1} \bm{C} \bm{f}$ from the mask in the current iteration. 
We can then compute the error map $\bm{e} = |\bm{u} - \bm{f}|^2$, partition 
the domain using a Delaunay triangulation with the current mask 
pixels as vertices, and then integrate the error map
within each triangle $\mathcal{T}_i$ to obtain the accumulated errors 
$e_{\mathcal{T}_i}$:
\begin{equation}
    e_{\mathcal{T}_i} = \sum_{j \in \mathcal{T}_i} e_{j} = 
    \sum_{j \in \mathcal{T}_i} |u_j - f_j|^2.
\end{equation}
Since we need to introduce multiple pixels in a single iteration, we 
distribute them so that we insert at most one per triangle, and we choose the 
triangles in order of decreasing triangle error $e_{\mathcal{T}_i}$. The 
pixels are placed at the location of highest pointwise error within the 
triangles. This allows us to avoid clumping while at the same time having a 
scale-adaptive partition. 
We iterate this process until we reach the desired mask density. 
\cref{alg:voronoi-densification} describes the Delaunay densification 
process, and \cref{fig:delaunay_dens_example} illustrates it.

%-----------------------------------------------------------------------------
\paragraph{Comparison to Spatial Optimization for Delaunay Based Inpainting}

\newtext{
Various methods rely on (at a first glance) similar strategies for 
greedily optimizing the inpainting mask. 
Specifically in the context of the used Delaunay partition, the work 
of Demaret et al.~\cite{DDI04}, Adams~\cite{Ad13}, and Marwood et 
al.~\cite{MMCB18} use 
linear splines over Delaunay triangulations for sparse image reconstruction. 
Their methods are specifically adapted to the linear 
splines Delaunay inpainting that is motivated from the desire of not storing 
the connectivity of a triangulation, while our method could easily be 
generalized for other inpainting methods. Our method is also more 
sophisticated, as we integrate the errors over the adaptive partition as a 
predictor for the potential incurred error from removing or adding a mask 
pixel. 
Last but not least, our method is also significantly faster, as their 
method requires approximately $4$ minutes for $23,092$ initial data points, 
which is about $3$ times lower than for a $256\times 256$ image.}

\begin{algorithm}[htb]
	\caption{Delaunay Densification} 
	\label{alg:voronoi-densification} 
    \textbf{Input     :} Original image $\bm{f}$,  
    number of iterations $n$, 
    number of mask pixels added in first iteration $m_0$, factor to increase the number of mask-pixels added per iteration $t$ \\
    \textbf{Output    :} Inpainting mask $\bm{c}$, reconstruction $\bm{u}$\\
    \textbf{Initialize:} Initial mask $\bm{c}$ with a small number of 
    mask pixels
    \begin{algorithmic}[1]
        \FOR{$i=1$ \TO $n-1$}
        % \FOR{$i=1, \dots ,n-1$}
            \STATE
            Construct the Delaunay triangulation $\{\mathcal{T}_{j}\}$ of the current mask pixels.
            \STATE
            Compute the inpainting $\bm{u} = \bm{A}^{-1} \bm{C} \bm{f}$
            and the error map $\bm{e} = |\bm{u} - \bm{f}|^2$. 
            \STATE
            Compute the triangle errors 
            $\forall j, \, e_{\mathcal{T}_j} = \sum_{k\in \mathcal{T}_j} e_k$.
            \STATE
            Compute the number of pixels to be added $m_i = t \cdot m_{i-1}$.
            \STATE
            Find the $m_i$ Delaunay triangles 
            $\{\mathcal{T}_{j_k}\}_{k=1}^{m_i}$ with the highest 
            errors $\{e_{\mathcal{T}_{j_k}}\}_{k=1}^{m_i}$.
            \STATE
            For each triangle in $\{\mathcal{T}_{j_k}\}_{k=1}^{m_i}$ find 
            the pixel with highest error and add it to $\bm{c}$.
        \ENDFOR
 \end{algorithmic}
\end{algorithm}

%-----------------------------------------------------------------------------

%-----------------------------------------------------------------------------

\subsection{Implementation Details}
\label{sec:spatial-implementation}

Here we discuss implementation details that improve the mask construction 
process from both a quality and runtime perspective.

%-----------------------------------------------------------------------------

\paragraph{Image-adaptive Initial Mask}
For the initial mask in the first iteration, both~\cite{DAW21} and~\cite{CW21} 
distribute the mask pixels uniformly at random. If we use only a very small 
amount of mask pixels in the initial mask, it does not have a substantial 
impact on the final densification result. However, as the initial mask density 
increases (i.e.\ as the number of densification steps decrease) the impact 
on the final result also increases. 
Generally, a uniform random mask is not optimal, as it is not adapted to image 
structures in any way. A much better distribution can be achieved by choosing 
the local mask density to be proportional to the Laplacian magnitude, as 
proposed by Belhachmi et al.~\cite{BBBW08}. 
They use a Floyd-Steinberg dithering~\cite{FS76} on the Laplacian magnitude 
of the reference image, however, such error diffusion algorithms are not well 
suited for GPU parallelization. Instead, 
\newtext{we use a simple random dithering on 
the Laplacian magnitude} - 
we do a biased coin flip for every pixel to decide whether it should become a 
mask pixel - with pixel probabilities proportional to the Laplacian magnitude. 
While by itself this results in a worse reconstruction quality compared to 
Floyd-Steinberg dithering, the quality degradation in a Delaunay densification 
process is negligible, and the greater efficiency more than pays off for it.

%-----------------------------------------------------------------------------

\paragraph{Number of New Pixels per Iteration}
In~\cite{DAW21}, the authors add only a small number of mask pixels per 
densification iteration, which results in a large number of iterations and 
thus a long runtime. Chizhov and Weickert~\cite{CW21} regard the number of 
iterations as a quality-efficiency trade-off parameter, and they fix it 
between $30$ and $100$ iterations depending on the intended goals. 
They introduce the exact same number of mask pixels in each iteration. 
We relax this restriction by allowing the number of added mask pixels per 
iteration to change during the densification. We specify the number of 
iterations $n$ and a multiplicative factor $t$, with which we decrease or 
increase the number of added mask pixels in each iteration.
For a fixed factor $t$ and a fixed number of iterations $n$, we compute the 
number of pixels added in the first iteration, such that the desired density 
is achieved after exactly $n$ iterations. 

By using our multigrid ORAS solver to compute the inpainting in each 
iteration, we are able to perform up to $20$ iterations in under $0.5$ seconds 
with our Delaunay densification.

%-----------------------------------------------------------------------------

\paragraph{Delaunay Triangulation Construction}
Besides the computation for the inpainting reconstructions, the runtime is 
also determined by the construction of the Delaunay partition. 
For an efficient implementation we generate the Delaunay triangulation from 
its corresponding Voronoi diagram. We can efficiently compute the latter on 
the GPU with the Jump Flood Algorithm~\cite{RT06} (JFA). While this does not 
necessarily give a pixel-perfect Voronoi tessellation, it generates a very 
good approximation. 

JFA uses several iterations to construct the Voronoi diagram. 
More specifically the number depends on the maximal distance between the mask 
points. As our densification progresses, the mask becomes denser and this 
distance decreases. This allows us to use fewer iterations in JFA, which 
speeds up the process.

%%%%%%%%%%%%%%%%%%%%%%%%%%%%%%%%%%%%%%%%%%%%%%%%%%%%%%%%%%%%%%%%%%%%%%%%%%%%%%

\section{Tonal Optimization with Domain Decomposition}
\label{sec:tonal}

\begin{figure}[tb]
	\begin{subfigure}[b]{0.325\linewidth}
		\centering
		\centerline{\includegraphics[width=1.0\linewidth]
			{images/trui/trui.png}}
		\subcaption{original image \textit{trui} }
		\label{fig:trui_tonal_a}
	\end{subfigure}
	\hfill
	\begin{subfigure}[b]{0.325\linewidth}
		\centering
		\centerline{\includegraphics[width=1.0\linewidth]
			{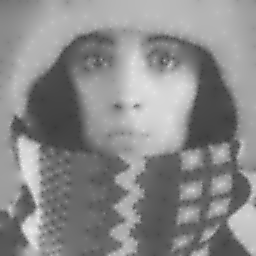}}
		\subcaption{\centering no tonal opt.: 141.48 MSE}
		\label{fig:trui_tonal_b}
	\end{subfigure}
	\hfill
	\begin{subfigure}[b]{0.325\linewidth}
		\centering
		\centerline{\includegraphics[width=1.0\linewidth]
			{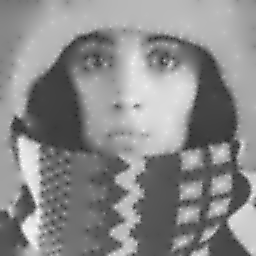}}
		\subcaption{\centering tonal opt.: 63.78 MSE}
		\label{fig:trui_tonal_c}
	\end{subfigure}
	\smallskip
	\caption{\textbf{Tonal Optimization Example on \textit{trui} with an 
	Optimized Mask of 2\% Density}. The tonal optimization enhances the 
	contrast and significantly improves the MSE.}
	\label{fig:trui_tonal}
\end{figure}

Homogeneous diffusion inpainting interpolates the values at the mask pixels 
from the reference image. By replacing the interpolation constraint with a best 
approximation condition over the whole image, we can obtain the optimal MSE 
given a specific image and mask. Solving this approximation problem is termed 
\textit{tonal optimization}. \cref{fig:trui_tonal} demonstrates the importance 
of tonal optimization for homogeneous diffusion inpainting. It significantly 
improves the sharpness of edges and leads to better grayscale or color values 
in homogeneous regions, which results in a huge improvement of the MSE.

Since the homogeneous diffusion inpainting operator 
$\bm{B} := \bm{A}^{-1} \bm{C}$ is a linear operator with respect to the 
grayscale or color values $\bm{f}$, we can formulate the tonal optimization
as a linear least squares optimization problem, given a fixed inpainting 
mask $\bm{c}$. 
We can replace the stored values at the mask $\bm{C}\bm{f}$ by $\bm{C}\bm{g}$, 
where we optimize $\bm{g}$ to give the best approximation w.r.t. the MSE:
\begin{equation}
\label{eq:tonal-opt-minimisation}
\underset{\bm{g} \in \mathbb{R}^N}{\argmin} \| \bm{f} - 
\bm{B} \, \bm{g}\|_2^2.
\end{equation}
Since grayscale or color values $\bm{g}_i$ at pixels $i \in \Omega \setminus K$ 
outside the inpainting mask $\bm{c}$ have no influence on the inpainting result 
we simply set them to $0$.
\newtext{The choice of the $2$-norm is especially appealing here (and rather 
standard in image processing) as it results in a linear least squares 
problem. Other options that result in linear least squares would 
include weighted $2$-norms $\|\cdot\|_{\bm{G}}$, however, in our case we want 
the reconstructed image $\bm{B}\bm{g}$ to look as close as possible to the 
original $\bm{f}$ overall, so introducing a weighting matrix does not 
provide any benefits. A weighted norm could be used potentially in 
a foveated rendering setting.} 

In the following subsection we discuss how we solve 
the resulting least squares problem efficiently.

%-----------------------------------------------------------------------------

%-----------------------------------------------------------------------------

\subsection{CGNR Solver for Tonal Optimization}
\label{sec:cgnr-tonal}

The global minima of the minimization problem 
(\ref{eq:tonal-opt-minimisation}) satisfy the normal equations, which in our 
case are given by the following positive semi-definite linear system of 
equations:
 \begin{equation}
 \label{eq:tonal-normal-equations}
  \bm{B}^\top \bm{B} \, \bm{g} = \bm{B}^\top \bm{f}.
\end{equation}
Since it is always symmetric and positive semi-definite, a standard conjugate 
gradient method (CG)~\cite{HS51, Sa03} can be used to solve it. Each iteration 
of the CG algorithm requires a matrix-vector multiplication and the evaluation 
of a few inner products.

%-----------------------------------------------------------------------------

\paragraph{Matrix-Vector Products}
While it is possible to compute this matrix explicitly, storing it requires 
memory that scales quadratically with the number of pixels. This becomes 
unfeasible for large images. To achieve high memory and computational 
efficiency, we keep our methods matrix-free and utilize the sparsity of the 
inpainting matrix $\bm{A}$. Therefore, we implement the matrix-vector products 
by solving linear systems involving $\bm{A}$ instead of precomputing 
$\textbf{B}$.

%\smallskip
\newtext{To compute the matrix-vector product $\bm{z}=\bm{B}^\top \bm{B}\bm{x}$ 
we first have to solve $\bm{A} \bm{y} = \bm{C} \bm{x}$ 
(for the step $\bm{y} = \bm{B}\bm{x} = \bm{A}^{-1}\bm{C}\bm{x}$), 
and then we have to solve $\bm{A}^\top \bm{C} \bm{z} = \bm{y}$ 
(for the step $\bm{z} = \bm{B}^\top\bm{y} = \bm{C}\bm{A}^{-\top}\bm{y}$).} 
This leads to a nested problem structure with the outer tonal optimization problem and the two inner inpainting problems in each iteration.

% This leads to a nested problem 
% structure, where the outer iterations optimize the color values at mask 
% points, while the inner iterations solve inpainting problems with the 
% matrix $\bm{A}$.

The linear system corresponding to the product $\bm{B}^\top\bm{w}$ involves 
the non-symmetric matrix $\bm{A}^\top$, which does not reduce to a symmetric 
system in the same way as $\bm{A}\bm{x}=\bm{b}$. Thus, we cannot solve it 
with the CG algorithm or the multigrid ORAS inpainting solver. 
As a remedy we use 
a symmetrized inpainting matrix $\tilde{\bm{A}}$ together with a modified 
$\tilde{\bm{C}}$, such that $\tilde{\bm{A}} \bm{u} = \tilde{\bm{C}} \bm{f}$ 
has the same solution as $\bm{A} \bm{u} = \bm{C} \bm{f}$. We can derive them 
with the following transformations, by using the fact that 
$\bm{C}\bm{u} = \bm{C}\bm{f}$, as the inpainting solution and the original 
image coincide at the mask pixels:
\begin{equation*}
\begin{aligned}
\left( \bm{C} + (\bm{I}-\bm{C}) \bm{L}\right)\bm{u} &= \bm{C}\bm{f} \\
\bm{C}\bm{u} + 
(\bm{I}-\bm{C})\bm{L}(\bm{I}-\bm{C}+\bm{C})\bm{u} &= 
\bm{C}\bm{f} \\
%\bm{C}\bm{u} + (\bm{I}-\bm{C})L(\bm{I}-\bm{C})\bm{u} + 
%\underbrace{(\bm{I}-\bm{C})\bm{L}\bm{C}\bm{u}}_{=(\bm{I}-\bm{C})\bm{L}\bm{C}\bm{f}}
% &= \bm{C}\bm{f}\\
\bm{C}\bm{u} + (\bm{I}-\bm{C})\bm{L}(\bm{I}-\bm{C})\bm{u} + 
(\bm{I}-\bm{C})\bm{L}\bm{C}\bm{u} &= \bm{C}\bm{f}\\
\bm{C}\bm{u} + (\bm{I}-\bm{C})\bm{L}(\bm{I}-\bm{C})\bm{u} + 
(\bm{I}-\bm{C})\bm{L}\bm{C}\bm{f} &= \bm{C}\bm{f}\\
% \bm{C}\bm{u} + (\bm{I}-\bm{C})\bm{L}(\bm{I}-\bm{C})\bm{u}  
%  &= \bm{C}\bm{f} - (\bm{I}-\bm{C})\bm{L}\bm{C}\bm{f}\\
\underbrace{\left(\bm{C} + 
	(\bm{I}-\bm{C})\bm{L}(\bm{I}-\bm{C})\right)}_{=:\tilde{\bm{A}}} \bm{u} &= 
\underbrace{\left(\bm{C}-(\bm{I}-\bm{C})\bm{L}\bm{C}\right)}_{=:\tilde{\bm{C}}} 
\bm{f}.
\end{aligned}
\end{equation*}
Consequently, we are able to replace the linear systems with their 
symmetrized variants, meaning we only have to solve symmetric linear systems. 
Solving both the outer problem and the inner symmetrized inpainting problems 
with a CG solver leads to the nested CG approach from Chizhov and 
Weickert~\cite{CW21}.

\newtext{We note that one could also apply other solvers for 
symmetric problems to solve $\tilde{\bm{A}}\bm{u}=\tilde{\bm{C}}\bm{f}$, 
such as the conjugate residual (CR) method, the minimal residual (MINRES) 
solver~\cite{PS75}, or the symmetric LQ (SYMMLQ) algorithm~\cite{PS75}. 
However, while those methods minimize the residual faster compared 
to conjugate gradients, and work for indefinite systems, CG remains the 
fastest w.r.t.\ minimizing the error to the solution $\bm{u}^*$ for 
our problem. 
In our case $\tilde{\bm{A}}$ is positive definite as long as 
$\bm{c}\ne\bm{0}$, so we leverage this fact by using CG.}

%-----------------------------------------------------------------------------
\paragraph{Improvements}
We improve the nested CG approach in two ways. First, we replace the inner CG 
solvers for the two inpainting problems by the multigrid ORAS inpainting 
solver, which improves the runtime by more than a factor of 4. 
Secondly, we propose 
to replace the outer CG solver with CGNR, a variant of CG for the normal 
equations~\cite{Bj96a, HS51, Sa03}. Besides better numerical stability 
compared to CG, CGNR also has the advantage that the original inpainting 
residual $\bm{r} = \bm{f} - \bm{A} \bm{u}$, and hence the MSE, is available 
in each iteration. 
Because we want to optimize the MSE, a stopping criterion based on the relative 
improvement of the MSE in an iteration is better suited for our purposes than
the standard stopping criterion based on the $2$-norm of the residual of the 
normal equations $\|\bm{B}^\top(\bm{f}-\bm{B}\bm{g})\|_2$.

%-----------------------------------------------------------------------------
\subsection{RAS Solver for Tonal Optimization}
\label{sec:ras-tonal}

Since the influence zone of each mask pixel is effectively local, as is seen 
in~\cite{H17}, this also holds for the system matrix of the normal equations.
The system matrix has non-negligible entries only for pairs of mask pixels 
that are near each other. All entries outside a local neighborhood are nearly 
zero and can be ignored without introducing large errors. This leads to a 
sparse matrix structure with localized connections, which is very suitable 
for domain decomposition methods. 

%-----------------------------------------------------------------------------

\paragraph{Domain Decomposition}
Taking into account the above, we solve the normal equations with a domain 
decomposition method instead of CGNR. We adapt a 
\textit{restricted additive Schwarz (RAS)} method~\cite{CS99}, 
which is similar to the one used in the multigrid ORAS inpainting solver. 
Similar to the inpainting solver, we 
subdivide the image domain $\Omega$ into $k\in \mathbb{N}$ overlapping 
subdomains $\Omega_1,...,\Omega_k \subset \Omega$, such that 
$\cup_{i=1}^k\Omega_i=\Omega$. The resulting RAS tonal optimization is 
given in \cref{alg:ras-tonal}. It consists of three steps that are iterated 
until the method is sufficiently converged.

%-----------------------------------------------------------------------------

\begin{algorithm}[htb]
	\caption{RAS for Tonal Optimization} 
	\label{alg:ras-tonal} 
	\begin{enumerate}
		\item
		Compute the global residual $\bm{r}^n \in \mathbb{R}^N$:
		%		\begin{equation}
			%		\bm{r}^n =  \bm{A}^{-1} \bm{C}  \bm{f} 
			%		- 
			%		((\bm{A}^{-1} \bm{C})^\top 
			%		(\bm{A}^{-1} \bm{C})) \, \bm{g}^n  
			%		\end{equation}
		\begin{equation}
			%		\bm{r}^n =  (\tilde{\bm{A}}^{-1}\tilde{\bm{C}})^\top  
			%\bm{f} - 
			%		(\tilde{\bm{A}}^{-1}\tilde{\bm{C}})^\top 
			%		(\tilde{\bm{A}}^{-1} \tilde{\bm{C}}) \, \bm{g}^n  
			\bm{r}^n =  \bm{B}^\top (\bm{f} -  \bm{B} \, \bm{g}^n)  
			%		\bm{r}^n =
			%		
			%(\tilde{\bm{A}}^{-1}\tilde{\bm{C}})^\top(\bm{f}-\bm{A}^{-1}\bm{C}\bm{g}^n)
		\end{equation}
		\item
		For $i \in \{1, ..., k\}$ solve for a local correction 
		$\bm{v}_i^n$:
		%		\begin{equation}
			%		((\bm{A}^{-1}_i \bm{C})_i^\top 
			%		(\bm{A}^{-1}_i \bm{C}_i)) \,
			%		\bm{v}_i^n = \bm{R}_i \bm{r}^n
			%		\end{equation}
		\begin{equation}
			%		(\tilde{\bm{C}}_i^\top \tilde{\bm{A}}^{-1}_i ) 
			%		(\tilde{\bm{A}}^{-1}_i \tilde{\bm{C}}_i) \,
			%		\bm{v}_i^n = \bm{R}_i \bm{r}^n
			\bm{B}^\top_i \bm{B}_i \,
			\bm{v}_i^n = \bm{R}_i \bm{r}^n
		\end{equation}
		\item 
		Update $\bm{g}^n$ with the weighted and extended local 
		corrections $\bm{v}_i^n$:
		\begin{equation}
			\bm{g}^{n+1} = \bm{g}^n + \sum_{i=1}^{k} 
			\bm{R}_i^\top \bm{D}_i \bm{v}_i^n
		\end{equation}
	\end{enumerate}
\end{algorithm}

%-----------------------------------------------------------------------------

\noindent
First we compute the global residual $\bm{r}^n$. This involves the 
application of the system matrix. As in \cref{sec:cgnr-tonal} we evaluate 
the matrix-vector products without explicitly forming the matrix, by solving 
two inner inpainting problems.
As the inpainting solution changes only slightly from one iteration to the 
next, it can be used as a good initialization for the next inpainting.
This allows us to skip the initialization phase in the full multigrid ORAS 
inpainting solver and start directly with a V-cycle at the finest resolution 
level, which improves the runtime of both inpaintings significantly. This is 
an advantage of RAS compared to CGNR, where we apply the system matrix only 
to conjugate vectors, which change more from one iteration to the next, and 
thus cannot be used as a good initialization. 

%-----------------------------------------------------------------------------

\paragraph{Local Normal Equations}
In the second step, we compute local corrections $\bm{v}_i^n$ to the residual 
for each subdomain, obtained by solving local versions of the global normal 
equations. The right-hand side for each subdomain $\Omega_i$ is given by 
restricting the residual $\bm{r}^n$ with the same restriction operator 
$\bm{R}_i$ as in the inpainting case. For the local system matrix, we 
individually restrict the symmetrized inpainting matrix $\tilde{\bm{A}}$ to 
$\tilde{\bm{A}}_i$ and the modified mask $\tilde{\bm{C}}$ to 
$\tilde{\bm{C}_i}$.
By restricting $\bm{C}$, $\bm{L}$, and $\bm{I}$ individually, we get 
homogeneous Neumann boundary conditions at all subdomain boundaries for the 
inner inpainting problems.

In the third step, we weight the local corrections $\bm{v}_i^n$ with weights
$\bm{D}_i$ and add them to the global solution $\bm{g}^n$. The weights 
$\bm{D}_i$ have to fulfill the partition of unity property in order to 
guarantee convergence.
\begin{figure}[tb]
%	\begin{minipage}[b]{0.49\linewidth}
    \centering
    \centerline{\includegraphics[width=1.0\linewidth]
        {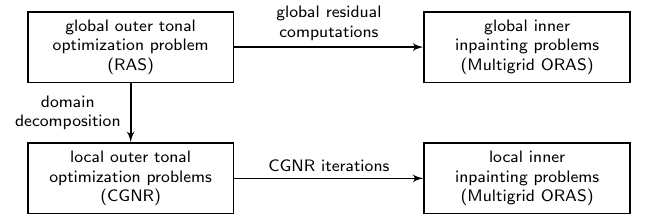}}
    
    \caption{\textbf{RAS Tonal Optimization Schematic.} 
        The RAS tonal optimization method consists of four problems 
            types, which are solved with different solvers.}
    \label{fig:tonal-ras-schematic}
\end{figure}
%

%-----------------------------------------------------------------------------
\paragraph{Solver Structure}

Overall our RAS method consists of four types of problems, which are shown in 
\cref{fig:tonal-ras-schematic}. Our \emph{global tonal optimization} 
problem is solved with RAS which includes two \emph{global inner inpainting} 
problems per iteration to compute the global residual. As with our spatial 
optimization method, we solve them with the multigrid ORAS inpainting 
solver. For each domain decomposition block in the RAS 
solver we also get \emph{local tonal optimization} problems which we solve 
with CGNR. These again include two \emph{local inner inpainting} problems per 
iteration that we solve with the same multigrid ORAS inpainting method.

% For the domain decomposition we use blocks of size $64\times64$, an
% overlap of $6$ pixels and a simple averaging for mask pixels in the overlap. 
% In practice, this ensures sufficient communication between the blocks such 
% that the method converges towards the global solution. 

\newtext{
For the domain decomposition we use blocks of size $64\times64$, an
overlap of $6$ pixels and a simple averaging for mask pixels in the overlap. 
The size of the blocks should be a multiple of $32$ due to the architecture of
our GPU in order to achieve a good performance. 
We found that a block size of $64$ and an overlap of $6$ pixels between them 
is optimal for our GPU, as the blocks are 
still quite small while the average number of mask pixels per block and the 
overlap is sufficiently large for a good convergence towards the global 
solution.}

%-----------------------------------------------------------------------------
\subsection{Voronoi Initialization}
\label{sec:voronoi-init}

Typically one would initialize CGNR or RAS for tonal optimization at the mask pixels with the values from the original image. While this is a reasonable 
first guess, an even better initialization can significantly improve the 
runtime of the tonal optimization. Therefore, we propose an initialization 
strategy inspired by an early predecessor of tonal optimization by Gali\'c 
et al.~\cite{GWWB08}. Along with the algorithmic improvements, we also 
provide a new theoretical perspective that subsumes both our method and the one of Gali\'c et al..

%-----------------------------------------------------------------------------
\paragraph{Error Balancing}
The approach by Gali\'c et al.~\cite{GWWB08} does not solve the full tonal 
optimization problem, but it already leads to a significant quality 
improvement, compared to the non-optimized inpainting result. The idea is to 
adjust the tonal value of each mask pixel, by adding the average signed 
inpainting error of its neighbors to it: 
\begin{equation}
\bm{g}_i = \bm{u}_i + \frac{1}{| \mathcal{N}(i) |}\sum_{j\in 
	\mathcal{N}(i)} (\bm{f}_j - \bm{u}_j) \quad \text{for all } i\in K,
\end{equation}
where $\mathcal{N}(i)$ denote all direct neighbors of mask pixel $i$, including 
the diagonal ones and itself. 
Since homogeneous diffusion inpainting suffers from logarithmic singularities 
at the mask pixels, their color values differ significantly from the color 
values of the surrounding pixels. This leads to a large error in the 
neighborhood, which this method tries to compensate for by modifying the 
value at the mask pixel. 

%-----------------------------------------------------------------------------
\paragraph{Interpolation of Local Averages}
We interpret the above method as performing a single step towards making 
mask pixels interpolate the averages over the neighborhoods. 
This is not unlike the local average constraints used in~\cite{JCW23}.
Taking this interpretation as a starting point we can rewrite the averaging 
around the mask pixels with a weight matrix 
$\bm{W} \in \mathbb{R}^{N \times N}$ to generalize the method. 
Instead of interpolating $\bm{f}$ at the mask pixels, we want to interpolate 
the averages $(\bm{W} \bm{f})_i$ around the mask pixels, which leads to the 
following constraints:
\begin{equation}
\label{eq:average-interpolation}
\bm{C}\bm{W}\bm{A}^{-1}\bm{C} \bm{g} =: \bm{C}\bm{W}\bm{u}
                                     =  \bm{C}\bm{W}\bm{f}.
\end{equation}
We solve (\ref{eq:average-interpolation}) with a modified Richardson 
iteration:
\begin{equation}
\bm{C}\bm{g}^{k+1} = \bm{C}\bm{g}^{k} + \tau (\bm{C}\bm{f} - 
\bm{C}\bm{W}\bm{A}^{-1} \bm{C}\bm{g}^{k}). 
\end{equation}
In our framework the original method by Gali\'c et al.\ corresponds to a 
single Richardson step with step size $\tau=1$, and a very simple $\bm{W}$. 
By performing multiple steps instead, we can further improve the MSE. 
We pick $\tau$ so that the scheme is stable in the $2$-norm: 
$\|\bm{C} - \tau \bm{C} \bm{W} \bm{A}^{-1}\bm{C} \|_2 < 1$. We propose to 
stop the iteration once the MSE starts to increase. This can happen, as the 
average interpolation problem is not the same as the tonal optimization 
problem, but is only an approximate surrogate for it.

%-----------------------------------------------------------------------------

\paragraph{Voronoi Diagram Average}
To get a closer surrogate to the tonal optimization problem, we propose to 
use a better approximation of the influence zone of each mask pixel. As 
discussed in \cref{sec:spatial}, the cells in the Voronoi diagram 
offer a reasonable approximation of the influence zones of the inpainting 
echoes.
Thus, we propose to replace the direct neighbors in $\bm{W}$ with the 
neighbors in the Voronoi cell of each mask pixel.
Furthermore, we also introduce a weighting, so that we down-weight pixels 
that are further away from the corresponding mask pixel and thus less 
important for the optimization. This leads to an improved approximation of 
the influence zones.

%%%%%%%%%%%%%%%%%%%%%%%%%%%%%%%%%%%%%%%%%%%%%%%%%%%%%%%%%%%%%%%%%%%%%%%%%%%%%%
\section{Experiments}
\label{sec:experiments}	

\begin{figure}[tb]
	\centering
	\setlength{\tabcolsep}{1mm}
	\begin{tabular}{ccc}
		\includegraphics[width=.32\linewidth]
		{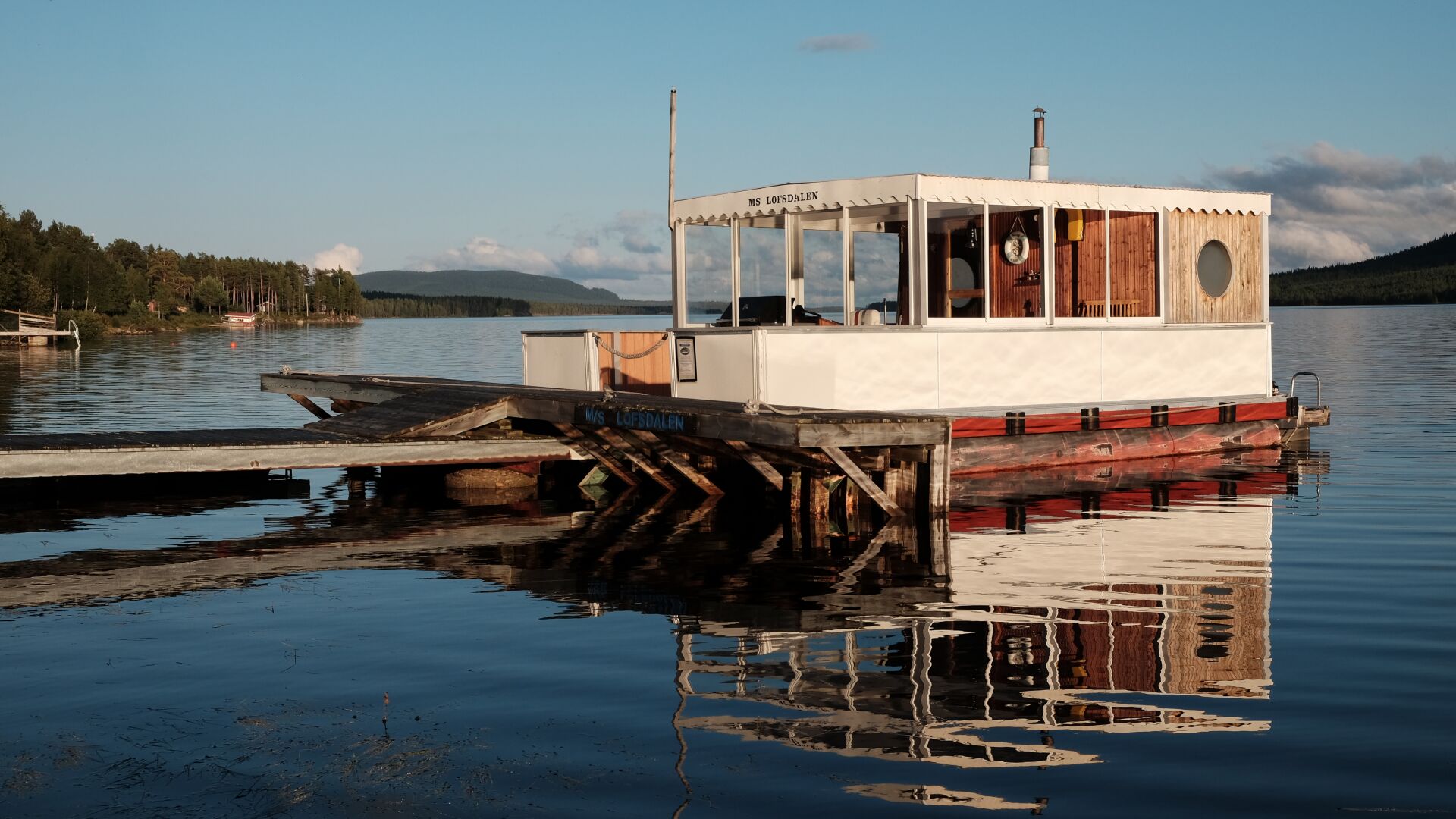} &
		\hfill
		\includegraphics[width=.32\linewidth]
		{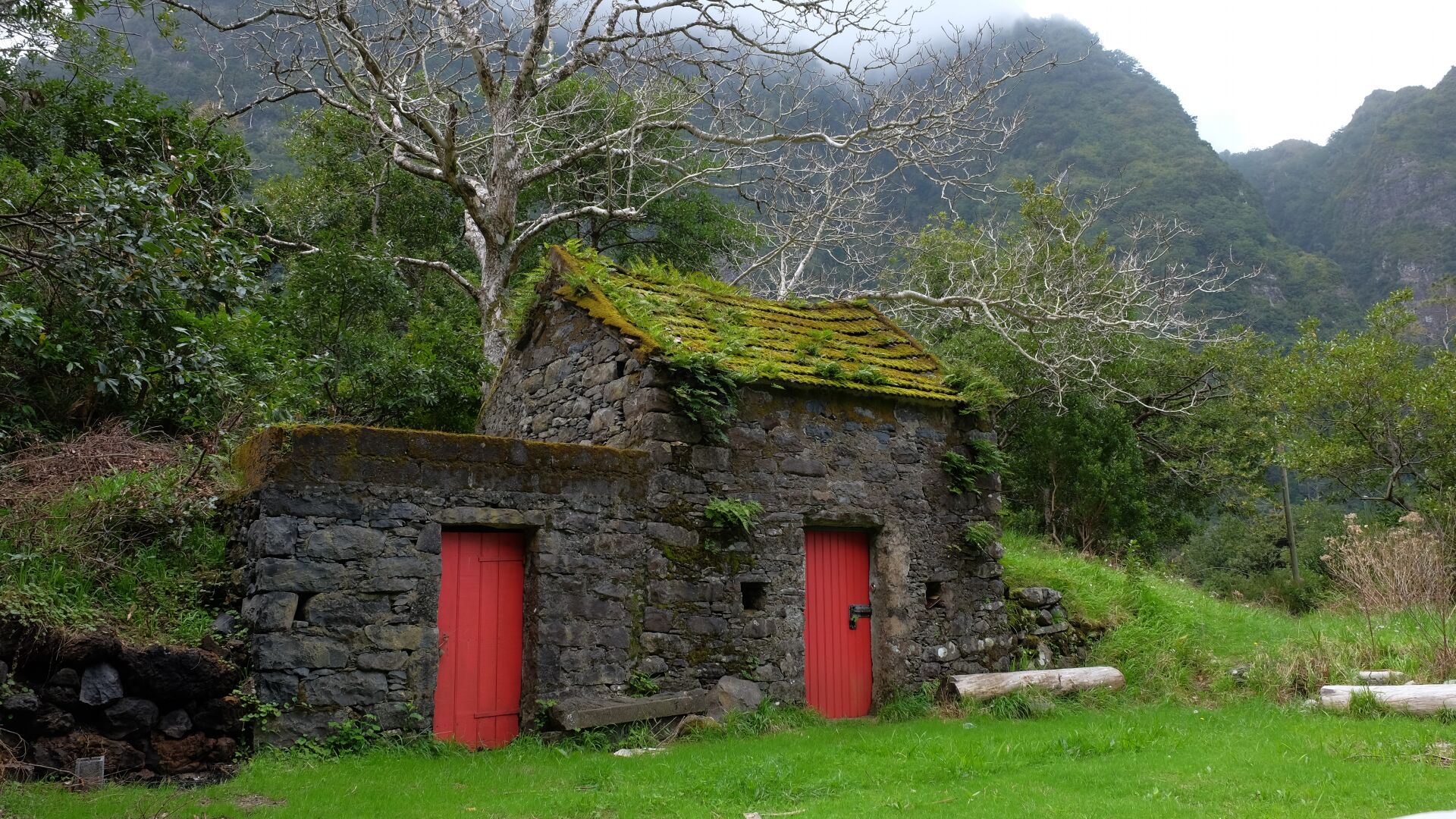} &
		\hfill
		\includegraphics[width=.32\linewidth]
		{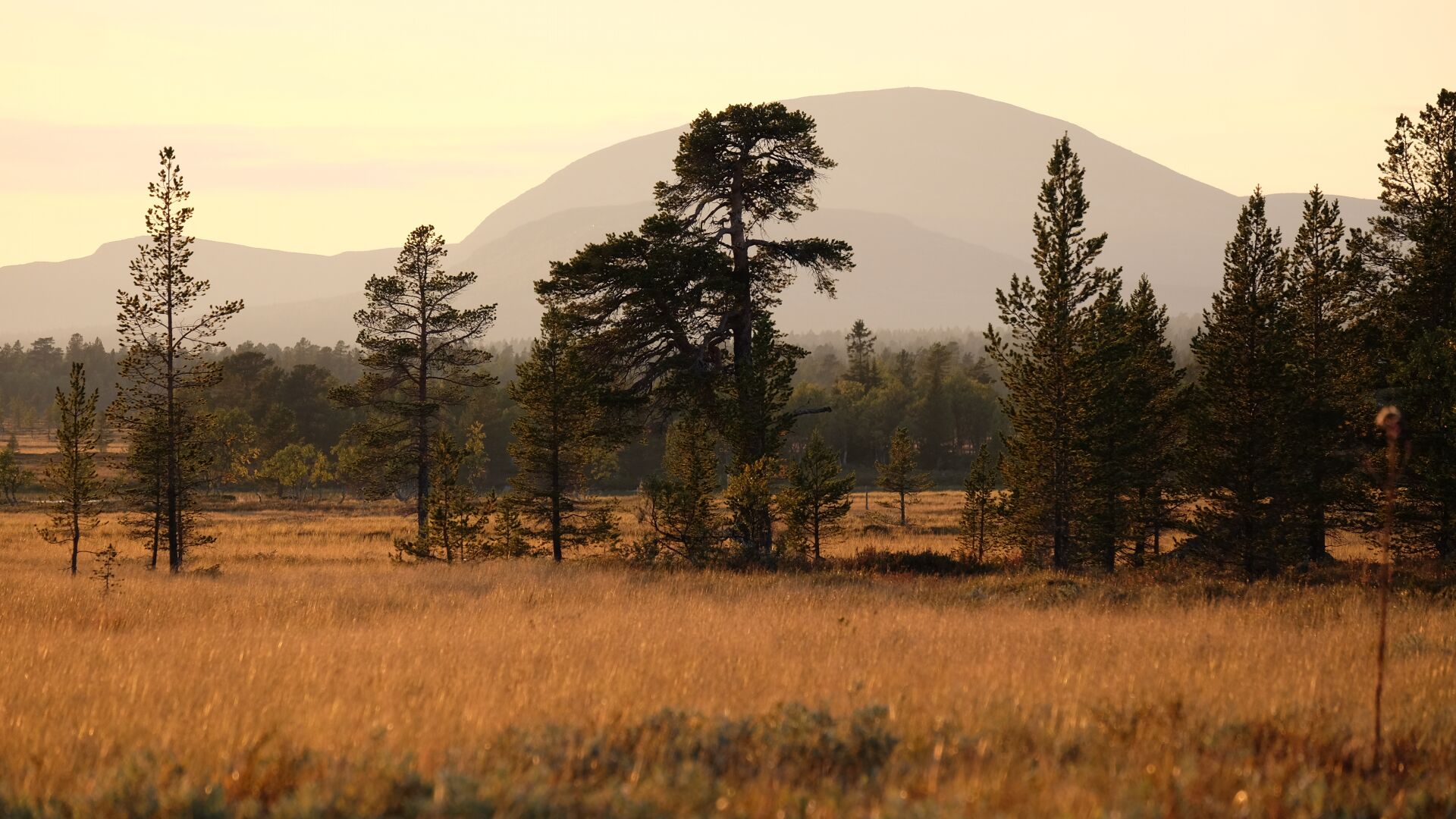} \\[1mm]
		
		\includegraphics[width=.32\linewidth]
		{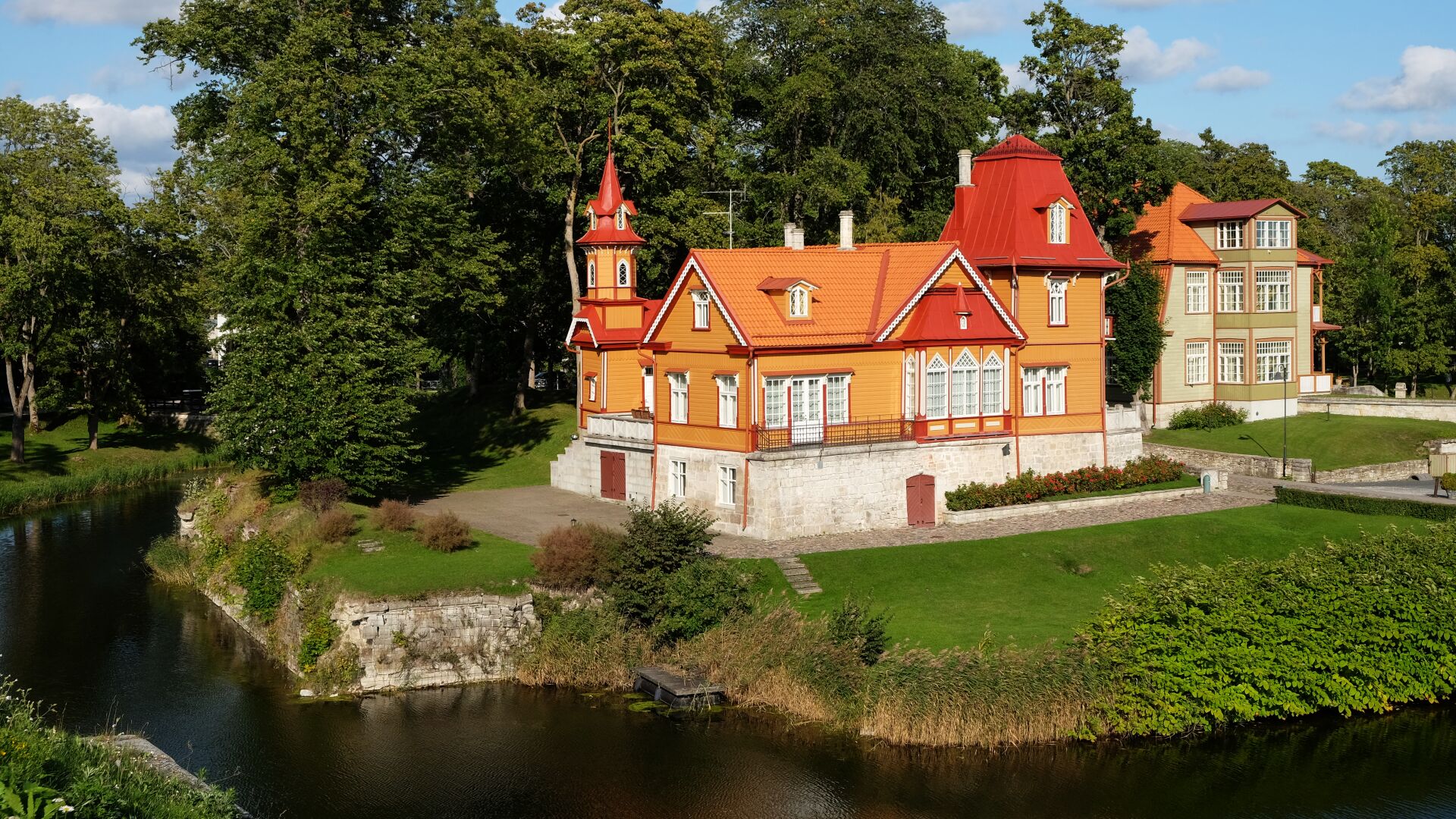} &
		\includegraphics[width=.32\linewidth]
		{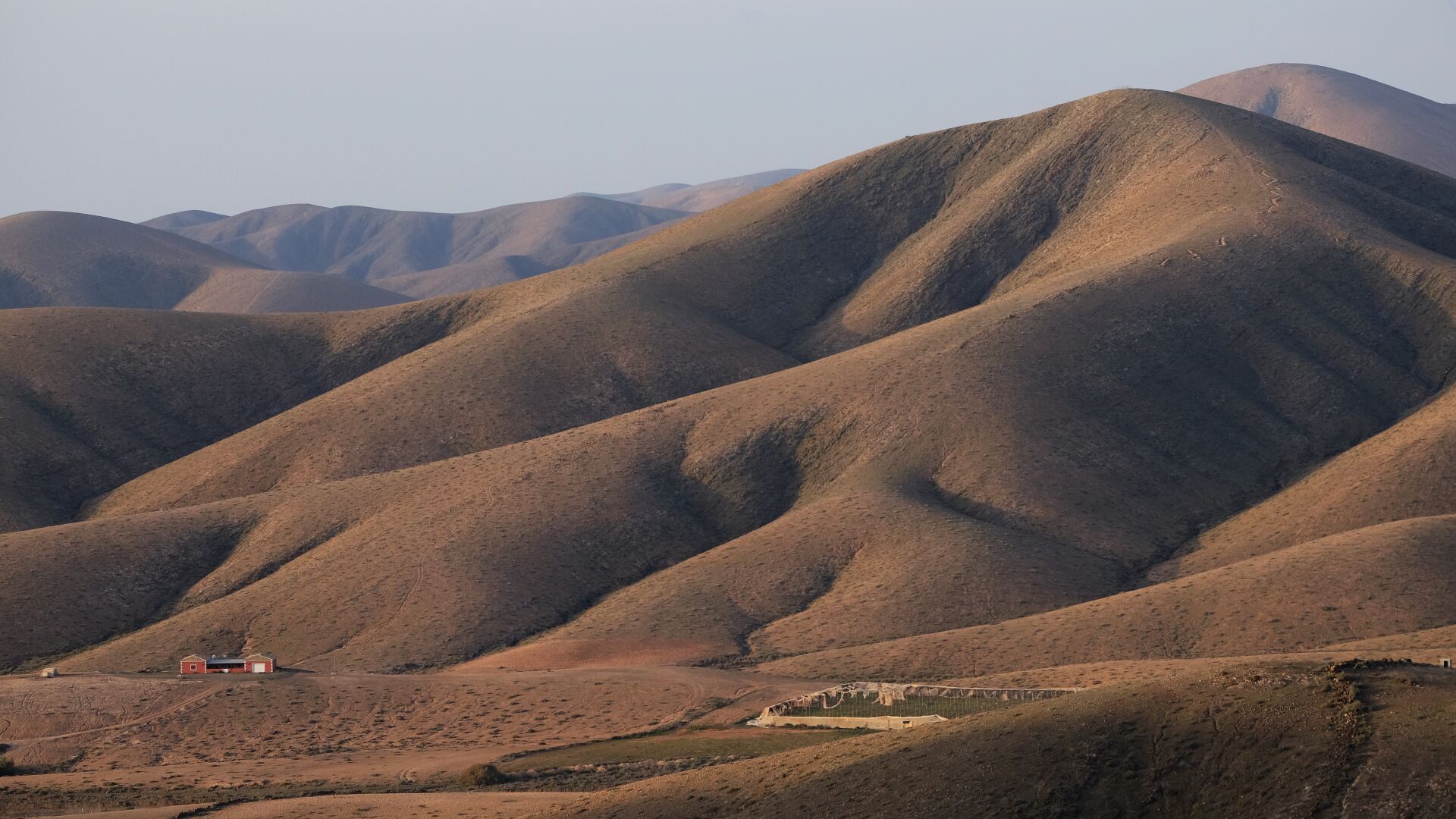} &
		\includegraphics[width=.32\linewidth]
		{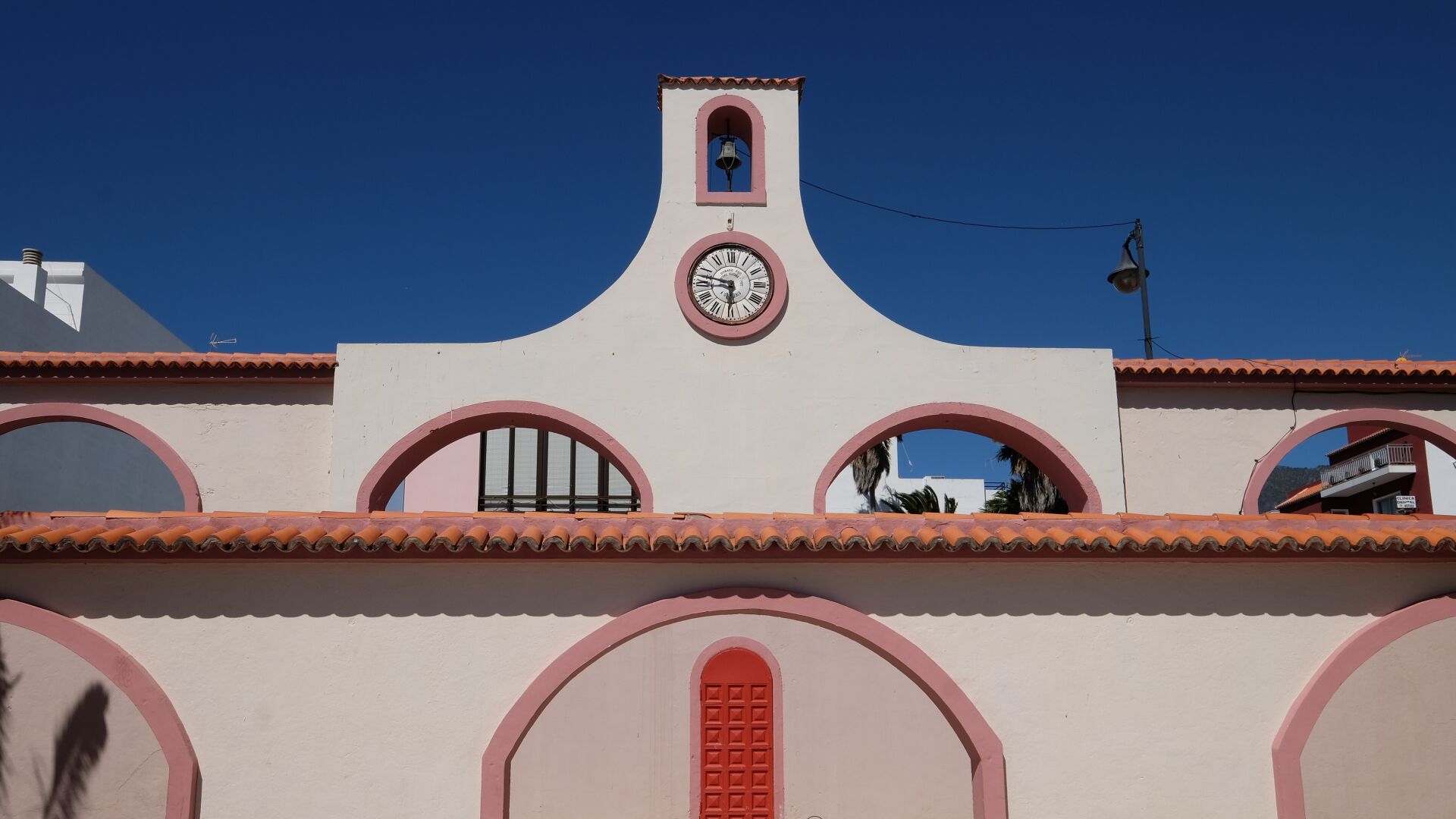} \\[1mm]
		
		\includegraphics[width=.32\linewidth]
		{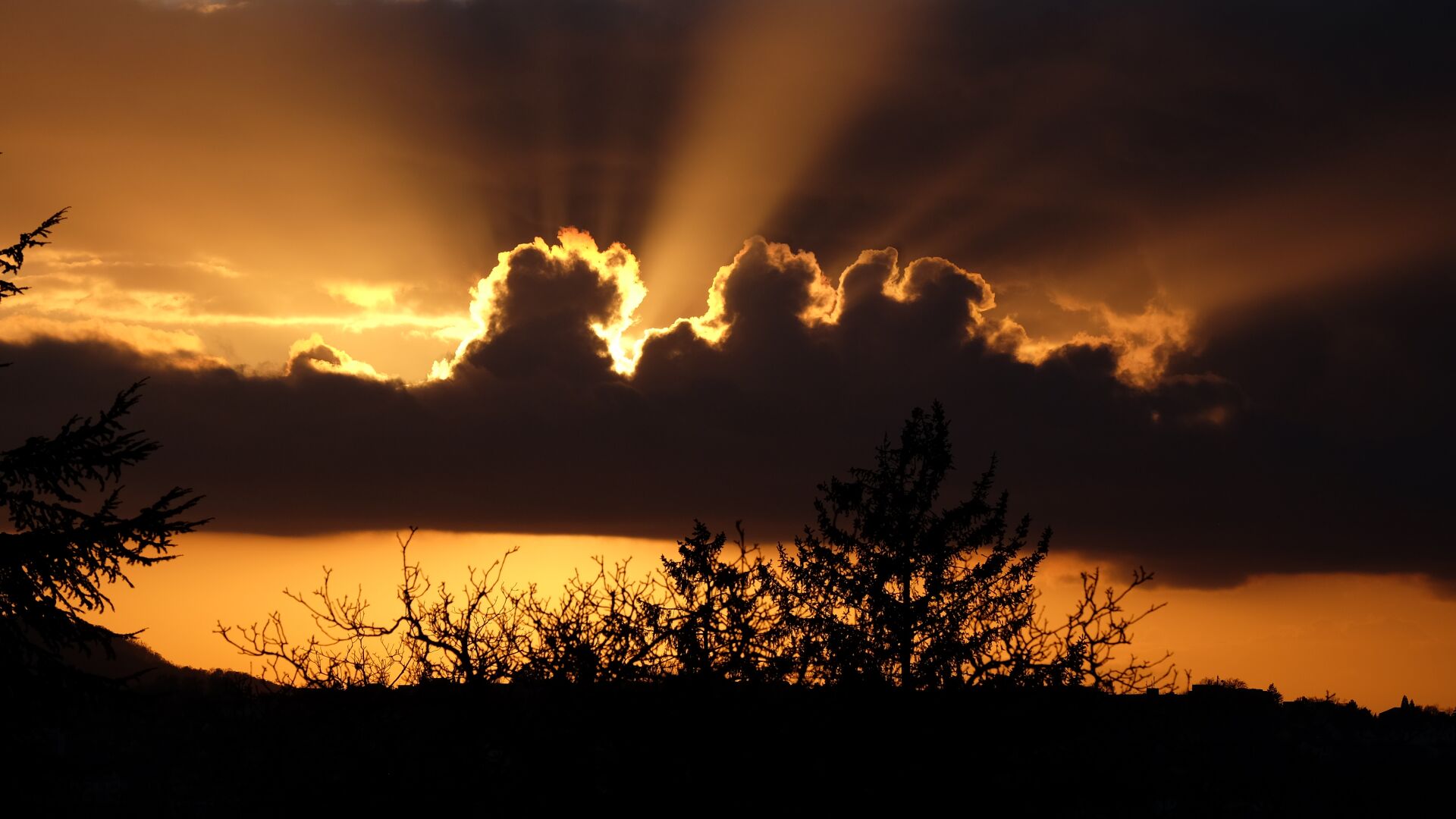} &
		\includegraphics[width=.32\linewidth]
		{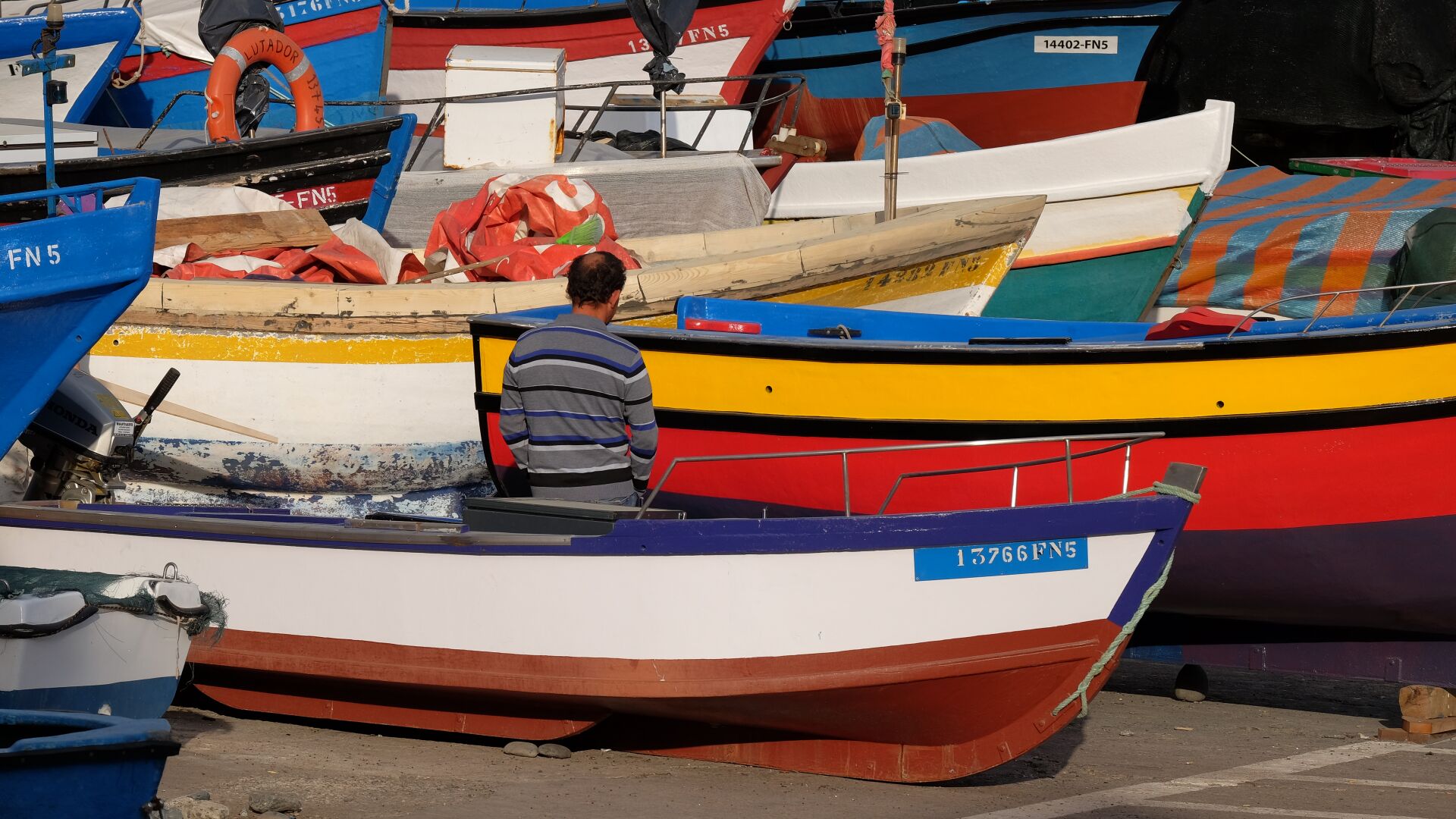} &
		\includegraphics[width=.32\linewidth]
		{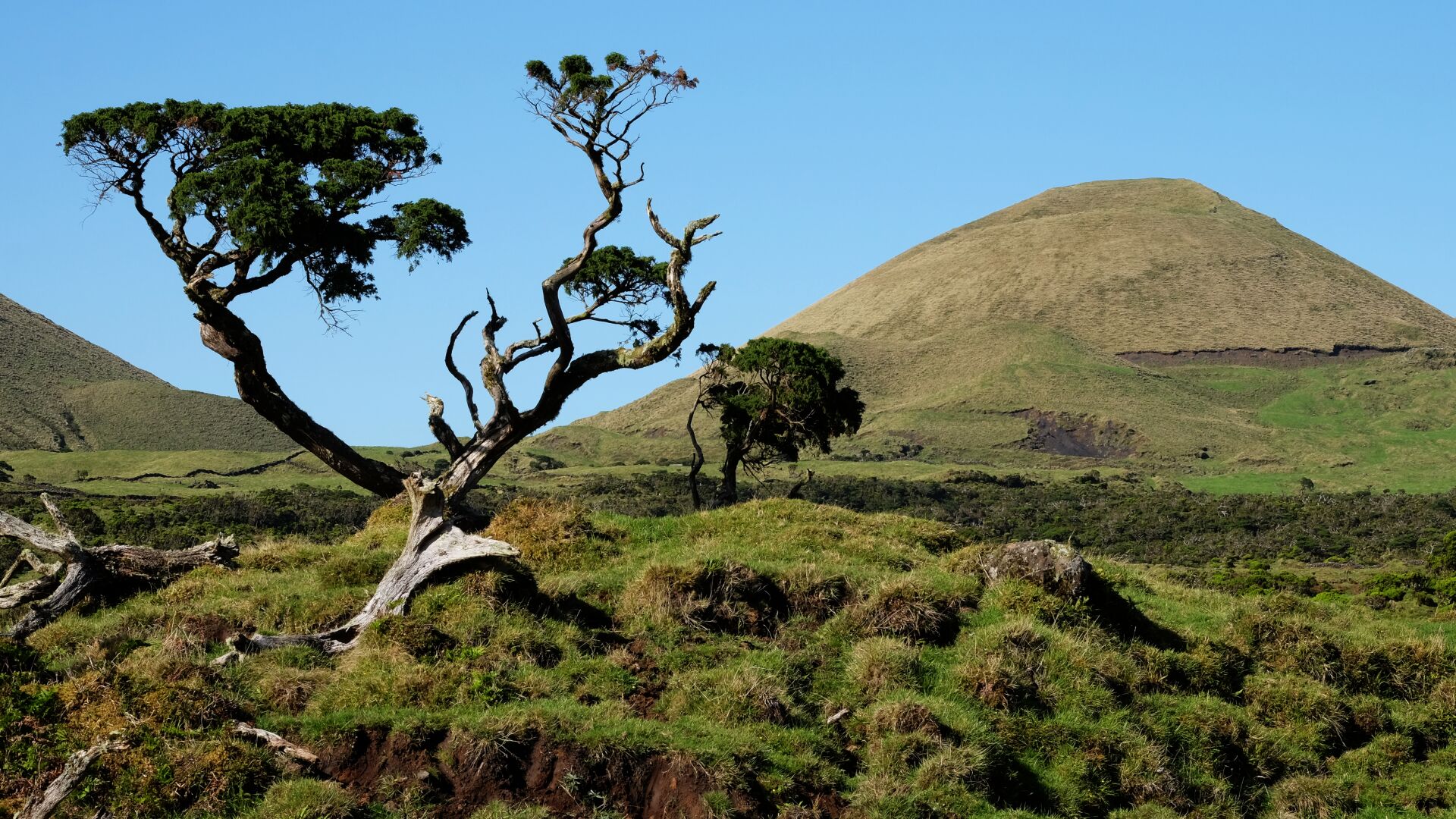} \\[1mm]
		
		\includegraphics[width=.32\linewidth
		]{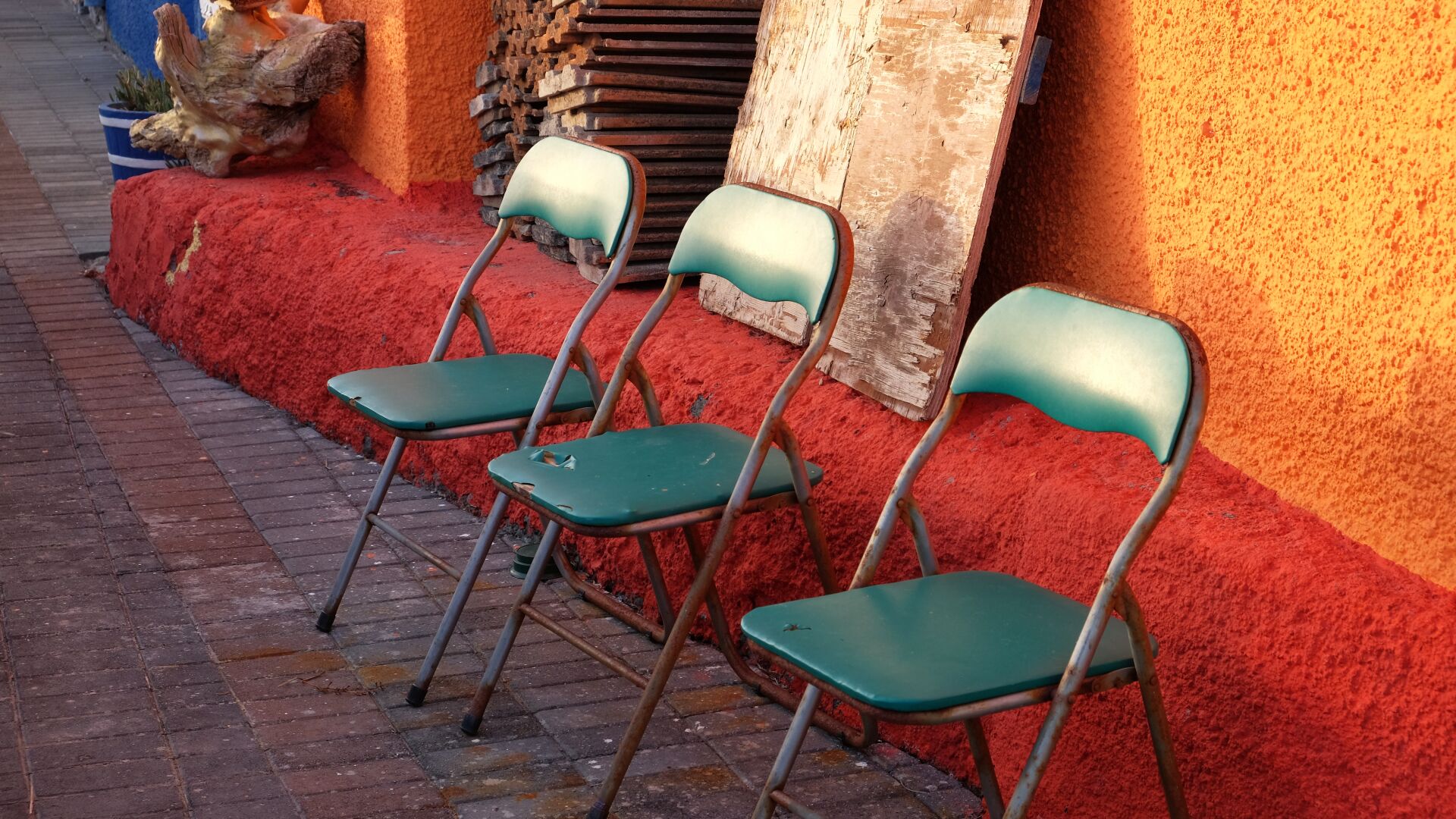} &
		\includegraphics[width=.32\linewidth]
		{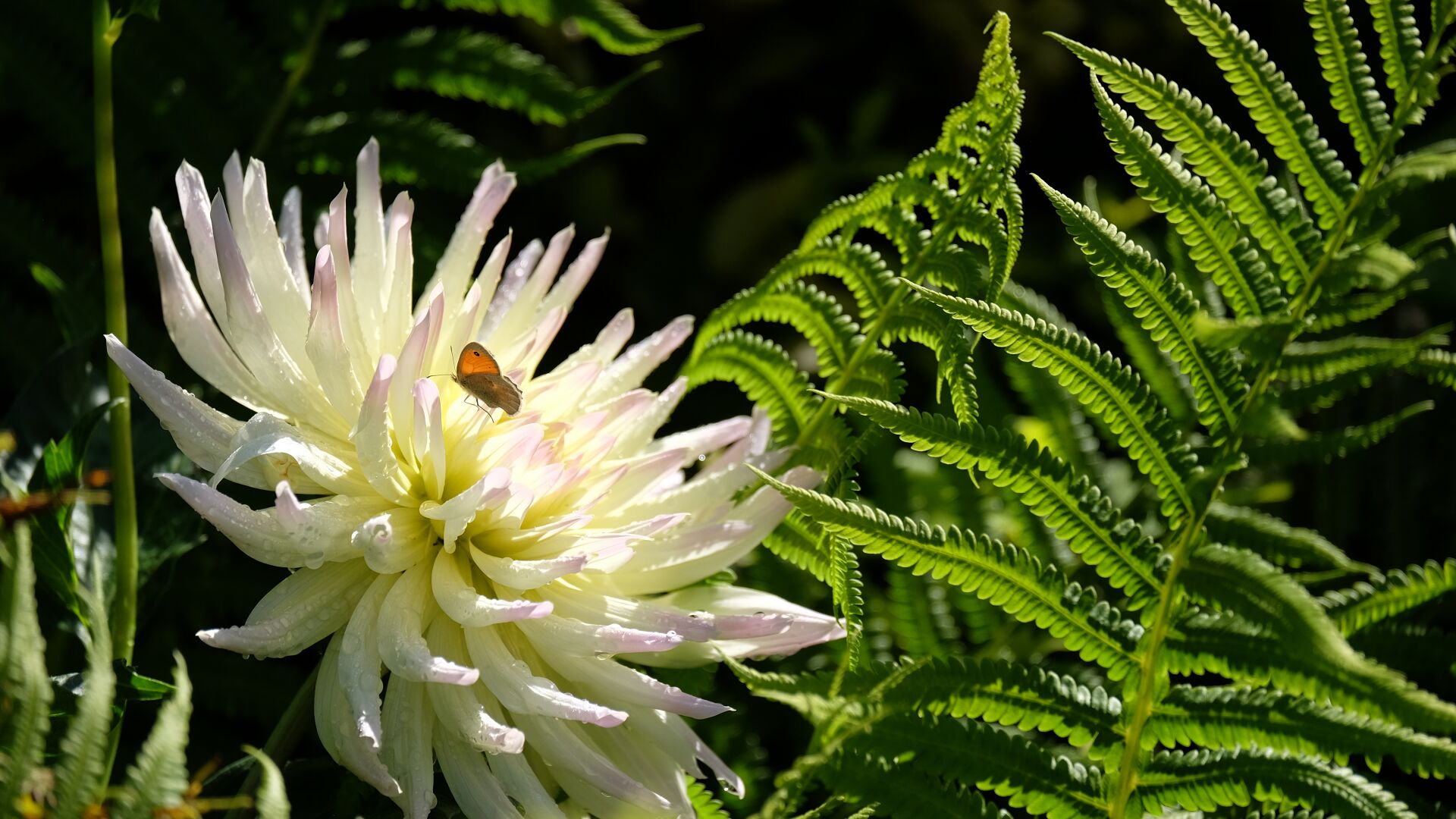} &
		\includegraphics[width=.32\linewidth]
		{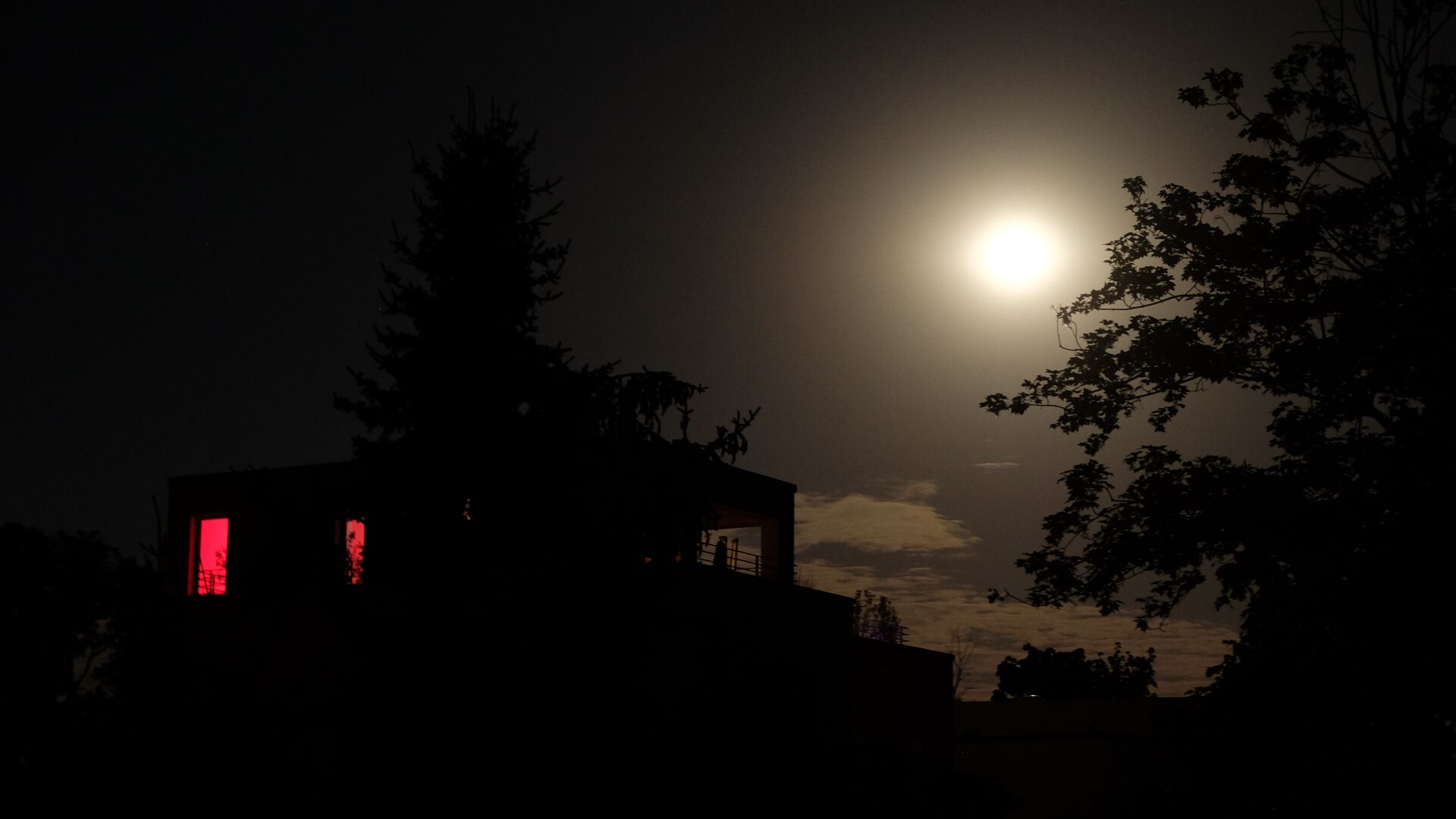}
		 
	\end{tabular}

	\caption{Our twelve test images of $4K$ resolution. Photos by J. 
 Weickert.}
	\label{fig:dataset}
\end{figure}

%-----------------------------------------------------------------------------

After mentioning the technical details of our experimental setup in 
\cref{sec:exp_setup}, we show the results of our spatial optimization method 
in \cref{sec:exp_spatial_opt} and the results for our domain decomposition 
tonal optimization framework in \cref{sec:exp_tonal_opt}.

%-----------------------------------------------------------------------------
\subsection{Experimental Setup}
\label{sec:exp_setup}

For the experimental evaluation of our Delaunay densification spatial 
optimization method, we compare against the analytic approach (AA) by 
Belhachmi et al.~\cite{BBBW08}, probabilistic 
sparsification (PS)~\cite{MHWT12}, with (PS+NLPE) and without 
nonlocal pixel exchange, and the most recent neural approach by 
Schrader et al.~\cite{SPKW23}. The analytic approach is implemented with 
Floyd-Steinberg dithering~\cite{FS76} of the Laplacian magnitude. 
To ensure fair comparisons, PS and NLPE are both implemented on the GPU with 
the same multigrid ORAS inpainting solver that we use for our methods.
For PS we use candidate fractions $p=0.3$ and $q=0.005$, and NLPE is run for 
five cycles. As the neural approach requires more than 20 GB of graphics 
memory for runtime tests, we exclude them in our runtime comparison. 

To evaluate our tonal optimization methods, we compare against a parallel GPU 
implementation of CGNR with CG inpainting, based on the nested CG approach by 
Chizhov and Weickert~\cite{CW21}. We adapt this approach by replacing CG 
with CGNR, and the finite elements with finite differences, as this is better 
suited for an efficient parallel GPU implementation. 
We exclude other tonal optimization approaches, such as the Green's function 
approach~\cite{H17} or the inpainting echo approach~\cite{MHWT12}, because 
they either use too much memory for 4K images or they are not well-suited 
for parallelization. 
All our experiments were conducted on an \textit{AMD Ryzen 5900X@3.7GHz} 
and an \textit{Nvidia GeForce GTX 1080 Ti} GPU.
Tests are performed on twelve representative 4K images containing scenes of 
nature with a varying amount of texture, and coarse and fine structures 
(see \cref{fig:dataset}).

%-------------------------------------------------------------------------------
\subsection{Spatial Optimization Experiments}
\label{sec:exp_spatial_opt}

\begin{figure}[tb]
	\begin{subfigure}[b]{0.49\linewidth}
		\centering
		\centerline{\includegraphics[width=1.0\linewidth]
			{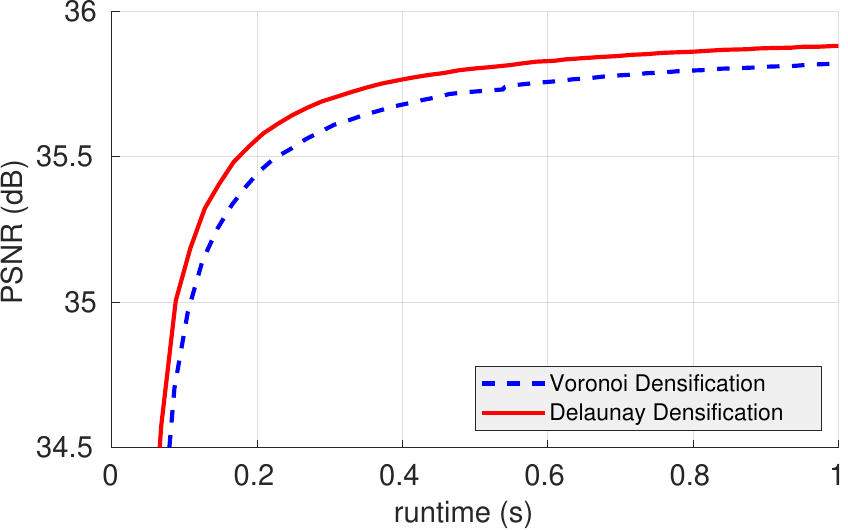}}
		\subcaption{5\% mask density}
	\end{subfigure}
	\hfill
	\begin{subfigure}[b]{0.49\linewidth}
		\centering
		\centerline{\includegraphics[width=1.0\linewidth]
			{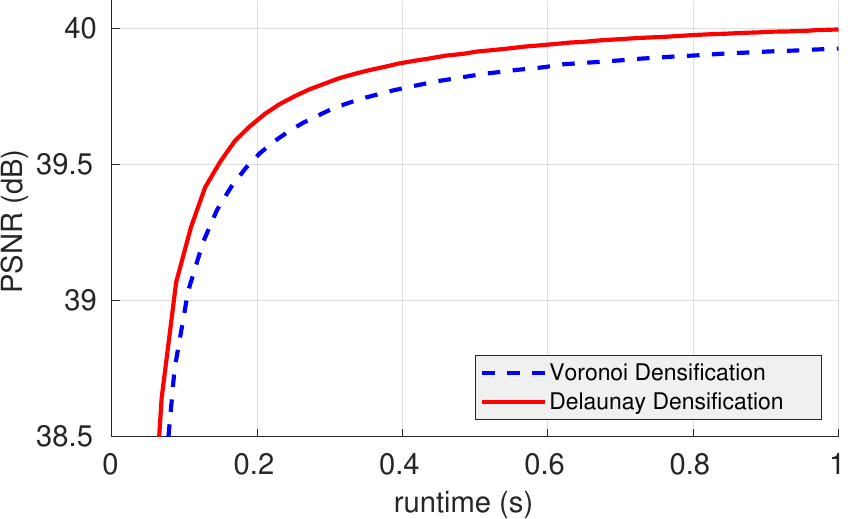}}
		\subcaption{10\% mask density}
	\end{subfigure}
    \smallskip
	\caption{\textbf{Comparison between Voronoi Densification 
        and Delaunay Densification.} Results for a mask density of 5\% (a) and 10\% (b). The Delaunay-based approach clearly outperforms the Voronoi-based one at all times by about $0.1$ dB.}
	\label{fig:spatial_vd_dd}
\end{figure}

In \cref{fig:spatial_vd_dd} we compare a Voronoi-based densification with a 
Delaunay-based one for mask densities of 5 and 10\%, by varying the number 
of densification iterations. Since Voronoi and Delaunay densification have 
a different runtime per iteration, we compare the PSNR against the total  
runtime instead of the number of iterations. We observe that for both 
densities, the Delaunay densification outperforms Voronoi densification 
by approximately $0.1$ dB. This suggests that the Delaunay triangulation 
is in general better suited for a densification approach than 
the Voronoi tessellation.

%-----------------------------------------------------------------------------

%-----------------------------------------------------------------------------

\begin{figure}[tb]
    \begin{subfigure}[b]{0.49\linewidth}
		\centering
		\centerline{\includegraphics[width=1.0\linewidth]
			{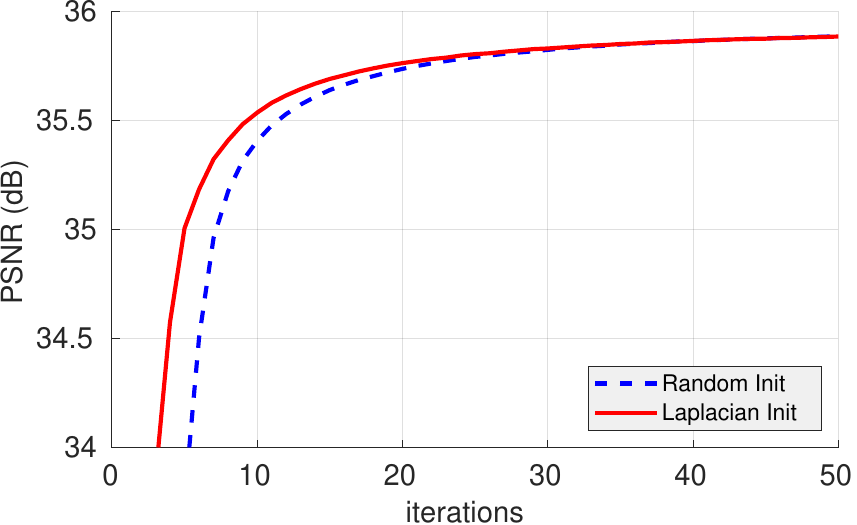}}
		\subcaption{PSNR vs iterations}
		\label{fig:spatial_vd_init_iter}
	\end{subfigure}
	\hfill
	\begin{subfigure}[b]{0.49\linewidth}
		\centering
		\centerline{\includegraphics[width=1.0\linewidth]
			{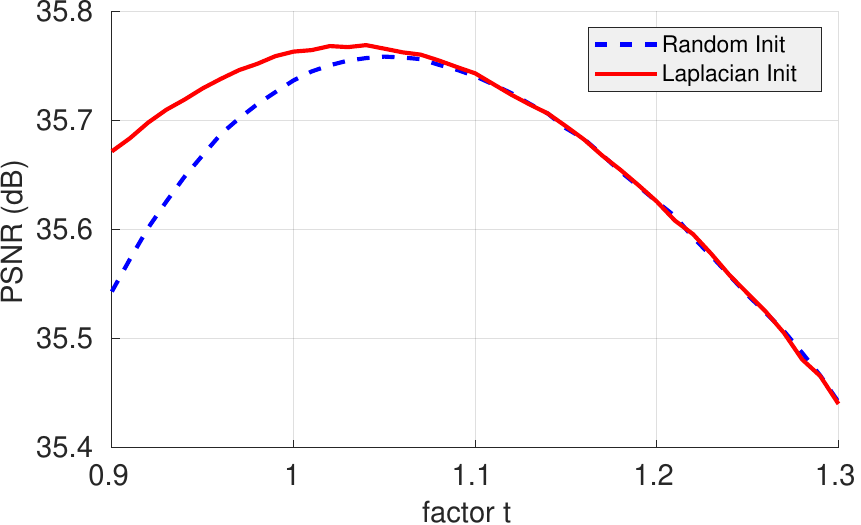}}
		\subcaption{PSNR vs factor}
		\label{fig:spatial_vd_init_factor}
	\end{subfigure}
	\smallskip
	\caption{\textbf{Comparison of Delaunay Densification 
 with Different Initial Iterations.} Results for a mask density of 5\%. 
 \textbf{(a)} A dithered Laplacian magnitude as an initial mask improves 
 the PSNR significantly compared to a uniform random initialization, 
 especially for a low number of iterations. 
 \textbf{(b)} The random initialization benefits clearly from increasing 
 the number of mask pixels added per iteration, while for the Laplacian 
 initialization a constant number of mask pixels is very close to the 
 optimum.}
	\label{fig:spatial_vd_init}
\end{figure}

\paragraph{Initial Mask and Iteration Count}
In \cref{fig:spatial_vd_init_iter} we evaluate the impact of the choice 
of the initial mask on our Delaunay densification. We compare a uniform 
random initial mask with a Laplacian magnitude sampled initial mask. 
We use the same number of initial mask pixels and add a constant amount 
of new pixels in each iteration for both masks. We observe that for a 
small amount of iterations the Laplacian magnitude approach results in 
a significantly higher PSNR, while the difference becomes smaller the more 
iterations are used. 
This can be explained with the fact that fewer mask pixels are 
introduced per iteration if we have a larger numbers of iterations, 
which reduces the impact of the initial mask.
It also shows that the improvement per iteration becomes quite small for 
iterations beyond $20$, while the additional runtime for each iteration 
is approximately constant as each requires exactly one inpainting. 
Thus $20$ iterations seem to be a good trade-off between runtime and 
reconstruction quality. 

%-----------------------------------------------------------------------------

\paragraph{Number of Mask Pixels per Iteration}
\Cref{fig:spatial_vd_init_factor} shows the effect of increasing or 
decreasing the number of mask pixels that are added per iteration during the 
densification with a constant factor $t$. For a uniform random initial mask 
we observe that adding a constant amount of pixels in each iteration is not 
the optimal strategy. By increasing the number of added pixels by a factor 
of $t=1.08$ or $8$\% in each iteration, we can improve the PSNR by $0.05$ dB. 
With a sampled Laplacian magnitude as an initial mask, however, the optimal 
factor is only $t=1.03$ and the improvement compared to a constant number
of pixels is less than $0.01$ dB. 
These results also reinforce the conclusion that the Laplacian magnitude 
yields a suitable initial mask and suggests only slightly increasing the 
number of pixels added per iteration. Since the difference between adding 
a constant number of pixels and the optimal strategy is negligible, for 
simplicity we use a constant number of pixels for the rest of our experiments.

%-----------------------------------------------------------------------------

%-----------------------------------------------------------------------------

\begin{figure}[tb]
	\begin{subfigure}[b]{0.49\linewidth}
		\centering
		\centerline{\includegraphics[width=1.0\linewidth]
			{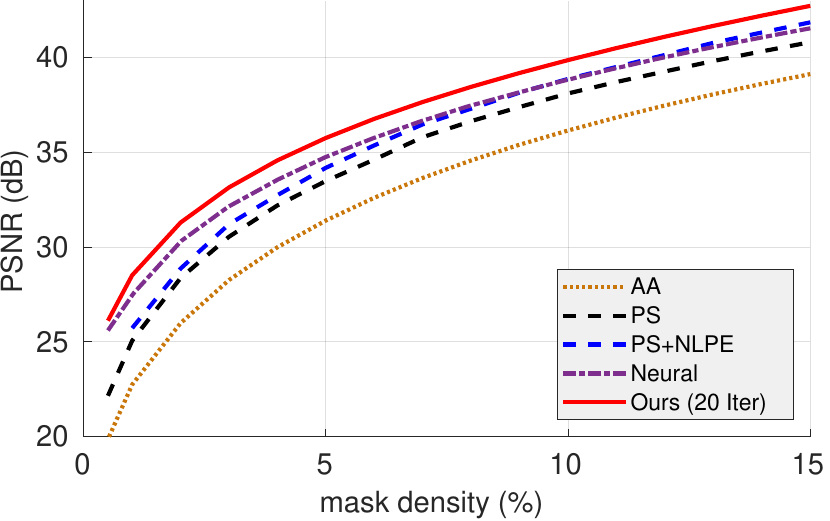}}
		\caption{quality \label{fig:spatial-comp-quality}}
	\end{subfigure}
	\hfill
	\begin{subfigure}[b]{0.49\linewidth}
		\centering
		\centerline{\includegraphics[width=1.0\linewidth]
			{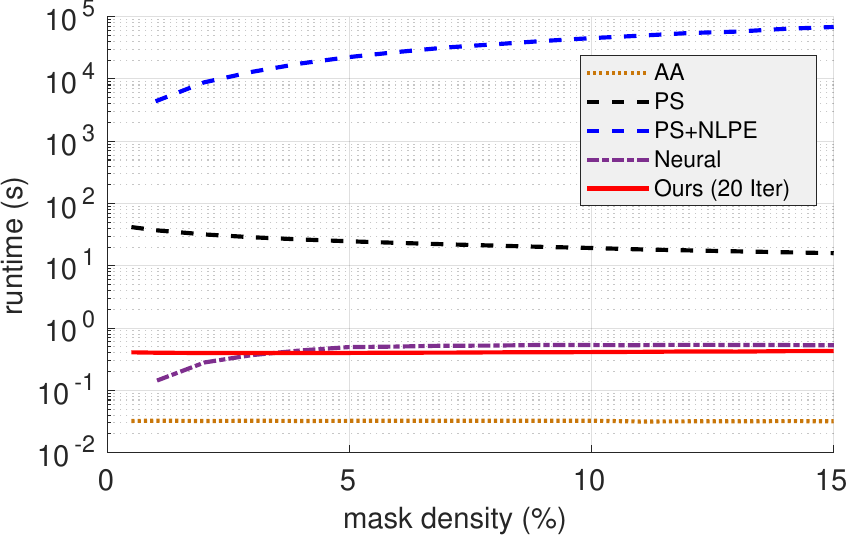}}
		\caption{runtime\label{fig:spatial-comp-runtime}} 
	\end{subfigure}
	\smallskip
	\caption{\textbf{Spatial Optimization Comparison}. 
		Whilst the analytical approach~\cite{BBBW08} is the fastest, its 
		generated masks are clearly inferior. Our Delaunay densification 
            method is only slightly slower, but consistently outperforms all 
            its competitors in terms of quality.}
	\label{fig:spatial-comp-density}
\end{figure}

\paragraph{Performance Evaluation}
We evaluate the runtimes in \cref{fig:spatial-comp-runtime}. While the 
inferior analytic approach is clearly the fastest, our method is the second 
fastest with an average runtime of around $0.4$ seconds. We achieve this by 
restricting our densification to $20$ iterations, which still produces masks 
of very high quality.
Compared to PS and PS+NLPE we are several orders of magnitude faster, since 
they both require thousands of inpaintings. 
For the neural approach we cannot generate any runtime results on our GPU due 
to memory restrictions, but the runtimes reported by 
Schrader et al.~\cite{SPKW23} are quite similar to the results of our 
approach, albeit a bit faster for densities below $3$\%. As they are obtained 
with a significantly more powerful GPU, our method should be faster on 
a similar GPU. Furthermore, our method has the advantage of needing 
significantly less GPU memory. 

%-----------------------------------------------------------------------------

\paragraph{Qualitative Evaluation}
% The qualitative evaluation of the spatial optimization methods in 
% \cref{fig:spatial-comp-quality} shows that our Delaunay densification 
% approach (DD) consistently outperforms all its competitors on all tested mask 
% densities, even with just $20$ iterations. It even surpasses the recent 
% neural approach by Schrader et al.~\cite{SPKW23} by at least $0.75$ dB. 
% Compared to the model-based competitors our method is especially well-suited 
% for lower mask densities, which are the most interesting for inpainting-based 
% compression.
\newtext{The qualitative evaluation of the spatial optimization methods in 
\cref{fig:spatial-comp-quality} shows that with just $20$ 
densification iterations our Delaunay densification approach (DD) already 
consistently outperforms all its competitors on all tested mask densities.} 
It even surpasses the recent 
neural approach by Schrader et al.~\cite{SPKW23} by at least $0.75$ dB. 
Compared to the model-based competitors our method is especially well-suited 
for lower mask densities, which are the most interesting for inpainting-based 
compression.

\textbf{Here our method outperforms the analytic approach by more than $6$ dB, 
PS by more than $3.8$, and PS+NLPE by more than $2.6$ dB.}
This suggests that densification together with an adaptation to the local 
mask density with for example the Delaunay triangulation should be preferred 
over the probabilistic sparsification approach, as it significantly 
outperforms it in terms of quality as well as runtime.

%-----------------------------------------------------------------------------

%-----------------------------------------------------------------------------
\begin{figure}[tb]

	\centering
	\centerline{\includegraphics[width=0.5\linewidth]
		{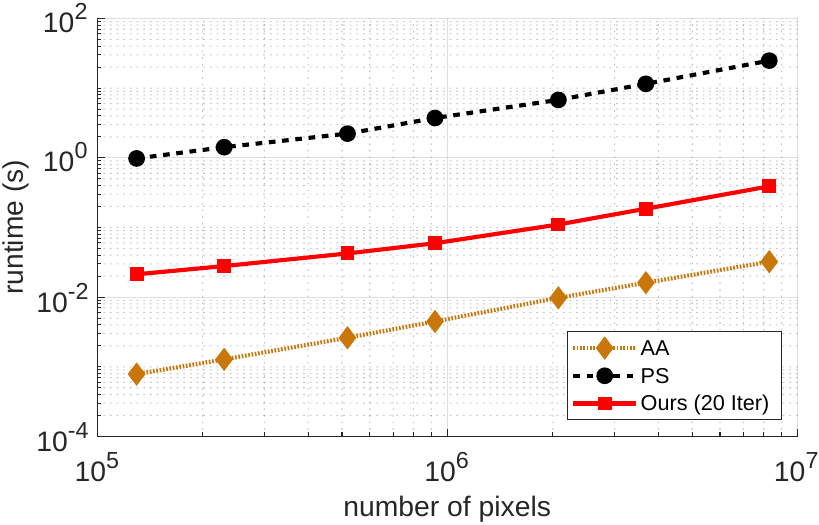}}

	\smallskip
	\caption{\textbf{Spatial Optimization Resolution Scaling Comparison}. 
		Runtime depending on the image resolution for a $5$\% mask density. 
		While the analytic approach is clearly the fastest, our method at least 
		$50$ times faster than PS for all image resolutions.}
	\label{fig:spatial-comp-scaling}
\end{figure}
%-----------------------------------------------------------------------------

\paragraph{Resolution Runtime Scaling}
We also evaluate the spatial optimization methods over different image 
resolutions ranging from $480 \times 270$ to $3840 \times 2160$. The 
results are shown in \cref{fig:spatial-comp-scaling}. 

While our Delaunay-based densification method is one order of magnitude slower 
than the qualitatively inferior analytic approach, it is at least $50$ times 
faster than PS for all image resolution.
As for the inpainting (see~\cite{KCW23}), all spatial optimization methods also
show a nearly linear behavior with respect to the number of image pixels in 
the double logarithmic plot. As the slope is approximately $1$ for all 
methods, we observe an ideal linear scaling behavior. 

%-----------------------------------------------------------------------------

\begin{figure}[hp]
\centering
\begin{tabular}{cc}
\small image \textit{lofsdalen} & \small zoom to region \\
\includegraphics[height=0.162\linewidth]{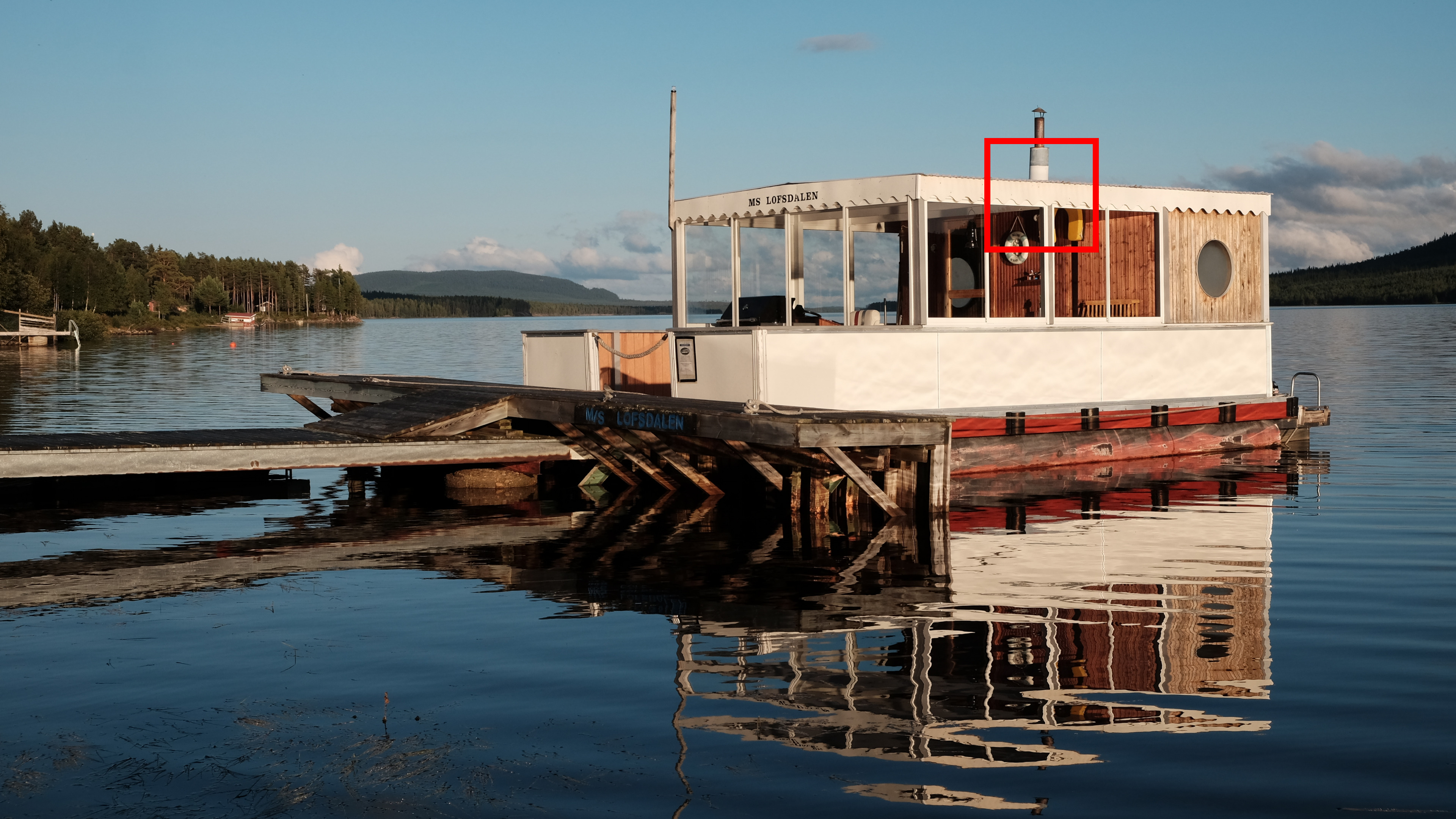}
\hspace{1mm} &
\includegraphics[height=0.162\linewidth]{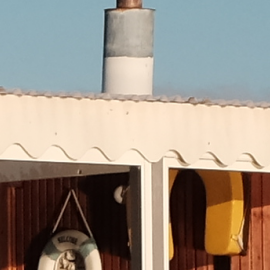}
 
\end{tabular}
\\[1mm]
\setlength{\tabcolsep}{0.8mm}
\begin{tabular}{cccccc}
 & \small mask & \small inpainted &\small  mask (crop) & \small  inp. (crop) & 
 \\

\centered{\rotatebox{90}{\small AA~\cite{BBBW08}}} & 
\centered{\includegraphics[height=0.162\linewidth]
  {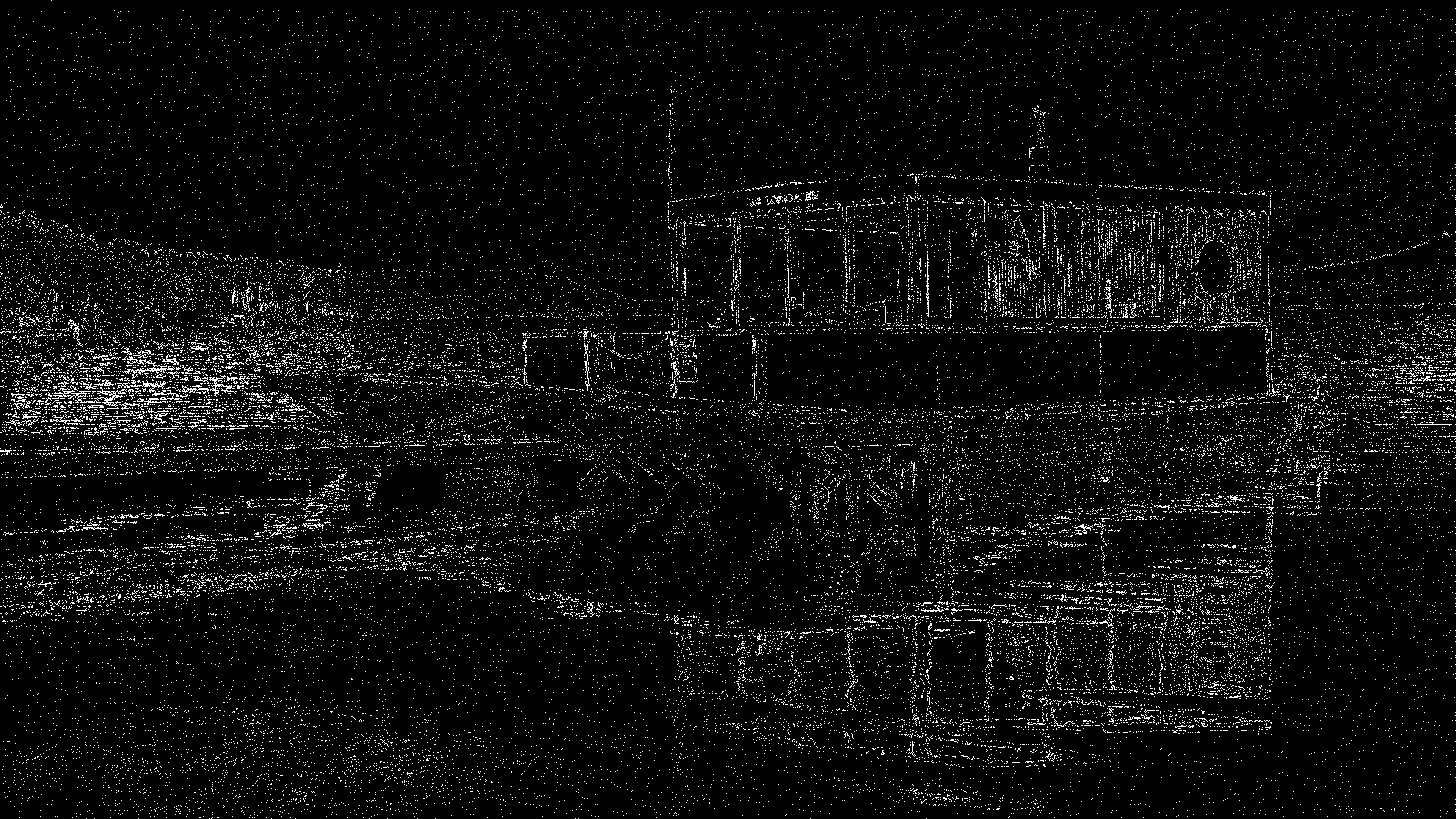}} &
\centered{\includegraphics[height=0.162\linewidth]
  {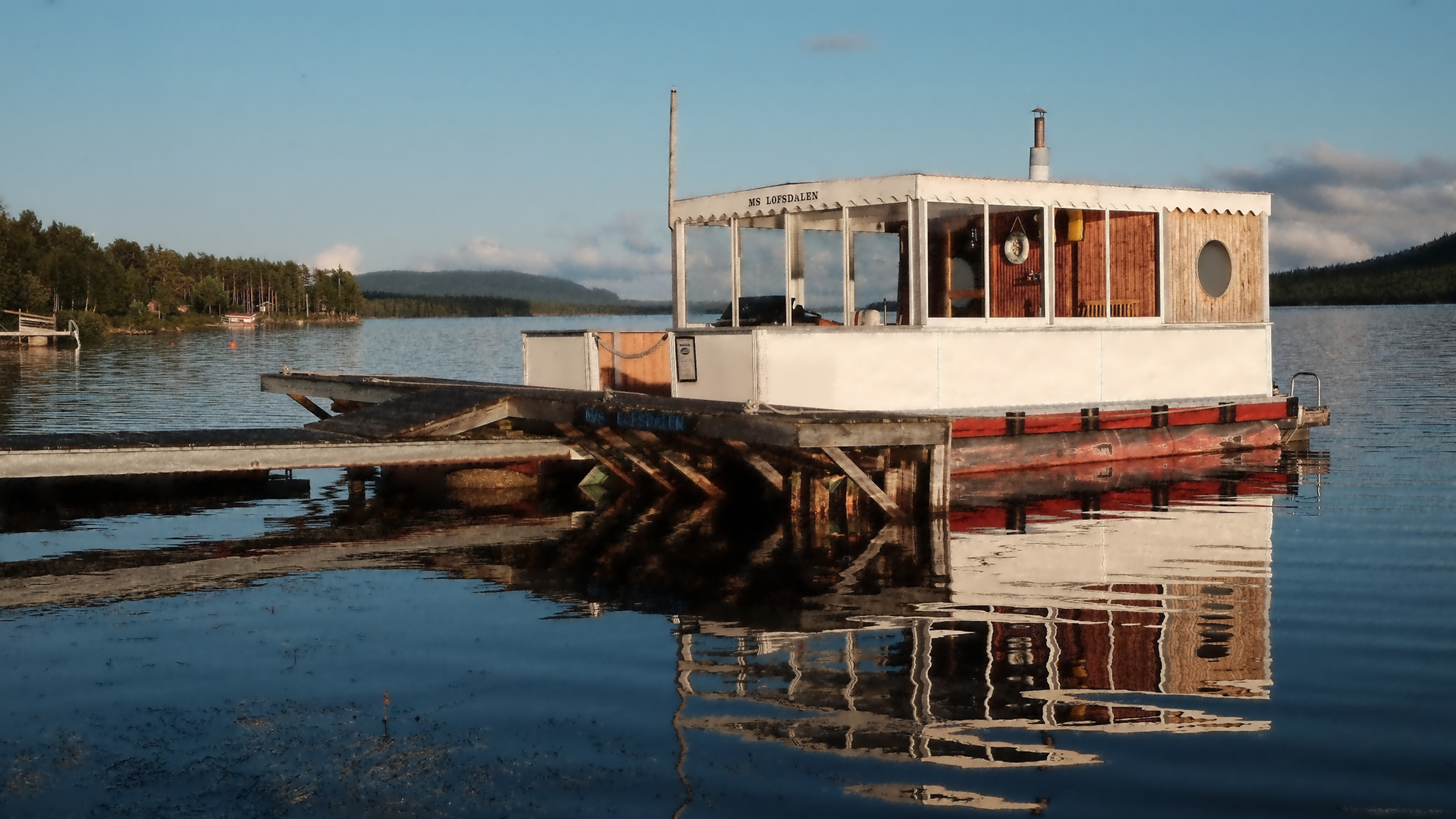}} &
\centered{\includegraphics[height=0.162\linewidth]
  {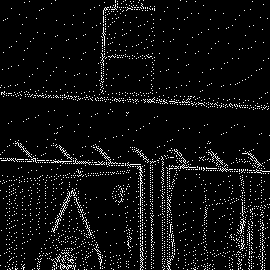}} &
\centered{\includegraphics[height=0.162\linewidth]
  {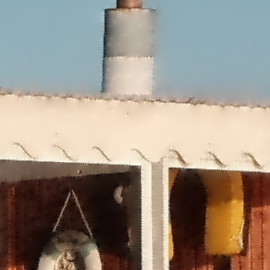}} &
\centered{\rotatebox{90}{\small PSNR: 30.42}} \\[0mm] 
  
\centered{\rotatebox{90}{\small PS~\cite{MHWT12}}} & 
\centered{\includegraphics[height=0.162\linewidth]
  {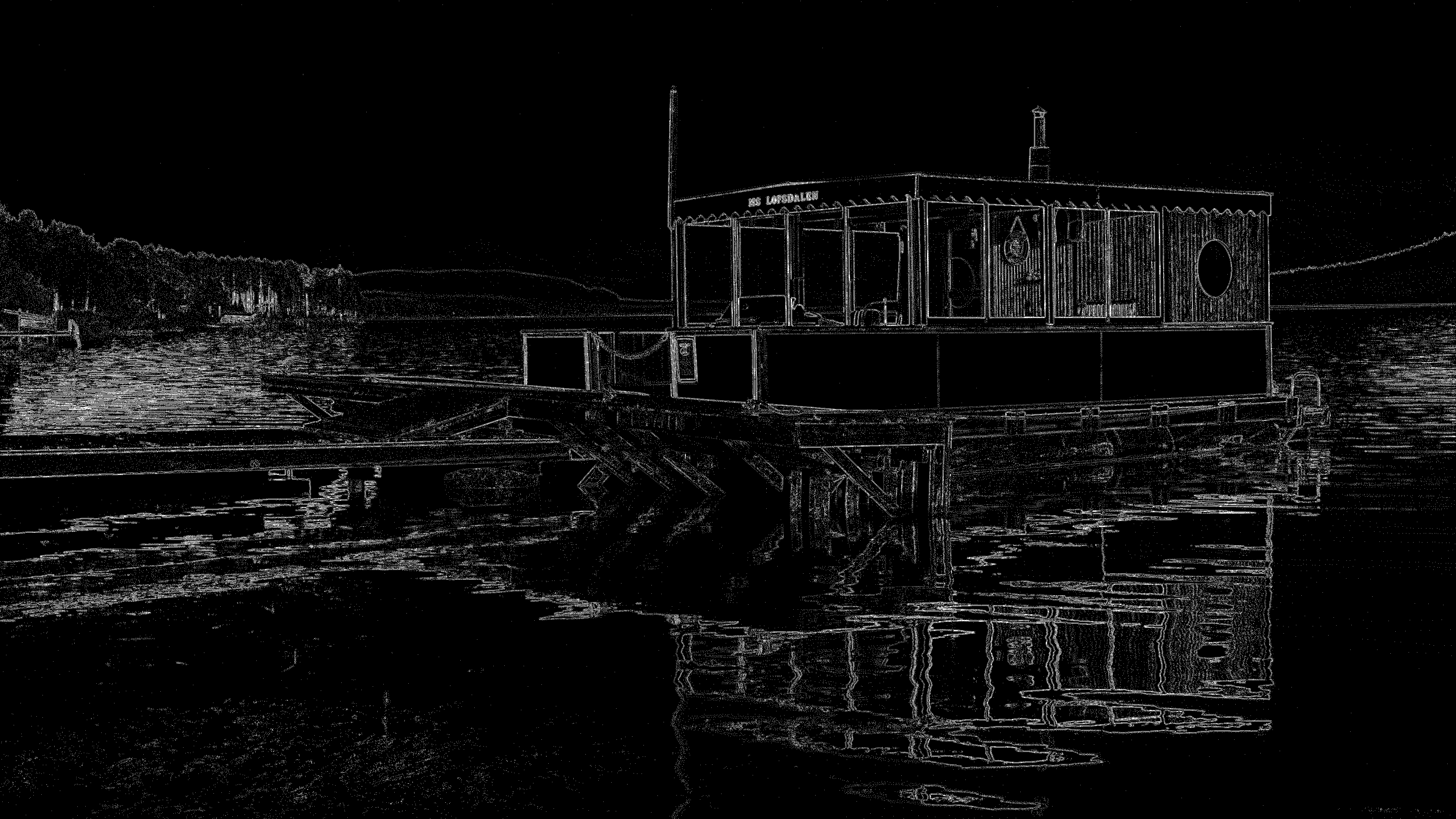}} &
\centered{\includegraphics[height=0.162\linewidth]
  {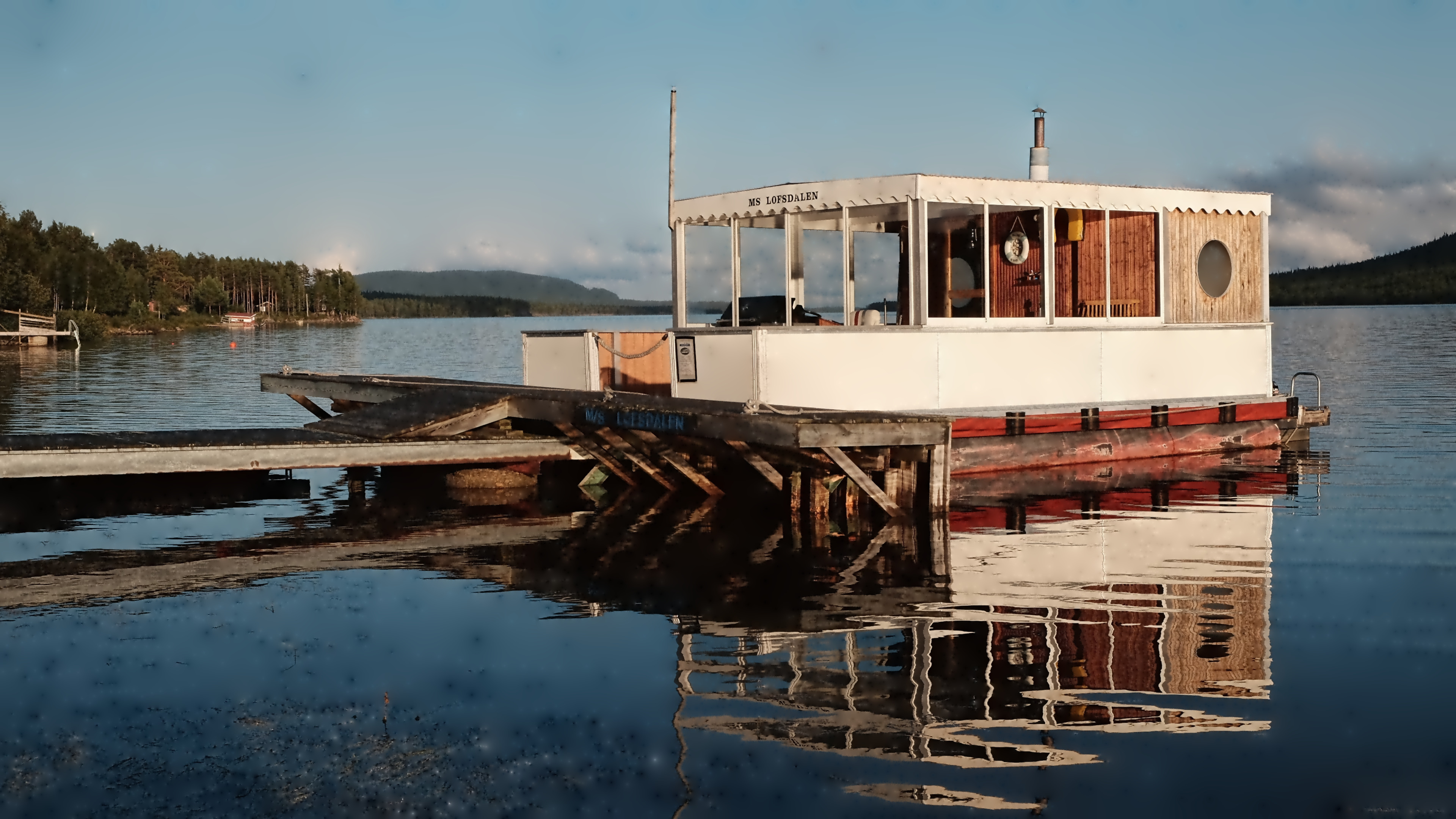}} &
\centered{\includegraphics[height=0.162\linewidth]
  {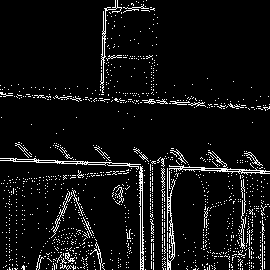}} &
\centered{\includegraphics[height=0.162\linewidth]
  {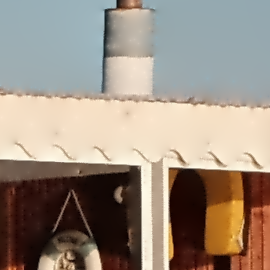}} & 
\centered{\rotatebox{90}{\small PSNR: 31.17}} \\[0mm] 

\centered{\rotatebox{90}{\small PS+NLPE~\cite{MHWT12}}} & 
\centered{\includegraphics[height=0.162\linewidth]
  {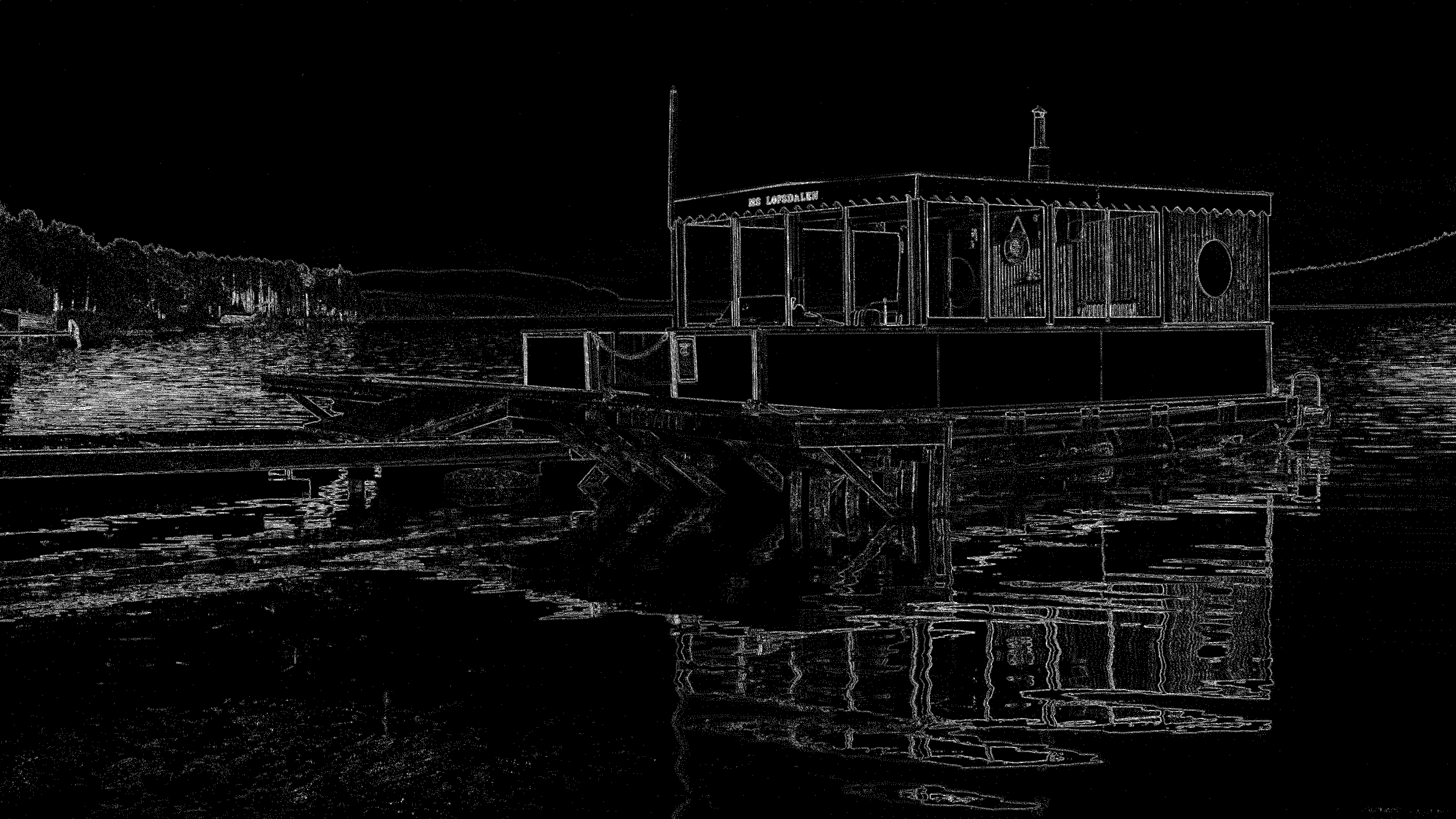}} &
\centered{\includegraphics[height=0.162\linewidth]
  {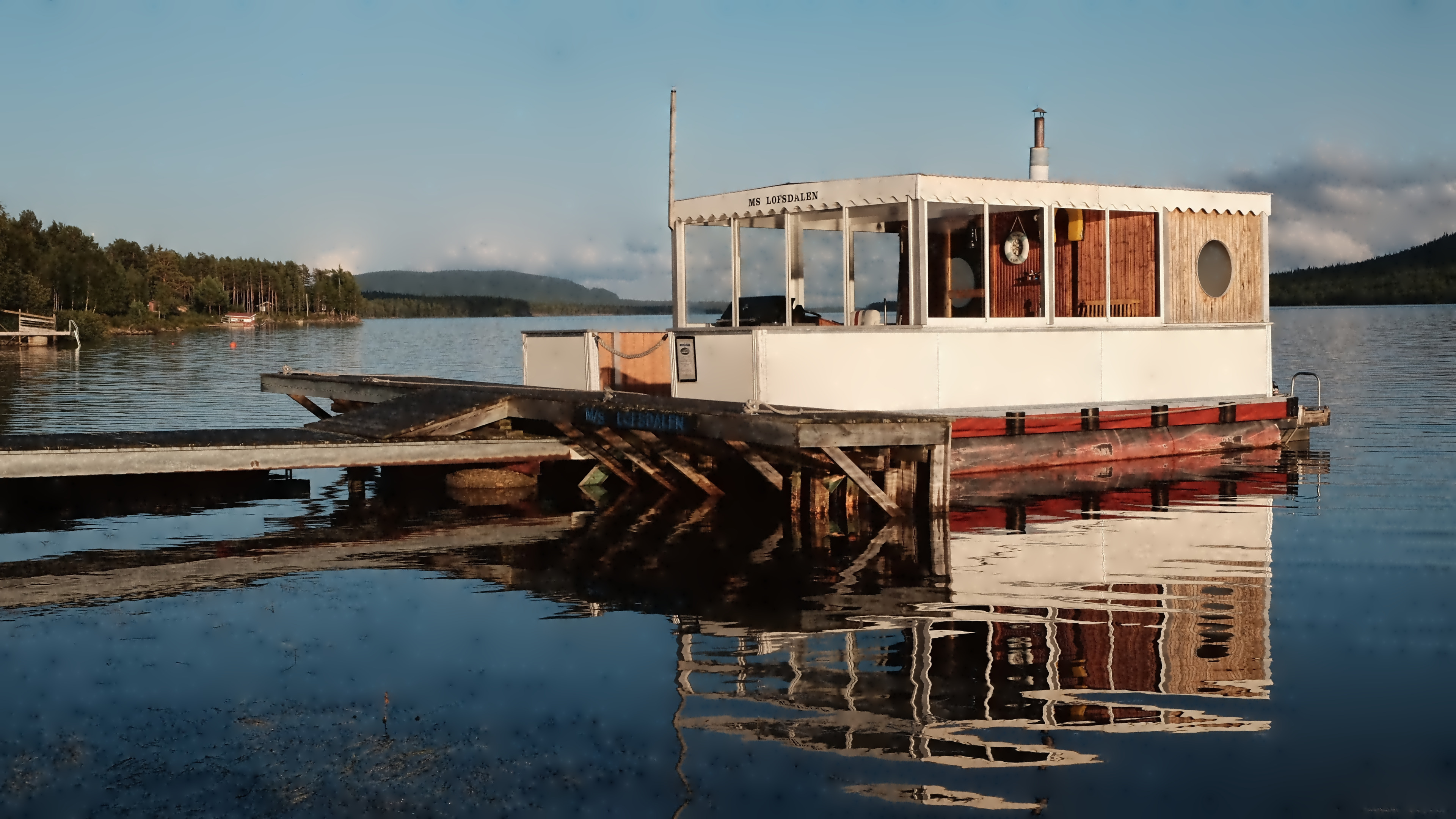}} &
\centered{\includegraphics[height=0.162\linewidth]
  {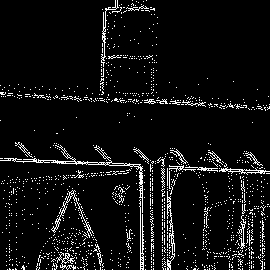}} &
\centered{\includegraphics[height=0.162\linewidth]
  {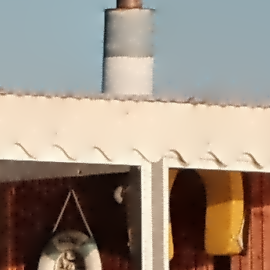}} & 
\centered{\rotatebox{90}{\small PSNR: 31.88}}\\[0mm] 

\centered{\rotatebox{90}{\small  Neural~\cite{SPKW23}}} & 
\centered{\includegraphics[height=0.162\linewidth]
{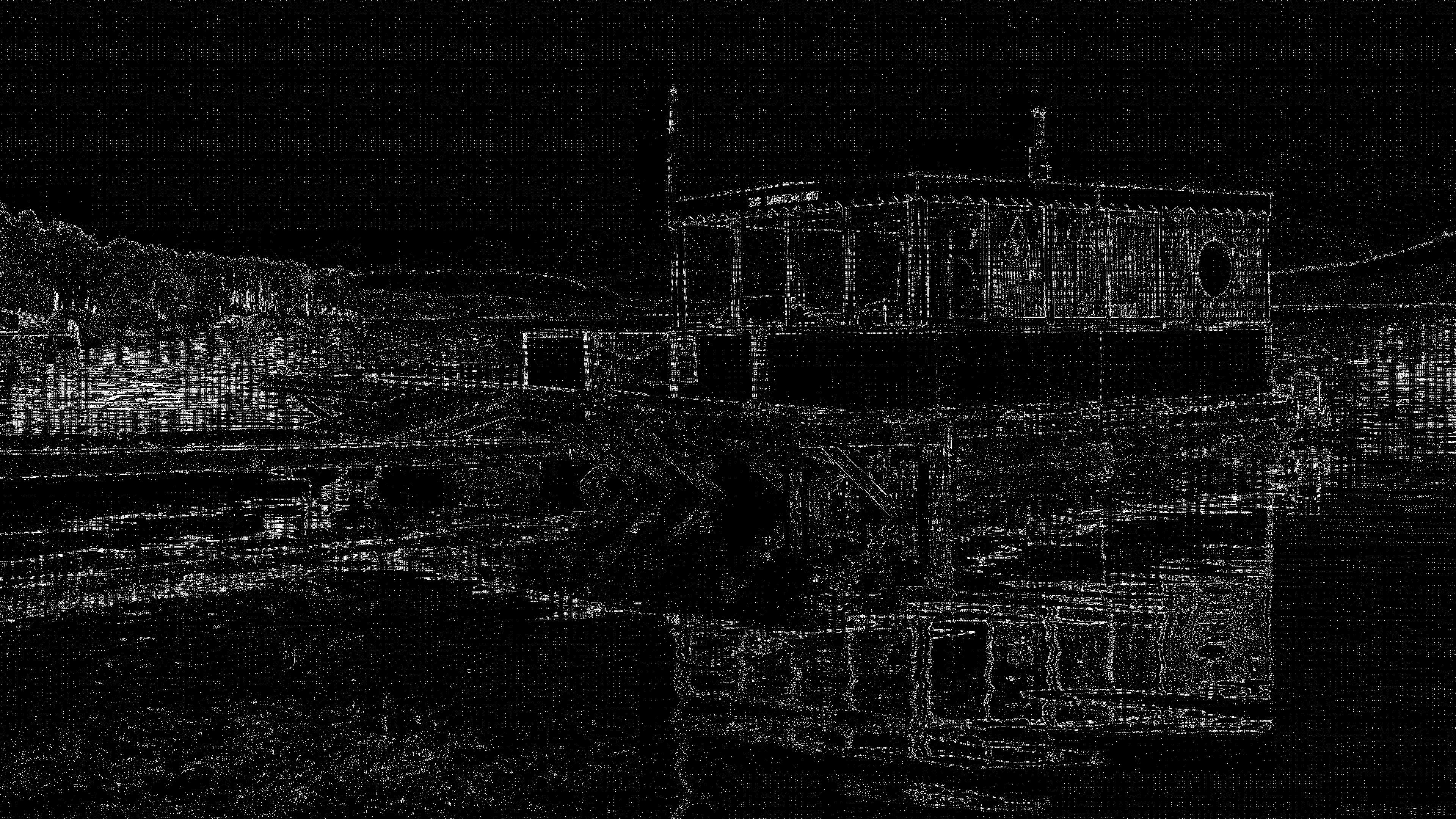}} &
\centered{\includegraphics[height=0.162\linewidth]
{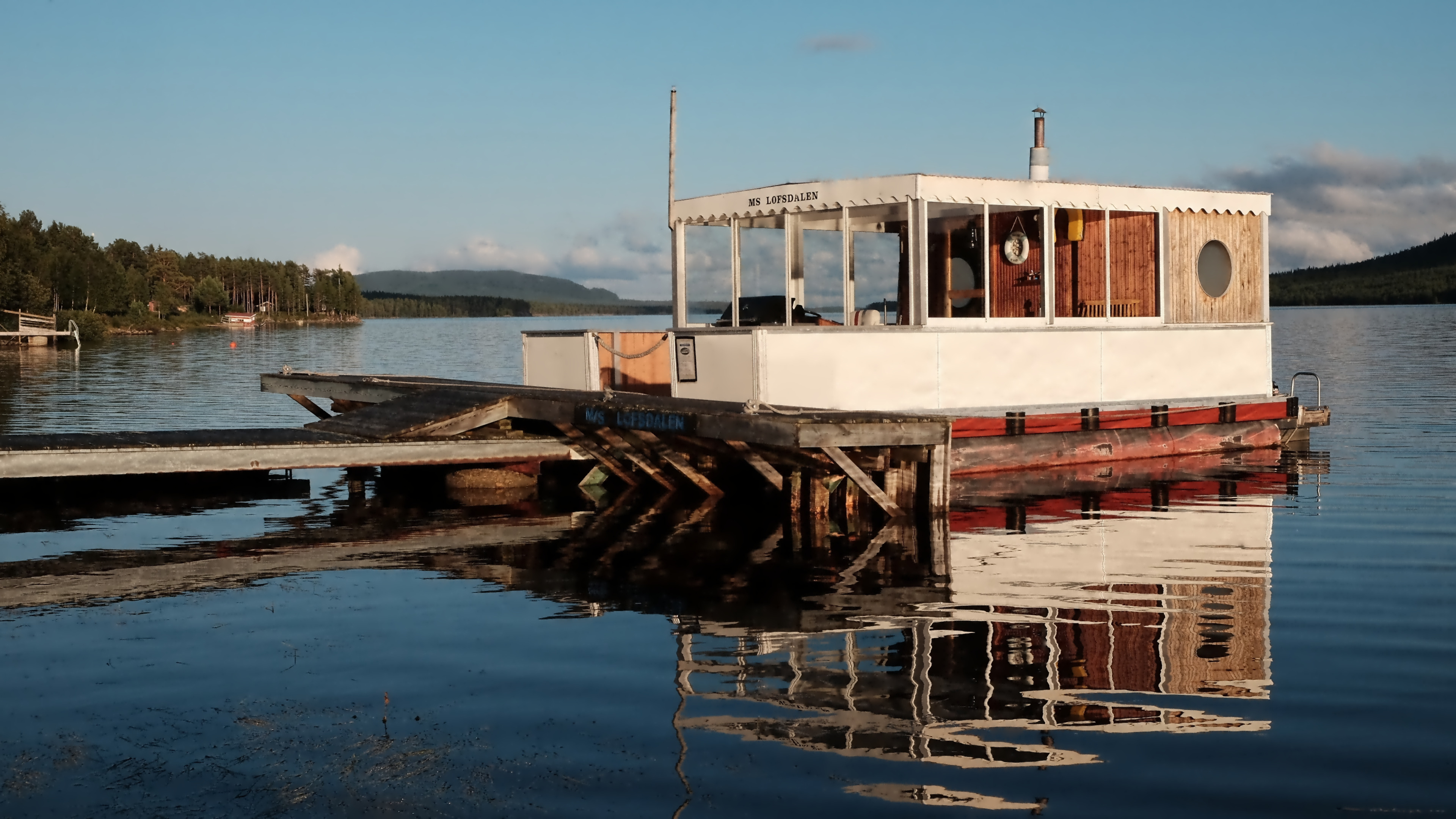}} &
\centered{\includegraphics[height=0.162\linewidth]
{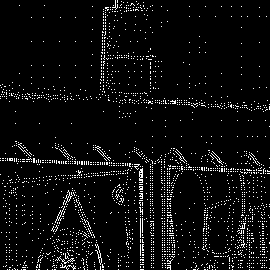}} &
\centered{\includegraphics[height=0.162\linewidth]
{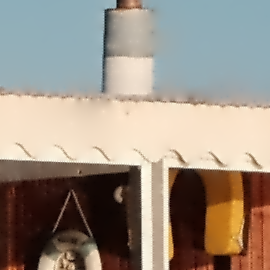}} & 
\centered{\rotatebox{90}{\small PSNR: 33.36}} \\[0mm] 
  
\centered{\rotatebox{90}{\small \textbf{Ours (20 Iter)}}} & 
\centered{\includegraphics[height=0.162\linewidth]
  {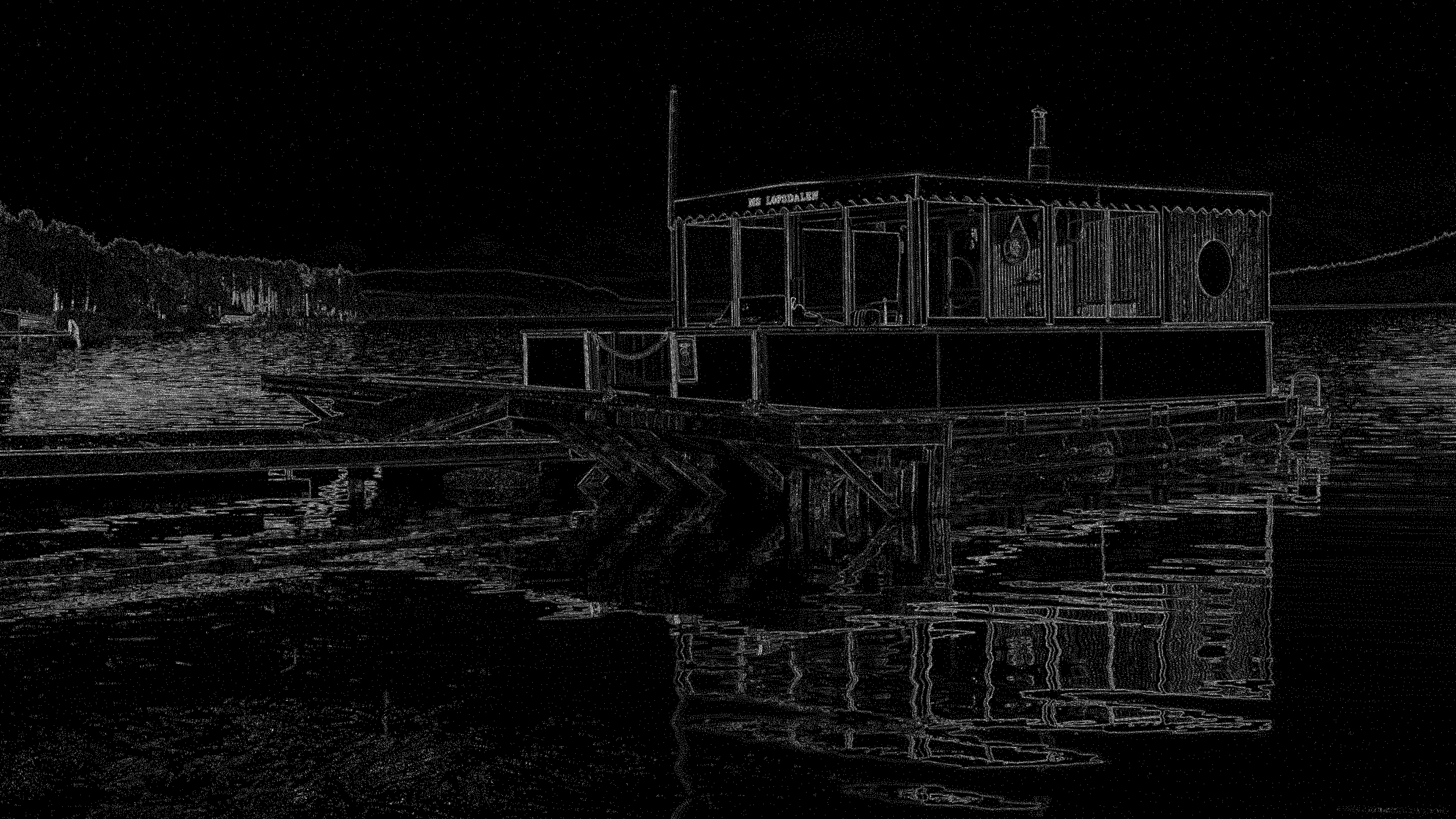}} &
\centered{\includegraphics[height=0.162\linewidth]
  {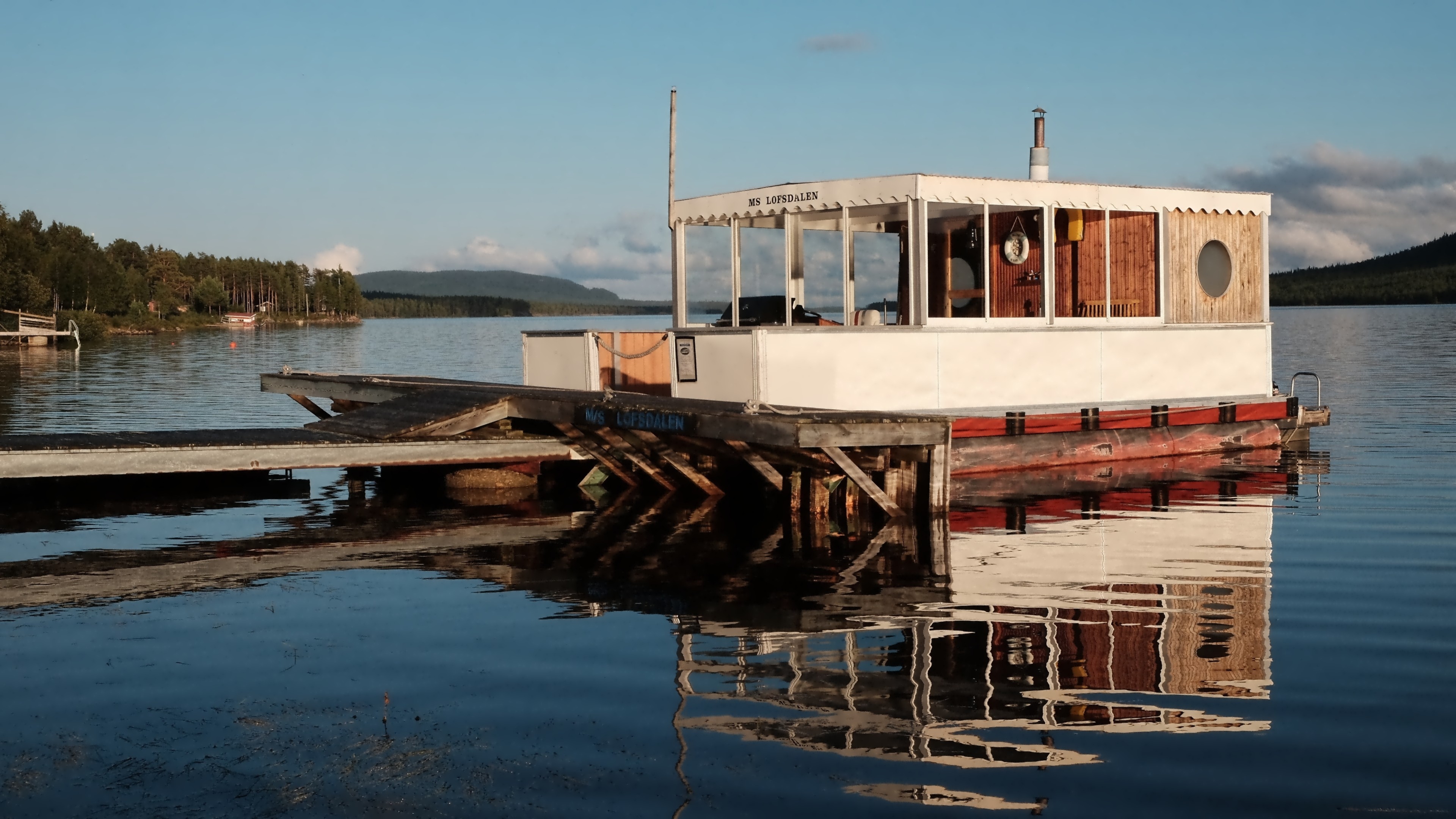}} &
\centered{\includegraphics[height=0.162\linewidth]
  {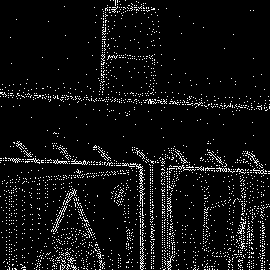}} &
\centered{\includegraphics[height=0.162\linewidth]
  {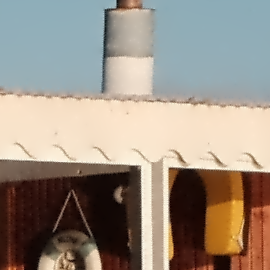}} & 
\centered{\rotatebox{90}{\small \textbf{PSNR: 34.31}}} \\[0mm] 

\end{tabular}\vspace{-1mm}
\caption{\textbf{Visual Comparison of Spatial Optimization for 5\% Mask Density 
on Image 
\textit{lofsdalen}.} 
PSNRs are for the whole image. Notice the blurry edges for the analytic 
approach and the discoloration present in the sky for PS and PS+NLPE.  
\label{fig:spatial_visual_comp}}
% \vspace{1.5cm}
\vspace{2cm}
\end{figure}

\paragraph{Visual Comparison}
A visual comparison of the different spatial optimization methods for a 
$5$\% mask density on the image \textit{lofsdalen} is given in 
\cref{fig:spatial_visual_comp}. We observe that the analytic and to some 
extent also the neural approach lead to slightly blurrier edges compared to 
the other methods. While PS and PS+NLPE result in similarly sharp edges as 
our Delaunay densification, they can also lead to wrong colors in homogeneous 
areas, e.g.\ in the sky. This happens as PS generates too few mask-pixels in 
homogeneous areas, due to only considering pixel-wise errors.

%-----------------------------------------------------------------------------

%-----------------------------------------------------------------------------
\subsection{Tonal Optimization Experiments}
\label{sec:exp_tonal_opt}

\begin{figure}[tbh!]
	\centering
	\setlength{\tabcolsep}{1mm}
	\begin{tabular}{ccc}
		\small original & \small inpainted & 
        \small \textbf{tonally optimized} \\
		
		\includegraphics[width=0.32\linewidth]
		{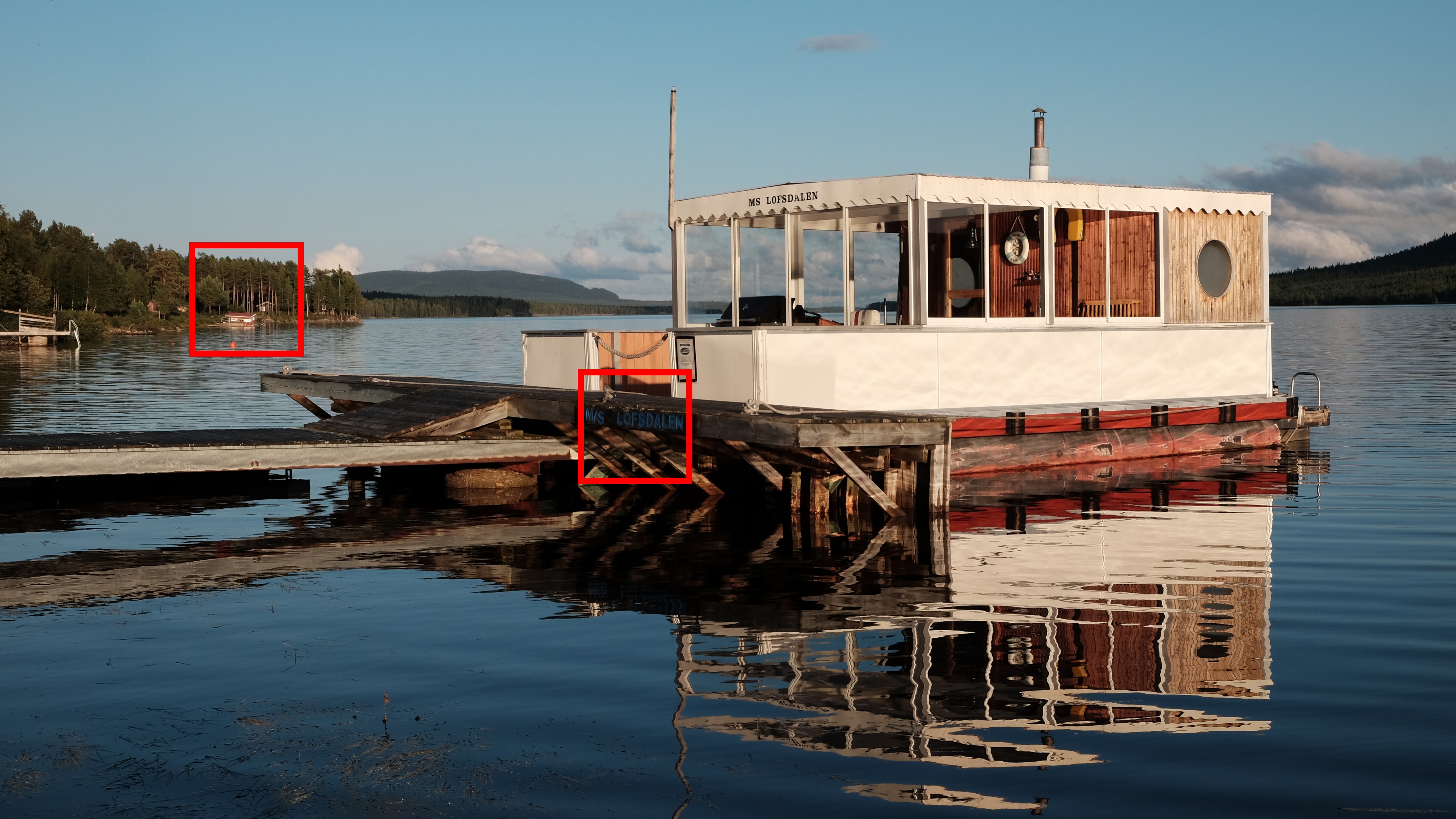} &
		\includegraphics[width=0.32\linewidth]
		{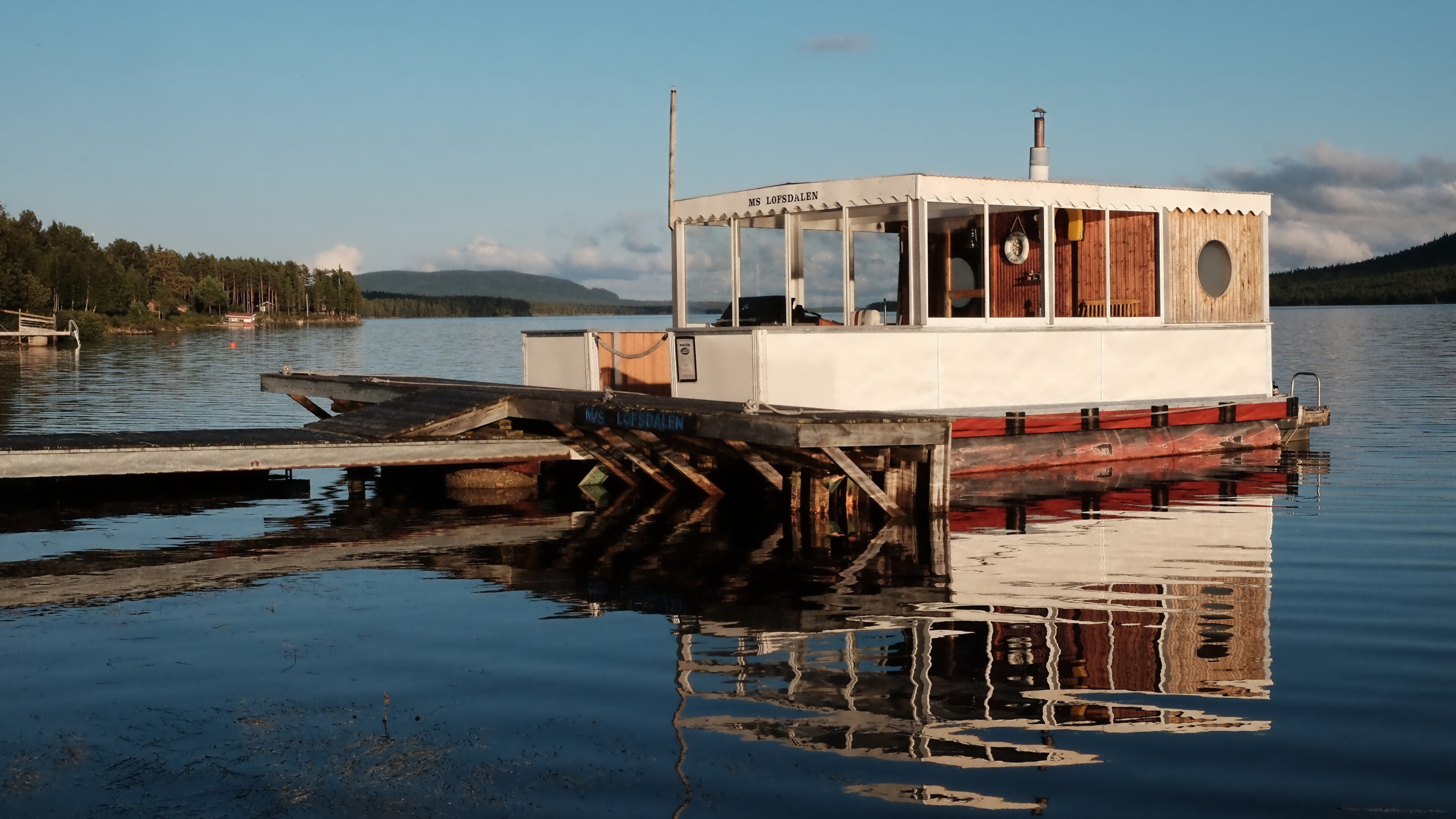} &
		\includegraphics[width=0.32\linewidth]
		{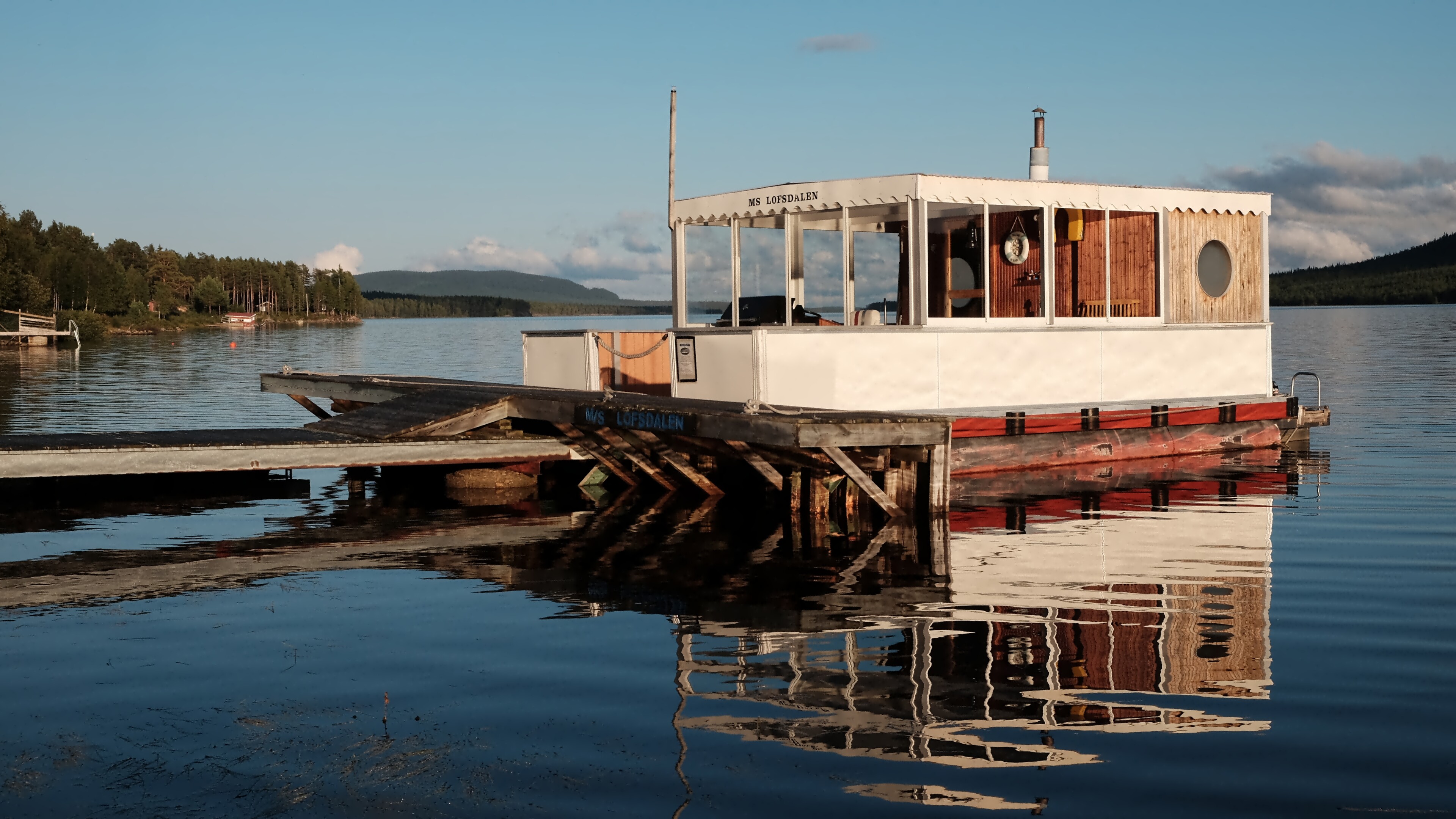} 
		\\[0.5mm] 
		
		\includegraphics[width=0.32\linewidth]
		{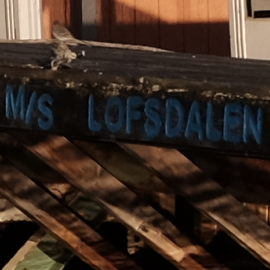} &
		\includegraphics[width=0.32\linewidth]
		{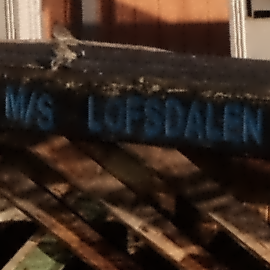} &
		\includegraphics[width=0.32\linewidth]
		{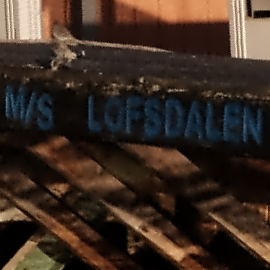} 
		\\[0mm] 
		
		\includegraphics[width=0.32\linewidth]
		{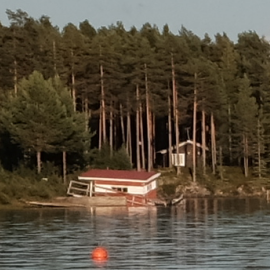} &
		\includegraphics[width=0.32\linewidth]
		{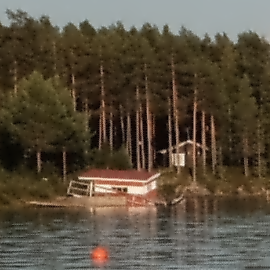} &
		\includegraphics[width=0.32\linewidth]
		{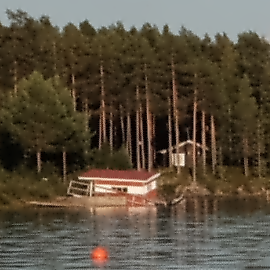} 
		\\[0mm] 

		 & \small PSNR: 34.31 &\small  \textbf{PSNR: 35.66} \\
		
	\end{tabular}\vspace{-1mm}
 	\caption{\textbf{Visual Comparison of Tonal Optimization for 
        $5$\% Mask Density on Image \textit{lofsdalen}.} 
	PSNRs are for the whole image. Notice the enhanced contrast around the 
    letters and the improved colors in the tonally optimized inpainting. 
		\label{fig:tonal_visual_comp}}
\end{figure}

A visual example of tonal optimization is shown in 
\cref{fig:tonal_visual_comp} for the 4K image \textit{lofsdalen} with a 
$5$\% inpainting mask obtained with our Delaunay densification spatial 
optimization (see \cref{fig:spatial_visual_comp}). We can observe that the 
tonal optimization increases the overall contrast, which leads to an 
improvement of more than $1$ dB.

%-----------------------------------------------------------------------------

\begin{figure}[tb]
	\begin{subfigure}[b]{0.49\linewidth}
		\centering
		\centerline{\includegraphics[width=1.0\linewidth]
			{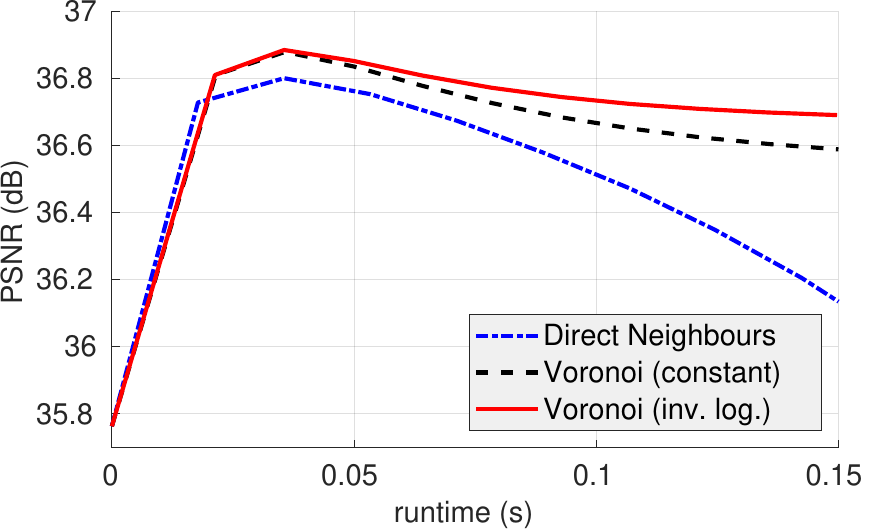}}
		\caption{tonal initialization \label{fig:tonal-comp-init}}
	\end{subfigure}
	\hfill
	\begin{subfigure}[b]{0.49\linewidth}
		\centering
		\centerline{\includegraphics[width=1.0\linewidth]
			{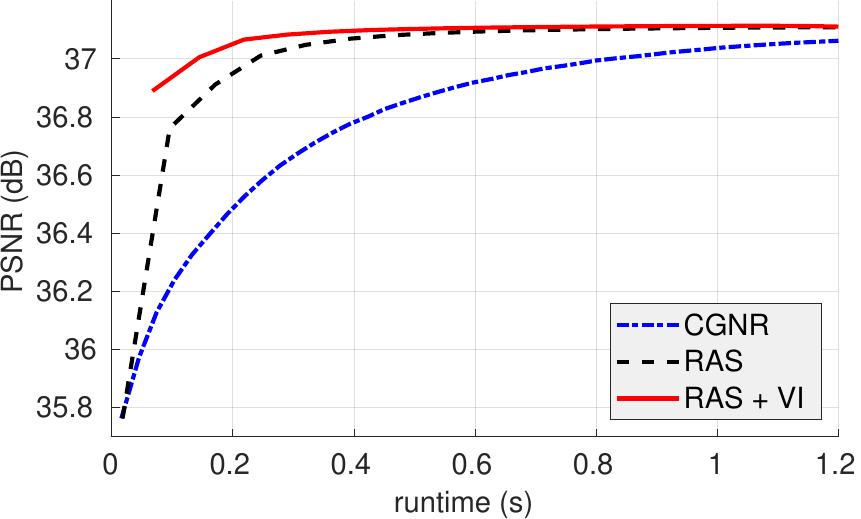}}
		\caption{full tonal optimization \label{fig:tonal-comp-full}} 
	\end{subfigure}
	\smallskip
	\caption{\textbf{Tonal Optimization Comparison for a $5$\% Mask Density}. 
		\textbf{(a)} The Voronoi-based initialization approach clearly 
		outperforms the neighbor-based approach. Using an inverse logarithmic 
		weighting improves the quality slightly compared to a constant one. 
		\textbf{(b)} Our RAS domain decomposition solver converges 
		significantly faster than the simpler CGNR solver. Together with our 
		Voronoi initialization (VI) RAS shows the best performance.}
	\label{fig:tonal-comp}
\end{figure}

%-----------------------------------------------------------------------------
\paragraph{Voronoi Initialization}
Before we present results for the full tonal optimization framework, we 
evaluate the performance of our Voronoi initialization. 
\cref{fig:tonal-comp-init} shows a comparison with the iterated direct 
neighbor initialization based on~\cite{GWWB08}. 
For our Voronoi initialization 
we show results for two different weighting functions: a constant weighting 
and an inverse logarithmic distance weighting. We observe that all methods 
significantly improve the PSNR within a few iterations by around $1$ dB. 
We can see that the PSNR starts to decrease after a few iterations, since we 
only solve a surrogate problem. Thus, it is necessary to stop the iteration 
once the PSNR begins to decrease and use the tonal values from the best 
iteration as the final result. 
Each iteration needs exactly one inpainting that takes around $15.4$ ms, 
except for the very first iteration of the Voronoi approaches which needs to 
construct the Voronoi diagram, which takes around $8$ ms. 
We can see that both Voronoi approaches outperform the direct neighbor 
approach by about $0.1$ dB, while being $8$ ms slower. The difference 
between the constant weighting and the inverse logarithmic distance weighting 
is, however, small. Since the weighting has no effect on the runtime, 
we choose the inverse logarithmic distance weighting. Overall, we can observe 
that the Voronoi diagram seems to be a better approximation of the mask pixel 
influence areas than a simple direct neighborhood.

%-----------------------------------------------------------------------------

%-----------------------------------------------------------------------------

In order to quantify the effect of our Voronoi approach on the 
tonal optimization problem, we perform an experiment using our RAS tonal 
optimization method with and without the Voronoi initialization. 
Additionally, we compare against a CGNR solver with the same multigrid ORAS 
inpainting. 
\cref{fig:tonal-comp-full} shows the result for a $5$\% mask density.
For a better visualization, we have omitted the results for the CGNR 
solver with CG inpainting, as each iteration needs approximately $0.8$ 
seconds, which makes it clearly inferior to the other methods.
We can observe that RAS converges significantly faster to the 
optimal PSNR compared to the CGNR method. When we use RAS together 
with the Voronoi initialization, we can stop the tonal optimization much 
earlier, even though the initial iteration takes longer. As the tonal 
optimization is a convex least squares problem, all tested methods 
eventually converge to the same optimal PSNR. Therefore, we omit a 
comparison of the reconstruction quality, as all methods end up with the 
same solution. The experiment also shows that the Voronoi initialization 
alone almost reaches the optimal PSNR up to $0.2$ dB. \textbf{This proves 
that the Voronoi initialization is an essential tool for tonal optimization.}

\begin{figure}[tb]
	\begin{subfigure}[b]{0.49\linewidth}
		\centering
		\centerline{\includegraphics[width=1.0\linewidth]
			{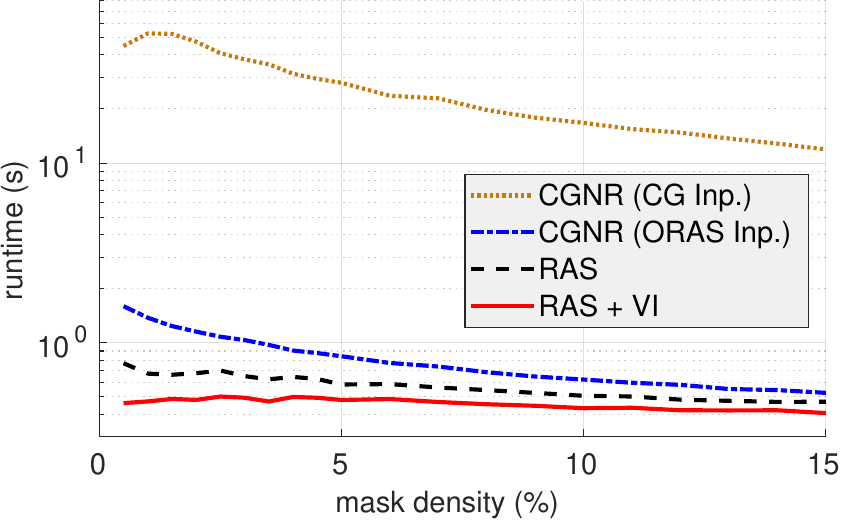}}
		\caption{runtime vs densities \label{fig:tonal-comp-density}}
	\end{subfigure}
	\hfill
	\begin{subfigure}[b]{0.49\linewidth}
		\centering
		\centerline{\includegraphics[width=1.0\linewidth]
			{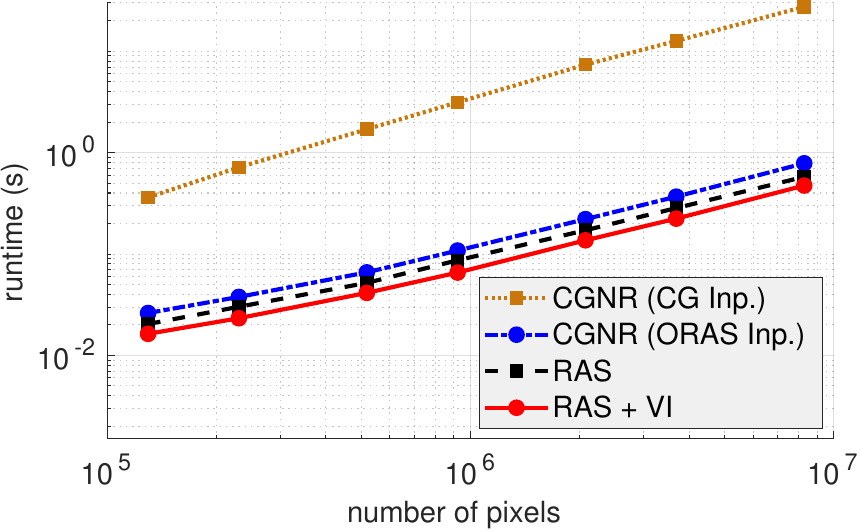}}
		\caption{runtime vs resolution\label{fig:tonal-comp-scaling}} 
	\end{subfigure}
	\smallskip
	\caption{\textbf{Runtime Comparisons of the Tonal Optimization}. 
		\textbf{(a)} Runtime depending on the mask density. Tonal optimization 
		for lower mask densities takes longer for all methods without our 
		Voronoi initialization (VI). RAS+VI is more than one order of 
		magnitude faster compared to nested CG for all densities.
		\textbf{(b)} Runtime depending on the number of pixels (double 
		logarithmic plot).}
	\label{fig:tonal-comp-scaling-density}
\end{figure}

%-----------------------------------------------------------------------------
\paragraph{Runtime Scaling}
We evaluate our tonal optimization methods on multiple mask densities ranging 
from $0.5$\% up to $15$\%, which covers the density range for which 
homogeneous diffusion inpainting is practically relevant. As a stopping 
criterion for the tonal optimization, we use the relative improvement of the 
MSE. We stop when the MSE improves by less than $0.1$\%. The results are 
shown in \cref{fig:tonal-comp-density}.
We see that for all mask densities our RAS method with a Voronoi 
initialization outperforms CGNR with CG inpainting by more than one order of 
magnitude, and is clearly faster than CGNR with ORAS inpainting and the RAS 
method without the Voronoi initialization.

It also shows that all methods need more time for lower densities than for 
higher ones. This has two reasons: The runtime of the inpaintings within the 
tonal optimization depends on the mask density, as seen in~\cite{KCW23}; but 
we also need fewer tonal optimization iterations the larger the density 
becomes, since the influence of pixels is more localized.

We also evaluate the tonal optimization methods over different image 
resolutions ranging from $480 \times 270$ to $3840 \times 2160$. The 
results are shown in \cref{fig:tonal-comp-scaling}. 
We see that for all image resolutions our RAS method with Voronoi 
initialization is at least one order of magnitude faster compared to CGNR 
with CG inpainting and around $60$\% faster than CGNR with ORAS multigrid 
inpainting.
As for the inpainting (see~\cite{KCW23}) and also the spatial optimization 
methods, all tonal optimization methods show a nearly linear behavior with 
respect to the number of image pixels in the double logarithmic plot with a 
slope of approximately $1$.
\textbf{This reveals that the tonal optimization problem has an underlying 
linear time complexity.}

%-----------------------------------------------------------------------------

%%%%%%%%%%%%%%%%%%%%%%%%%%%%%%%%%%%%%%%%%%%%%%%%%%%%%%%%%%%%%%%%%%%%%%%%%%%%%%
\section{Conclusions and Outlook}
\label{sec:conclusions}

% Besides the computation of the inpainting itself, which we already tackled in 
% the first part of our companion paper, inpainting-based compression also 
% requires the optimization of the inpainting data. 
Inpainting-based compression does not only require a good inpainting method but also methods for the optimization of the inpainting data.
The optimization of the 
spatial data as well as the tonal data for homogeneous diffusion inpainting 
is a challenging task, and previous approaches were either slow or of 
inferior quality.
As a remedy, we developed new methods for the data optimization of 
homogeneous diffusion inpainting that outperform previous approaches by a 
wide margin. Furthermore, most prior approaches could handle only relatively 
small images, while we are able to work with images of 4K resolution and higher 
with optimal linear resolution to runtime scaling.
We achieved this by carefully adapting some of the most successful numerical 
concepts and transferring them to an image compression application.
For our spatial optimization method we combined a mask densification approach 
with error-map dithering ideas using a Delaunay triangulation. 
For the tonal optimization we adopted a domain decomposition method that 
solves the corresponding normal equations in a matrix-free fashion.
Additionally we proposed a Voronoi-based initialization strategy that allows 
us to stop the tonal optimization much earlier.
This allows us to create high quality inpainting masks and optimized tonal 
values in a runtime of less than a second on a contemporary GPU. This 
constitutes a significant improvement over any comparable previous approaches 
and suggests a similar order of runtime efficiency as JPEG2000, while solving 
a numerically much more challenging problem. \newtext{We note, however, 
that the runtimes in our approach are for the data optimization and 
do not include an encoding step currently.
Together with a real-time capable inpainting (see~\cite{KCW23}),} 
we presented a basis for inpainting-based compression 
methods that may serve as serious alternatives to classical transform-based 
codecs, both in terms of quality as well as runtimes.

% In our ongoing research we are working on extending our spatial and tonal 
% optimization approaches to more sophisticated inpainting operators, such as 
% anisotropic nonlinear diffusion. The latter has been shown to qualitatively 
% outperform homogeneous diffusion.

\newtext{
So far we only considered the data optimization for homogeneous 
diffusion inpainting. However, many of our proposed methods can be 
extended to more sophisticated operators, such as anisotropic nonlinear
diffusion that are able to qualitatively outperform homogeneous diffusion.
Our Delaunay-based spatial optimization and our Voronoi-based 
initialization strategy for the tonal optimization are fairly general and 
can be used with most inpainting operators. 
In our ongoing research we investigate how well these methods 
perform with different inpainting operators and how they can be improved.}

%%%%%%%%%%%%%%%%%%%%%%%%%%%%%%%%%%%%%%%%%%%%%%%%%%%%%%%%%%%%%%%%%%%%%%%%%%%%%%

% ---- Bibliography ----
\bibliographystyle{siamplain}
\bibliography{myrefs,additional_refs}
	
\end{document}